\documentclass[acmsmall,screen,nonacm]{acmart}

\usepackage[utf8]{inputenc}
\usepackage[T1]{fontenc}  
\usepackage{natbib}
\usepackage[inline]{enumitem}

\usepackage{algorithmic}
\usepackage{graphicx}
\usepackage{xcolor}
\usepackage{xspace}
\usepackage{listings}

\makeatletter
\let\old@lstKV@SwitchCases\lstKV@SwitchCases
\def\lstKV@SwitchCases#1#2#3{}
\makeatother
\usepackage{lstlinebgrd}
\makeatletter
\let\lstKV@SwitchCases\old@lstKV@SwitchCases

\lst@Key{numbers}{none}{%
    \def\lst@PlaceNumber{\lst@linebgrd}%
    \lstKV@SwitchCases{#1}%
    {none:\\%
     left:\def\lst@PlaceNumber{\llap{\normalfont
                \lst@numberstyle{\thelstnumber}\kern\lst@numbersep}\lst@linebgrd}\\%
     right:\def\lst@PlaceNumber{\rlap{\normalfont
                \kern\linewidth \kern\lst@numbersep
                \lst@numberstyle{\thelstnumber}}\lst@linebgrd}%
    }{\PackageError{Listings}{Numbers #1 unknown}\@ehc}}
\makeatother

\usepackage[capitalise]{cleveref}

\usepackage{mdframed}

\usepackage{tikz}
\usetikzlibrary{arrows,positioning,calc} 
\tikzset{
    >=stealth',
    punkt/.style={
           rectangle,
           rounded corners,
           draw=black, 
           minimum height=2em,
           text centered},
    pil/.style={
           ->,
           thick,
           shorten <=2pt,
           shorten >=2pt
       },
  parambox/.style = {
    inner sep=0ex,
    minimum height=1.4em,
    text centered
  },
    rect/.style={
           draw, 
           rectangle, 
           minimum height=1.4em,
           text centered}
}

\usepackage{mathpartir}

\renewcommand{\paragraph}[1]{\smallskip\noindent\emph{\textbf{#1.}}}

\definecolor{colorrefines}{HTML}{1b9e77}
\definecolor{coloruses}{HTML}{e7298a}
\definecolor{colorextends}{HTML}{e7298a}
\definecolor{colorfaithful}{HTML}{d95f02}
\definecolor{colorspec}{HTML}{e6f5c9}
\definecolor{colorOPspec}{HTML}{fff2ae}
\definecolor{colorcontrib}{HTML}{d0def5}
\definecolor{colorIMP}{HTML}{fcd9dd}
\definecolor{colorBILss}{HTML}{fcd9dd}
\definecolor{colorBILInf}{HTML}{fcecd9}
\definecolor{colorBD}{HTML}{f7bae7}
\definecolor{colorRS}{HTML}{AFEEEE}
\colorlet{colorexamples}{colorcontrib}
\colorlet{colorrsinstantiation}{colorBILInf}
\colorlet{colorrsproofs}{colorexamples}
\colorlet{colorrsabstract}{colorOPspec}
\colorlet{colorbil}{colorspec}
\colorlet{colorspecbil}{colorBILss}
\colorlet{colorarch}{colorrsinstantiation}
\colorlet{colormem}{colorexamples}
\colorlet{colorparser}{colorBILss}
\colorlet{colorother}{colorBD}
\definecolor{colorcurl}{HTML}{ffd1cc}

\tikzset{
  BDsec/.style={draw=black, align=center, rounded corners, fill=colorBD, minimum size=3.6ex},
  specRS/.style={draw=black, align=center, rounded corners, fill=colorRS, minimum size=3.6ex},
  BILss/.style={draw=black, align=center, rounded corners, fill=colorBILss, minimum size=3.6ex},
  spec/.style={draw=black, align=center, rounded corners, fill=colorspec},
  specOP/.style={draw=black, align=center, rounded corners, fill=colorOPspec},
  BILInf/.style={draw=black, align=center, rounded corners, fill=colorBILInf},
  IMP/.style={draw=black, align=center, rounded corners, fill=colorIMP},
  app/.style={draw=black, align=center, rounded corners, fill=colorcontrib},
  curl/.style={draw=black, align=center, rounded corners, fill=colorcurl},
  impl/.style={draw=black, dashed, align=center},
  exec/.style={draw=black, align=center, rounded corners},
  refines/.style={thick,-latex, black, every node/.style={color=colorrefines}},
  extends/.style={-latex, dashed, thick, colorextends, every node/.style={color=colorextends}},
  extendsl/.style={dashed, thick, colorextends, every node/.style={color=colorextends}},
  faithful/.style={-latex, colorfaithful, dotted, ultra thick, every node/.style={color=colorfaithful}},
  uses/.style={-latex, thick, coloruses, every node/.style={color=coloruses}},
}

\usepackage{enumitem} 
\usepackage{wrapfig}
\newtheorem{assumption}{Assmuption}

\usepackage{footnote}

\usepackage{color, colortbl}
\definecolor{Gray}{gray}{0.88}
\definecolor{DarkGray}{gray}{0.6}

\usepackage{pdfpages}

\newtheorem{example}{Example}[section]

\crefname{section}{\S\!}{\S\!}
\Crefname{section}{\S\!}{\S\!}
\crefname{appendix}{\S\!}{\S\!}
\Crefname{appendix}{\S\!}{\S\!}

\newcommand{\bd}[1]{{\color{red} BD: #1}}

\makeatletter
\newcommand\footnoteref[1]{\protected@xdef\@thefnmark{\ref{#1}}\@footnotemark}
\makeatother

\usepackage[inference]{semantic}

\usepackage{stmaryrd}

\usepackage{pifont}

\usepackage{mathtools}

\usepackage{wasysym}

\usepackage{multicol}
\usepackage{multirow}

\lstdefinelanguage{bir}{
	morekeywords={when, mem, with, goto, call, return, specfence, sub, noreturn},
    sensitive=false, 
    morecomment=[l]{//}, 
    morecomment=[s]{/*}{*/}, 
    morestring=[b]" 
} %

\lstdefinelanguage{bil}{
	morekeywords={jmp, when, mem, signed, el, be, with},
    sensitive=false, 
    morecomment=[l]{//}, 
    morecomment=[s]{/*}{*/}, 
    morestring=[b]" 
} %

\lstdefinelanguage{birn}{
	morekeywords={when, mem, with, goto, call, return, specfence, sub, noreturn},
    sensitive=false, 
    morecomment=[l]{//}, 
    morecomment=[s]{/*}{*/}, 
    morestring=[b]", 
    basicstyle=\tt\normalsize
} %

\newcommand{\isabil}{{\sc IsaBIL}\xspace}
\newcommand{\IsaBIL}{\isabil}

\newcommand{\BIL}{\textsc{BIL}\xspace}

\newcommand{\BILa}{$\BIL_{{\sf ADT}}$\xspace}
\newcommand{\sublocale}{\subseteq}

\newcommand{\nat}{\mathbb{N}}

\newcommand{\InferenceName}[1]{{\sc #1}\xspace}

\newcommand{\FunctionStyle}[1]{{\tt #1}\xspace}

\newcommand{\Locale}[1]{\textbf{\texttt{#1}}\xspace}

\newcommand{\FindSymbolSmallStep}{\xrightarrow[FS]{}}
\newcommand{\FindSymbolLocale}{\Locale{find\_symbol}}
\newcommand{\IsSymbol}{\FunctionStyle{is\_symbol}}

\newcommand{\AvRuleLocaleE}[1]{\Locale{AV\_rule\_#1}}
\newcommand{\AvRuleLocale}{\AvRuleLocaleE{*}}
\newcommand{\AvRuleProofLocaleE}[1]{\Locale{AV\_rule\_#1\_proof}}
\newcommand{\AvRuleProofLocale}{\AvRuleProofLocaleE{*}}

\newcommand{\code}[1]{\texttt{#1}} 

\newcommand{\StackAllocLemmaName}{\InferenceName{stack\_alloc\_lemma}}
\newcommand{\StackDeallocLemmaName}{\InferenceName{stack\_dealloc\_lemma}}

\newcommand{\inarr}[1]{
  \begin{array}[t]{@{}l@{}}
    #1
  \end{array}}

\newcommand{\inarrC}[1]{
  \begin{array}[c]{@{}l@{}}
    #1
  \end{array}}

\newcommand{\DefaultSize}{{\it sz}}
\newcommand{\AddressSize}{{\it sz_{addr}}}
\newcommand{\ValueSize}{{\it sz_{val}}}
\newcommand{\TypeImmediateE}[1]{\textbf{imm}\langle#1\rangle}
\newcommand{\TypeImmediate}{\TypeImmediateE{\DefaultSize}}
\newcommand{\TypeMemoryE}[2]{\textbf{mem}\langle#1,\ #2\rangle}
\newcommand{\TypeMemory}{\TypeMemoryE{\AddressSize}{\ValueSize}}
\newcommand{\TypeFunctionE}[1]{{\texttt{type}(#1)}}

\newcommand{\WordE}[2]{#1 :: #2}
\newcommand{\Word}{\WordE{nat}{\DefaultSize}}
\newcommand{\WordExtractE}[3]{\textbf{ext}\ #1 \backsim \textbf{hi} : #2 \backsim \textbf{lo} : #3}
\newcommand{\WordExtract}{\WordExtractE{w}{sz_1}{sz_2}}

\newcommand{\ValueImmediate}{{\it w}}
\newcommand{\ValueMemoryE}[4]{#1[#2 \leftarrow #3,\ #4]}
\newcommand{\ValueMemory}{\ValueMemoryE{v}{w}{v'}{\ValueSize}}
\newcommand{\ValueUnknownE}[2]{\textbf{unknown}[#1] : #2}
\newcommand{\ValueUnknown}{\ValueUnknownE{str}{t}}

\newcommand{\CastHigh}{\textbf{high}}
\newcommand{\CastLow}{\textbf{low}}
\newcommand{\CastUnsigned}{\textbf{unsigned}}
\newcommand{\CastSigned}{\textbf{signed}}
\newcommand{\ExpressionLoadE}[4]{#1[#2, #3] : #4}
\newcommand{\ExpressionLoad}{\ExpressionLoadE{e_1}{e_2}{ed}{sz}}

\newcommand{\ExpressionStoreE}[5]{#1\ \textbf{with}\ [#2, #3] : #4 \leftarrow #5}
\newcommand{\ExpressionStore}{\ExpressionStoreE{e_1}{e_2}{ed}{sz}{e_3}}

\newcommand{\ExpressionLetE}[3]{\textbf{let}\ #1 = #2\ \textbf{in} \ #3}
\newcommand{\ExpressionLet}{\ExpressionLetE{var}{e_1}{e_2}}
\newcommand{\ExpressionIteE}[3]{\textbf{ite}\ #1\ #2\ #3}
\newcommand{\ExpressionIte}{\ExpressionIteE{e_1}{e_2}{e_3}} 
\newcommand{\ExpressionExtractE}[3]{\textbf{extract}\colon#1\colon#2[#3]}
\newcommand{\ExpressionExtract}{\ExpressionExtractE{sz_1}{sz_2}{e}} 
\newcommand{\ExpressionConcatE}[2]{#1\ \text{@}\ #2}
\newcommand{\ExpressionConcat}{\ExpressionConcatE{e_1}{e_2}}
\newcommand{\ExpressionBinOpE}[3]{#1\ #2\ #3}
\newcommand{\ExpressionBinOp}{\ExpressionBinOpE{e_1}{bop}{e_2}}
\newcommand{\ExpressionUnOpE}[2]{#1\ #2}
\newcommand{\ExpressionUnOp}{\ExpressionUnOpE{uop}{e}}
\newcommand{\ExpressionCastE}[3]{#1 : #2[#3]}
\newcommand{\ExpressionCast}{\ExpressionCastE{cast}{sz}{e}}

\newcommand{\EndianLittle}{\textbf{el}}
\newcommand{\EndianBig}{\textbf{be}}
\newcommand{\BILVariables}{\Delta}

\newcommand{\BILStepE}[3]{#1 \vdash #2 \leadsto #3 }
\newcommand{\BILStep}{\BILStepE{\BILVariables}{e}{e'}}
\newcommand{\BILStepsE}[3]{#1 \vdash #2 \leadsto^* #3 }
\newcommand{\BILSteps}{\BILStepsE{\BILVariables}{e}{e'}}

\newcommand{\Succ}[1]{\texttt{succ}(#1)}

\newcommand{\TypingContext}{\Gamma}
\newcommand{\TypingRuleIsOk}[1]{#1\ \textbf{is ok}}

\newcommand{\TypeIsOkE}[1]{\TypingRuleIsOk{#1}}
\newcommand{\TypeIsOk}{\TypeIsOkE{t}}
\newcommand{\TypingContextIsOkE}[1]{\TypingRuleIsOk{#1}}
\newcommand{\TypingContextIsOk}{\TypingContextIsOkE{\Gamma}}

\newcommand{\StatementMoveE}[2]{#1 := #2}
\newcommand{\StatementMove}{\StatementMoveE{var}{exp}}
\newcommand{\StatementJmpName}{\textbf{jmp}\xspace}
\newcommand{\StatementJmpE}[1]{\StatementJmpName\ #1}

\newcommand{\StatementCpuExnName}{\textbf{cpuexn}\xspace}
\newcommand{\StatementCpuExnE}[1]{\StatementCpuExnName(#1)}

\newcommand{\StatementSpecialName}{\textbf{special}\xspace}
\newcommand{\StatementSpecialE}[1]{\StatementSpecialName(#1)}

\newcommand{\StatementWhileName}{\textbf{while}\xspace}
\newcommand{\StatementWhileE}[2]{\StatementWhileName\ (#1)\ #2}

\newcommand{\StatementIfName}{\textbf{if}\xspace}
\newcommand{\StatementIfThenElseE}[3]{\StatementIfName\ (#1)\ #2\ \textbf{else}\ #3}

\newcommand{\StatementIfThenElse}{\StatementIfThenElseE{e}{seq_1}{seq_2}}
\newcommand{\StatementIf}{\StatementIfThenElse}
\newcommand{\StatementIfThenE}[2]{\StatementIfName\ (#1)\ #2}
\newcommand{\StatementIfThen}{\StatementIfThenE{e}{seq}}

\newcommand{\StatementStepE}[3]{#1 \vdash #2 \leadsto #3}

\newcommand{\SequenceStepE}[5]{(#1,#2) \vdash #3 \leadsto (#4,#5)}

\newcommand{\ListConcat}[2]{#1\ \#\ #2}
\newcommand{\TypingContextConcat}{\ListConcat{(str : t)}{\TypingContext}}

\newcommand{\InsnTuple}{\llparenthesis~ \textbf{addr},\ \textbf{size},\ \textbf{code}~\rrparenthesis}
\newcommand{\InsnExpandedE}[3]{\llparenthesis~ \textbf{addr} = #1, \textbf{size} = #2, \textbf{code} = #3~\rrparenthesis}

\newcommand{\BILMachineStateE}[3]{(#1, #2, #3)}
\newcommand{\BILMachineState}{\BILMachineStateE{\BILVariables}{pc}{mem}}

\newcommand{\BILDecodeTransition}{\mapsto}
\newcommand{\BILTransition}{\leadsto}
\newcommand{\BILDecodeInstructionE}[2]{#1 \BILDecodeTransition #2}

\newcommand{\BILMachineStateStepInitialE}[2]{#1 \BILTransition #2}
\newcommand{\BILMachineStateStepE}[3]{\BILMachineStateStepInitialE{\BILMachineState}{\BILMachineStateE{#1}{#2}{#3}}}
\newcommand{\BILMachineStateStep}{\BILMachineStateStepE{\BILVariables'}{pc'}{mem'}}

\newcommand{\BIRCarryFlag}{\texttt{CF}}
\newcommand{\BIROverflowFlag}{\texttt{OF}}
\newcommand{\BIRAdjustFlag}{\texttt{AF}}
\newcommand{\BIRParryFlag}{\texttt{PF}}
\newcommand{\BIRSizeFlag}{\texttt{SF}}
\newcommand{\BIRZeroFlag}{\texttt{ZF}}
\newcommand{\BIRAccumulatorRegister}{\texttt{AX}}
\newcommand{\BIRBaseRegister}{\texttt{BX}}
\newcommand{\BIRCounterRegister}{\texttt{CX}}
\newcommand{\BIRDataRegister}{\texttt{DX}}
\newcommand{\BIRStackPointerRegister}{\texttt{SP}}
\newcommand{\BIRStackBaseRegister}{\texttt{BP}}
\newcommand{\BIRSourceRegister}{\texttt{SI}}
\newcommand{\BIRDestinationRegister}{\texttt{DI}}
\newcommand{\BIRRegisterSixFour}{\texttt{R}}
\newcommand{\BIRAccumulatorRegisterSixFour}{\BIRRegisterSixFour\BIRAccumulatorRegister}
\newcommand{\BIRBaseRegisterSixFour}{\BIRRegisterSixFour\BIRBaseRegister}
\newcommand{\BIRCounterRegisterSixFour}{\BIRRegisterSixFour\BIRCounterRegister}
\newcommand{\BIRDataRegisterSixFour}{\BIRRegisterSixFour\BIRDataRegister}
\newcommand{\BIRStackPointerRegisterSixFour}{\BIRRegisterSixFour\BIRStackPointerRegister}
\newcommand{\BIRStackBaseRegisterSixFour}{\BIRRegisterSixFour\BIRStackBaseRegister}
\newcommand{\BIRSourceRegisterSixFour}{\BIRRegisterSixFour\BIRSourceRegister}
\newcommand{\BIRDestinationRegisterSixFour}{\BIRRegisterSixFour\BIRDestinationRegister}
\newcommand{\BIRRegisterOneTwoEight}{\texttt{XMM}}
\newcommand{\BIRRegisterOneTwoEightE}[1]{\BIRRegisterOneTwoEight\texttt{#1}}
\newcommand{\BIRRegisterTwoFiveSix}{\texttt{YMM}}
\newcommand{\BIRRegisterTwoFiveSixE}[1]{\BIRRegisterTwoFiveSix\texttt{#1}}
\newcommand{\BIRRegisterFiveOneTwo}{\texttt{ZMM}}
\newcommand{\BIRRegisterFiveOneTwoE}[1]{\BIRRegisterFiveOneTwo\texttt{#1}}
\newcommand{\BIRRegisterSixFourE}[1]{\BIRRegisterSixFour\texttt{#1}}

\newcommand{\OHearnTriple}[3]{\left[#1\right]#2\left[#3\right]}
\newcommand{\HoareTriple}[3]{\{#1\}#2\{#3\}}
\newcommand{\BigStep}{\mathbin{\Rightarrow}}
\newcommand{\HolBigStep}{\mathbin{\xRightarrow[{\sc IMP}]{}}}
\newcommand{\BILBigStep}{\mathbin{\xRightarrow[{\sc BIL}]{}}}
\newcommand{\CorrectnessLocale}{\Locale{correctness}}
\newcommand{\IncorrectnessLocale}{\Locale{incorrectness}}
\newcommand{\InferenceLocale}{\Locale{inference\_rules}}

\newcommand{\Command}{\mathcal{C}}
\newcommand{\offset}{{\it offset}}

\newcommand{\XESSF}{{\tt x86\_64}}
\newcommand{\RiscVsf}{{\tt riscv64}}

\newcommand{\DoubleFreeLocaleE}[1]{\Locale{double\_free\_{#1}}}
\newcommand{\DoubleFreeGoodLocale}{\DoubleFreeLocaleE{good}}
\newcommand{\DoubleFreeBadLocale}{\DoubleFreeLocaleE{bad}}
\newcommand{\DoubleFreeLocale}{\DoubleFreeLocaleE{*}}
\newcommand{\DoubleFreeBinaryLocale}{\DoubleFreeLocale}
\newcommand{\DoubleFreeProofLocaleE}[1]{\Locale{double\_free\_{#1}\_proof}}
\newcommand{\DoubleFreeGoodProofLocale}{\DoubleFreeProofLocaleE{good}}
\newcommand{\DoubleFreeBadProofLocale}{\DoubleFreeProofLocaleE{bad}}
\newcommand{\DoubleFreeProofLocale}{\DoubleFreeProofLocaleE{*}}

\newcommand{\RiscVauipcNameOnly}{{\tt auipc}\xspace}
\newcommand{\RiscVauipcE}[2]{\RiscVauipcNameOnly \ #1,\ #2}
\newcommand{\RiscVauipc}{\RiscVauipcE{rd}{imm}}
\newcommand{\RiscVldNameOnly}{{\tt ld}\xspace}
\newcommand{\RiscVldE}[3]{\RiscVldNameOnly\ #1,\ #3(#2)}
\newcommand{\RiscVld}{\RiscVldE{rd}{rs1}{\offset}}
\newcommand{\RiscVjalrNameOnly}{{\tt jalr}\xspace}
\newcommand{\RiscVjalrE}[3]{\RiscVjalrNameOnly\ #1, \ #2, \ #3}
\newcommand{\RiscVjalr}{\RiscVjalrE{rd}{rs1}{\offset}}
\newcommand{\RiscVaddiNameOnly}{{\tt addi}\xspace}
\newcommand{\RiscVaddiE}[3]{\RiscVaddiNameOnly \ #1,\ #2,\ #3}
\newcommand{\RiscVaddi}{\RiscVaddiE{rd}{rs1}{imm}}
\newcommand{\RiscVsdNameOnly}{{\tt sd}\xspace}
\newcommand{\RiscVsdE}[3]{\RiscVsdNameOnly \ #1,\ #2(#3)}
\newcommand{\RiscVsd}{\RiscVsdE{rs2}{\offset}{rs1}}
\newcommand{\RiscVretNameOnly}{{\tt ret}\xspace}
\newcommand{\RiscVret}{\RiscVretNameOnly}

\newcommand{\prestate}{\sigma}
\newcommand{\poststate}{\tau}

\newcommand{\BilSmallStep}{\xrightarrow[BIL]{}}
\newcommand{\BilInferenceLocale}{\Locale{BIL\_inference}}

\newcommand{\BILSmallStep}{\BilSmallStep}

\newcommand{\PltStubLemmaName}{{\sc plt\_stub\_lemma}}
\newcommand{\RiscV}{RISC\nobreakdash-V\xspace}

\newcommand{\SymbolTable}{{\tt sym\_table}}
\newcommand{\ProgramDomain}{{\tt addr\_set}} 
\newcommand{\Decode}{\BILDecodeTransition}
\newcommand{\DecodeProg}{\BILDecodeTransition_{\textsf{prog}}}
\newcommand{\CommandBIL}{\lstinline[language=isabellen]{BIL}\xspace}
\newcommand{\CommandwithSR}{\lstinline[language=isabellen]{with_subroutines}\xspace}
\newcommand{\CommandBILfile}{\lstinline[language=isabellen]{BIL_file}\xspace}

\newcommand{\BilLocale}{\Locale{BIL\_specification}}
\newcommand{\powerset}{\mathbb{P}}

\newcommand{\llpar}{\llparenthesis}
\newcommand{\rrpar}{\rrparenthesis}

\newcommand{\Allocator}{\mathcal{A}}
\newcommand{\MemoryTrace}{\omega_\Allocator}
\newcommand{\NextAddrAllocatorName}{\texttt{next\_ptr}\xspace}
\newcommand{\NextAddrAllocator}{\NextAddrAllocatorName(\MemoryTrace)}
\newcommand{\FreePredicateName}{\texttt{is\_free}\xspace}
\newcommand{\FreePredicate}[1]{\FreePredicateName(#1)}
\newcommand{\AllocPredicateName}{\texttt{is\_alloc}\xspace}
\newcommand{\AllocPredicate}[1]{\AllocPredicateName(#1)}
\newcommand{\AllocatorLocale}{\Locale{allocation}}
\newcommand{\AllocationLocale}{\AllocatorLocale}
\newcommand{\GetFreedAddrName}{\texttt{get\_ptr}\xspace}
\newcommand{\GetFreedAddr}[1]{\GetFreedAddrName(#1)}
\newcommand{\GetSzName}{\texttt{get\_sz}\xspace}
\newcommand{\GetSz}[1]{\GetSzName(#1)}
\newcommand{\AllocSmallStep}{\xrightarrow[\Allocator]{}}

\newcommand{\AllocOpName}{\texttt{alloc}\xspace}
\newcommand{\AllocOpE}[2]{\AllocOpName(#1,#2)}
\newcommand{\AllocOp}{\AllocOpE{w}{sz}}
\newcommand{\FreeOpName}{\texttt{free}\xspace}
\newcommand{\FreeOpE}[1]{\FreeOpName(#1)}
\newcommand{\FreeOp}{\FreeOpE{w}}

\newcommand{\MopStateE}[1]{#1}
\newcommand{\MopState}{\MopStateE{\MemoryTrace}}
\newcommand{\memop}{{\it memop}}

\newcommand{\IsDF}{{\tt double\_free\_vuln}\xspace}

\newcommand{\IncorrectnessProof}[4]{\exists #3.\ \OHearnTriple{#1}{#2}{#3} \wedge (\forall \sigma.\ #3(\sigma) \implies #4(\sigma)) \wedge (\exists \sigma.\ #3(\sigma))}

\newcommand{\HoareProp}{\forall \sigma, \tau.\ ((c,\sigma) \BigStep \tau) \implies (P(\sigma) \implies Q(\tau))}
\newcommand{\OHearnProp}{\forall \tau.\ Q(\tau) \implies \exists \sigma.\ P(\sigma) \wedge ((c,\sigma) \BigStep \tau)}

\newcommand{\InitialStorage}{v}
\newcommand{\Variables}{\Delta}
\newcommand{\SmallStepExp}[3]{#1 \vdash #2 \leadsto #3}
\newcommand{\SmallStepExpLB}[3]{#1 \vdash #2 \leadsto \\\\ #3}
\newcommand{\BigStepExp}[3]{\BILStepsE{#1}{#2}{#3}}
\newcommand{\ReflValue}{w_{1}}
\newcommand{\ReflAddress}{w_{2}}
\newcommand{\ReflStorage}[5]{{\tt storage}_{#1}\left(\inarrC{#2, #3, #4, #5}\right)}
\newcommand{\ReflStorageEL}[4]{\ReflStorage{\EndianLittle}{#1}{#2}{#3}{#4}}
\newcommand{\ReflStorageBE}[4]{\ReflStorage{\EndianBig}{#1}{#2}{#3}{#4}}
\newcommand{\VarInE}[3]{(#1, #2) \in #3}
\newcommand{\VarIn}{\VarInE{var}{v}{\Variables}}
\newcommand{\ProgramStepRuleName}{{\sc step\_prog}\xspace}

\newcommand{\String}{{\it string}}

\newcommand{\SymbolicSolver}{\FunctionStyle{sexc}}
\newcommand{\SymbolicSolverBILI}{\FunctionStyle{sexc\_bilI}}
\newcommand{\TypeSolver}{\FunctionStyle{typec}}
\newcommand{\TypeSolverBILI}{\FunctionStyle{typec\_bilI}}

\newcommand{\AllocRule}{\InferenceName{alloc}}
\newcommand{\FreeRule}{\InferenceName{free}}
\newcommand{\ReallocRule}{\InferenceName{realloc}}
\newcommand{\ZeroReallocRule}{\InferenceName{zero\_realloc}}
\newcommand{\SkipRule}{\InferenceName{skip}}

\newcommand{\ReadData}{\FunctionStyle{read\_data}}
\newcommand{\SecRecv}{\FunctionStyle{sec\_recv}}
\newcommand{\Realloc}{\FunctionStyle{realloc}}
\newcommand{\Malloc}{\FunctionStyle{malloc}}
\newcommand{\KrbBuffer}{\code{krb5buffer}\xspace}

\newcommand{\ProgBad}{RD_{7.50.3}}
\newcommand{\ReallocLocale}{\Locale{reallocation}}
\newcommand{\ReadDataBadLocale}{\Locale{read\_data\_bad}}
\newcommand{\LibCurlLocale}{\Locale{lib\_curl}}
\newcommand{\SecRecvBadLocale}{\Locale{sec\_recv\_bad}}
\newcommand{\ChooseMechBadLocale}{\Locale{choose\_mech\_bad}}  
\newcommand{\CurlSecLoginBadLocale}{\Locale{Curl\_sec\_login\_bad}}

\newcommand{\CurlBadLocale}{\Locale{read\_data\_7\_50\_3}}

\newcommand{\CurlBadProofLocale}{\Locale{read\_data\_7\_50\_3\_proof}}
\newcommand{\ReallocPredicate}{\code{is\_realloc}}
\newcommand{\ReallocSymbol}{\mathcal{R}}
\newcommand{\ReallocSmallStep}{\mathbin{\xrightarrow[\ReallocSymbol]{}}}

\newcommand{\curlold}{7.50.3\xspace}
\newcommand{\curlfixed}{7.51.0\xspace}

\newcommand{\AgreementRuleName}{{\sc agreement}\xspace}
\newcommand{\DenialRuleName}{{\sc denial}\xspace}

\definecolor{isabdblue}{rgb}{0.00, 0.40, 0.60}
\definecolor{isablblue}{rgb}{0.00, 0.60, 0.99}
\definecolor{isabgreen}{rgb}{0.00, 0.60, 0.40}
\definecolor{isabred}{rgb}{0.99, 0.31, 0.31}
\definecolor{isabpurple}{rgb}{0.59, 0.40, 0.99}
\definecolor{isabcommentpurple}{rgb}{0.4, 0.0, 0.8}
\definecolor{isabcomment}{rgb}{0.80, 0.00, 0.00}
\definecolor{isabtextcomment}{rgb}{0.8, 0.4, 0.0}
\definecolor{isabmblue}{rgb}{0.05, 0.41, 0.49}
\definecolor{isabpink}{rgb}{0.99, 0.0, 0.8}
\definecolor{isabstring}{HTML}{DCDCDC}

\newcommand{\colorOnlyDisplay}[1]{\lst@ifdisplaystyle\color{#1}\fi} 
\newcommand{\isabkeystyleA}{\color{isabdblue}\bfseries} 
\newcommand{\isabproofkeystyleA}{\color{isabdblue}\bfseries} 
\newcommand{\isabkeystyleB}{\color{isabgreen}\bfseries} 
\newcommand{\isabkeystyleC}{\color{isabred}\bfseries}   
\newcommand{\isabkeystyleD}{\color{isabpurple}\bfseries} 
\newcommand{\isabkeystyleE}{\color{isablblue}\bfseries}  
\newcommand{\isabkeystyleF}{\color{isabmblue}\bfseries}  

\lstdefinelanguage{isabelle}{
    keywords={abbreviation,cpodef,pcpodef,datatype,default_sort,definition,domain,fun,instance,instantiation,lemma,next,lift,lift_definition,primrec,section,setup,setup_lifting,subsection,subsubsection,syntax,text,theorem,theory,thm,translations,typedef,type_synonym,fixrec,primrec,fun,function}, 
    keywordstyle=\isabkeystyleA,
    keywords=[2]{by,have,hence,in,proof,qed,using}, 
    keywordstyle=[2]\isabproofkeystyleA,
    keywords=[12]{and,assumes,obtains,for,begin,class,end,fixes,imports,infix,infixl,infixr,is,lazy,shows,where,defining,with_subroutines}, 
    keywordstyle=[12]\isabkeystyleB,
    keywords=[13]{apply,done,sorry},  
    keywordstyle=[13]\isabkeystyleC,
    keywords=[14]{in,only,'a,'b,'c,'d,'e,'f,'m,'M,'i,'o},  
    keywordstyle=[14]\isabkeystyleD,
    keywords=[15]{assume,case,fix,obtain,show,thus,of}, 
    keywordstyle=[15]\isabkeystyleE,
    keywords=[16]{CONST,else,if,LEAST,let,THE,then}, 
    keywordstyle=[16]\isabkeystyleF,
    sensitive=true,
    alsoletter=:'*,
    stringstyle=\color{isabstring},
    morekeywords={definition, BIL, BIL_file, locale, method},
    sensitive=true, 
    morecomment=[l]{//}, 
    morecomment=[s]{(*}{*)}, 
} %

\lstdefinelanguage{isabellen}{
    keywords={abbreviation,cpodef,pcpodef,datatype,default_sort,definition,domain,fun,instance,instantiation,lemma,next,lift,lift_definition,primrec,section,setup,setup_lifting,subsection,subsubsection,syntax,text,theorem,theory,thm,translations,typedef,type_synonym,fixrec,primrec,fun,function}, 
    keywordstyle=\isabkeystyleA,
    keywords=[2]{by,have,hence,in,proof,qed,using}, 
    keywordstyle=[2]\isabproofkeystyleA,
    keywords=[12]{and,assumes,obtains,for,begin,class,end,fixes,imports,infix,infixl,infixr,is,lazy,shows,where,defining,with_subroutines}, 
    keywordstyle=[12]\isabkeystyleB,
    keywords=[13]{apply,done,sorry},  
    keywordstyle=[13]\isabkeystyleC,
    keywords=[14]{in,only,'a,'b,'c,'d,'e,'f,'m,'M,'i,'o},  
    keywordstyle=[14]\isabkeystyleD,
    keywords=[15]{assume,case,fix,obtain,show,thus,of}, 
    keywordstyle=[15]\isabkeystyleE,
    keywords=[16]{CONST,else,if,LEAST,let,THE,then}, 
    keywordstyle=[16]\isabkeystyleF,
    sensitive=true,
    alsoletter=:'*,
    stringstyle=\color{isabstring},
    morekeywords={definition, BIL, BIL_file, locale, method},
    sensitive=true, 
    morecomment=[l]{//}, 
    morecomment=[s]{(*}{*)}, 
    basicstyle=\tt\normalsize
} %

\lstdefinelanguage{riscv}{
	morekeywords={auipc, ld, jalr, nop, addi, ret, sd, lui, jal, mv, lw},
    sensitive=true,
    morecomment=[l]{\#}, 
    morestring=[s]{<}{>}
} %

\definecolor{codegreen}{rgb}{0,0.6,0}
\definecolor{codegray}{rgb}{0.5,0.5,0.5}
\definecolor{codepurple}{rgb}{0.58,0,0.82}
\definecolor{backcolour}{rgb}{0.98,0.98,0.98}

\lstdefinestyle{mystyle}{
    backgroundcolor=\color{backcolour},   
    commentstyle=\color{codegreen},
    keywordstyle=\color{magenta},
    numberstyle=\tiny\color{codegray},
    stringstyle=\color{codepurple},
    basicstyle=\ttfamily\linespread{.95}\footnotesize,
    breakatwhitespace=false,
    breaklines=true,                 
    captionpos=b,                    
    keepspaces=true,                 
    numbers=left,                    
    numbersep=5pt,                  
    showspaces=false,                
    showstringspaces=false,
    showtabs=false,                  
    tabsize=2,
    mathescape=true,    
}

\begin{document}



\title{IsaBIL: A Framework for Verifying (In)correctness of Binaries
  in Isabelle/HOL (Extended Version)}

\author{Matt Griffin}
\email{matt.griffin@imperial.ac.uk}
\orcid{0000-0003-2703-0368}
\affiliation{%
  \institution{University of Surrey}
  \city{Guildford}
  \country{UK}
}
\author{Brijesh Dongol}
\email{b.dongol@surrey.ac.uk}
\orcid{0000-0003-0446-3507}
\affiliation{%
  \institution{University of Surrey}
  \city{Guildford}
  \country{UK}
}
\author{Azalea Raad}
\email{azalea.raad@imperial.ac.uk}
\orcid{0000-0002-2319-3242}
\affiliation{%
  \institution{Imperial College London}
  \city{London}
  \country{UK}
}

\begin{abstract}
  This paper presents \isabil, a binary analysis framework in
  Isabelle/HOL that is based on the widely used Binary Analysis
  Platform (BAP). Specifically, in \isabil, we formalise BAP's
  intermediate language, called BIL and integrate it with Hoare logic
  (to enable proofs of \emph{correctness}) as well as incorrectness logic
  (to enable proofs of \emph{incorrectness}). \isabil inherits the
  full flexibility of BAP, allowing us to verify binaries for a wide
  range of languages (C, C++, Rust), toolchains (LLVM, Ghidra) and
  target architectures (x86, RISC-V), and can also be used when the
  source code for a binary is unavailable.
  
  To make verification tractable, we develop a number of big-step rules
  that combine BIL's existing small-step rules at different levels of
  abstraction to support reuse. We develop high-level reasoning rules
  for RISC-V instructions (our main target architecture) to further
  optimise verification. Additionally, we develop Isabelle proof
  tactics that exploit common patterns in C binaries for RISC-V to
  discharge large numbers of proof goals (often in the 100s)
  automatically. \isabil includes an Isabelle/ML based parser for BIL
  programs, allowing one to {\em automatically} generate the
  associated Isabelle/HOL program locale from a BAP output. Taken
  together, \isabil provides a highly flexible proof
  environment for program binaries.  
  As examples, we prove correctness
  of key examples from the Joint Strike Fighter coding standards and
  the MITRE database.
\end{abstract}

\maketitle

\section{Introduction}
\label{sec:introduction}

Analysing binary code is highly challenging and is essential for
decompilation and verification when one does not have access to the
source code (e.g., proprietary code). 
It is also a key tool for finding security
vulnerabilities and exploits and is often the only way to (dis)prove
properties about programs~\cite{shoshitaishvili2016state}.  
Many tools for binary analysis such as
angr~\cite{DBLP:conf/secdev/WangS17}, LIEF~\cite{LIEF} and
Manticore~\cite{DBLP:conf/kbse/MossbergMHGGFBD19} lift machine code
to higher-level, human-readable representations. These tools have been developed for different goals
covering different languages (Rust, C++), hardware architectures
(e.g., x86, RISC-V), types of binaries (ELF, smart contracts)
and security analysis (malware analysis, reverse
engineering).  

The \emph{Binary Analysis Platform}
(BAP)~\cite{DBLP:conf/cav/BrumleyJAS11} provides a generic
approach to binary analysis. BAP natively supports several languages
(C, Python and Rust) and hardware architectures (x86, Arm, RISC-V,
MIPS) and additionally can be integrated with toolchains such as LLVM
and Ghidra to analyse programs on less common architectures such as
Atmel AVR~\cite{Atmel-manual}. BAP is well-supported and widely used
with an active user community. The platform can analyse
binaries derived from a particular source language such as C/C++ and
Rust, as well as closed-source functions. BAP provides a disassembler
that lifts binaries into a unified intermediate language called {\em
  BIL (Binary Instruction Language)}, thus enabling
architecture-agnostic analysis. While BAP also provides a set of
static analysis tools, they do not provide mechanisms for formal
verification, e.g., logics for (dis)proving correctness of programs.
Recent works have covered subsets of BAP (called AIR) in UCLID5~\cite{DBLP:conf/csfw/CheangRSS19} and
use an extended 
semantics to enable reasoning about an information flow hyper-property
known as {\em trace-property observational
  determinism}~\cite{DBLP:conf/csfw/CheangRSS19}.

We present \isabil, a verification framework for
low-level binaries encoded using the Isabelle/HOL theorem
prover. 
\isabil provides a {\em deep embedding} of BIL programs into
Isabelle/HOL, providing a {\em direct interface} between BAP and
Isabelle/HOL without using any additional tools. We leverage
the fact that BAP can generate BIL programs as abstract data
types, providing a convenient hook for \isabil's ML-based
parser.\footnote{Note that this is a significant advancement over the
  encoding by \citet{DBLP:conf/fm/GriffinD21}, which used an external
  Python tool to translate BAP outputs to
  Isabelle/HOL.} 
Since BAP can lift binaries into BIL both with and without the source
code being available, so can \isabil. Moreover, \isabil allows the
generated BIL program to be connected to {\em any} operational
semantics. We build on the semantics described by
\cite{DBLP:conf/cav/BrumleyJAS11} (see
\cref{sec:specification-of-bil}) to enable bit-precise
analysis. 
Our approach generates an Isabelle {\em
  locale}~\cite{kammuller1999locales}, which provides an extensible
interface that can be instantiated to different scenarios (see
\cref{fig:proofs}). We demonstrate this by both proving (using Hoare
logic~\cite{hoare1969axiomatic}) and disproving (using Incorrectness
logic~\cite{o2019incorrectness}) properties for a number of key
examples (see \cref{table:proof-stats}). 

\begin{wraptable}{r}{63mm}
\footnotesize
\vspace{-3ex}
\begin{tabular}{|l|c|c|c|c|}
\hline
\multicolumn{1}{|c|}{\multirow{2}{*}{Test}} & \multicolumn{4}{c|}{Size (LoC)}                                                                   \\ \cline{2-5} 
\multicolumn{1}{|c|}{}                      & C  & RISC-V & BIL & Isabelle \\ \hline 
AV Rule 17                                  & 13 &  34    & 80  & 94      \\ \hline 
AV Rule 19                                  & 10 &  16    & 36  & 64      \\ \hline 
AV Rule 20                                  & 18 &  52    & 118 & 393      \\ \hline 
AV Rule 21                                  & 16 &  45    & 105 & 210      \\ \hline 
AV Rule 23                                  & 11 &  35    & 81  & 90      \\ \hline 
AV Rule 24                                  & 15 &  40    & 107 & 74      \\ \hline 
AV Rule 25                                  & 24 &  99    & 233 & 578      \\ \hline 
DF (good)                                   & 27 &  19    & 44  & 366      \\ \hline 
DF (bad)                                    & 27 &  19    & 43  & 171      \\ \hline %
read\_data (bad)                            & 74 & 131    & 296 & 474      \\ \hline
sec\_recv (bad)                             & 38 & 122    & 272 & 752        \\ \hline
\textbf{Total}                           & \textbf{273} & \textbf{612} & \textbf{1415} & \textbf{3266}         \\ 
\hline
\end{tabular}
\caption{Our \IsaBIL examples and their sizes}
\label{table:proof-stats}
\vspace{-6ex}
\end{wraptable}

\paragraph{Scalability} Our approach is \emph{compositional} in that we can analyse and (dis)prove properties about a \emph{code fragment}, say a function $f$, \emph{in isolation} and then \emph{reuse} these properties (without reproving them) in bigger contexts where $f$ is used. 
To show this, we use our technique to detect (using incorrectness logic) a \emph{double-free vulnerability} in the cURL library (used primarily for transferring data over the internet). 
As we demonstrate in \cref{sec:double-free-curl}, although this vulnerability is ``hidden'' in an internal function (not exposed in a header file) and is not visible to client applications, it is accessible by a call chain (comprising four separate functions) that can be triggered from an external function, i.e.~it is accessible from within a client application and thus this vulnerability can be easily exploited. 
Indeed, this vulnerability 
has been previously documented as \href{https://cve.mitre.org/cgi-bin/cvename.cgi?name=2016-8619}{CVE-2016-8619}. 

\paragraph{Contributions} Our main contributions are as follows. 
\begin{enumerate*}[label=\bfseries(\arabic*)]
\item We develop \isabil, a flexible, semi-automated tool for
  verifying binaries (lifted to BIL) within the Isabelle/HOL theorem
  prover. Our mechanisation is complete with respect to the BIL language
  and semantics. During the mechanisation, we uncovered and fixed several
  inconsistencies in the official BAP documentation (see
  \cref{sec:specification-of-bil}).

\item We developed \isabil as a highly flexible and extensible system (see
  \cref{fig:proofs}) using Isabelle {\em
    locales}~\cite{DBLP:conf/types/Ballarin03,kammuller1999locales}.
  We incorporate logics for (in)correctness, thus obtain methods for
  both proving and disproving properties. 
  To the best of our knowledge, this is the first application of
  Incorrectness logic~\cite{o2019incorrectness} in the analysis of
  low-level binaries.

\item We extend \isabil with a number of automation
  techniques, including reusable high-level big-step theorems, proof
  tactics that automatically discharge large numbers of proof goals,
  and architecture-specific proof optimisations. We also include a
  native parser for \BILa programs, written in Isabelle/ML, to
  \emph{automatically} transpile BAP outputs to an Isabelle/HOL
  locale suitable for
  verification. 

\item We apply our methods to a number of examples (see \cref{table:proof-stats}), including key
  tests from the Joint Strike Fighter (JSF) C++ Coding
  Standards~\cite{JSF2005}.

\item We show the scalability of \isabil by detecting
      a double-free vulnerability in a large example (253 assembly LoC) in the cURL library. 

\end{enumerate*}


\paragraph{Supplementary material} We have mechanised the proofs of all theorems stated in this paper in Isabelle/HOL (see \cite{IsabilZenodo}).

\section{Motivation and Workflow}
\label{sec:overview-and-motivation}

In this section, we motivate our work and provide an overview of our
overall approach using two running examples. We give the C program and
their equivalent BIL in~\cref{sec:c-program-its}. We present our
workflow and an overview of the Isabelle/HOL mechanisation in
\cref{sec:workfl-mech}.

\subsection{Two C programs and their BIL representations}
\label{sec:c-program-its}

\begin{figure}[t]
  \begin{minipage}[b]{0.49\columnwidth}
    \begin{mdframed}[backgroundcolor=backcolour,hidealllines=true,%
                     innerleftmargin=0,innerrightmargin=0,
                     innertopmargin=-0.2cm,innerbottommargin=-4.3ex]
\begin{lstlisting}[
    language=c,linerange={14-21},caption={Double Free (Bad)},label={DF-bad},
    linebackgroundcolor={%
    \ifnum\value{lstnumber}=5
        \color{red!10}
    \else
        \color{backcolour}
    \fi},
    basicstyle=\tt\fontsize{7}{7.5}\selectfont]
void bad() {
    int *p = malloc(42);
    if (MyTrue) {
        free(p);
        // ...
    }
    free(p);
}
\end{lstlisting}
    \end{mdframed}
    \vspace{-10pt}
  \end{minipage}
  \hfill
  \begin{minipage}[b]{0.47\columnwidth}
    \begin{mdframed}[backgroundcolor=backcolour,hidealllines=true,%
                     innerleftmargin=0,innerrightmargin=0,
                     innertopmargin=-0.2cm,innerbottommargin=-4.3ex]
\begin{lstlisting}[
    language=c,linerange={5-12},firstnumber=1,caption={Double Free (Good)},label={DF-good},
    linebackgroundcolor={%
    \ifnum\value{lstnumber}=5
        \color{green!10}
    \else
        \color{backcolour}
    \fi},
    basicstyle=\tt\fontsize{7}{7.5}\selectfont]
void good() {
    int *p = malloc(42);
    if (MyTrue) {
        free(p);
        return;
    }
    free(p);
}
\end{lstlisting}
    \end{mdframed}
    \vspace{-10pt}
  \end{minipage}
\end{figure}

\begin{figure}[t]
  \begin{minipage}[b]{0.49\columnwidth}
    \begin{mdframed}[backgroundcolor=backcolour,hidealllines=true,%
                     innerleftmargin=0,innerrightmargin=0,
                     innertopmargin=-0.2cm,innerbottommargin=-4.7ex]
\begin{lstlisting}[
    language=bir,
    linerange={1-17},
    caption={Pretty printed BIL (aka BIR) of \cref{DF-bad}},
    label={DF-bad-bil},
    escapeinside={(*}{*)},
    linebackgroundcolor={%
    \ifnum\value{lstnumber}=11
        \color{red!10}
    \else
        \color{backcolour}
    \fi},
    basicstyle=\tt\fontsize{7}{7.5}\selectfont]
sub bad(bad_result)

X10 := 0x2A
call @malloc with return (*\ref{line:malloc-return-bad}*)
mem := mem with [X8 - 0x18, el]:u64 <- X10 (*\label{line:malloc-return-bad}*)
X15 := mem[X3 - 0x7C8, el]:u32
when (X15 = 0) goto (*\ref{line:block-2-bad}*)
goto (*\ref{line:block-1-bad}*)

X10 := mem[X8 - 0x18, el]:u64(*\label{line:block-1-bad}*)
call @free with return (*\ref{line:block-2-bad}*)

X10 := mem[X8 - 0x18, el]:u64(*\label{line:block-2-bad}*)
call @free with return (*\ref{line:end-bad}*)(*\label{line:call-free-2-bad}*)

call X1 with noreturn(*\label{line:end-bad}*)
\end{lstlisting}
    \end{mdframed}
  \end{minipage}
  \hfill
  \begin{minipage}[b]{0.47\columnwidth}
    \begin{mdframed}[backgroundcolor=backcolour,hidealllines=true,%
                     innerleftmargin=0,innerrightmargin=0,
                     innertopmargin=-0.2cm,innerbottommargin=-4.7ex]
\begin{lstlisting}[
    language=bir,
    linerange={18-37},
    caption={Pretty printed BIL (aka BIR)  of \cref{DF-good}},
    label={DF-good-bil},
    escapeinside={(*}{*)},
    linebackgroundcolor={%
    \ifnum\value{lstnumber}=11
        \color{green!10}
    \else
        \color{backcolour}
    \fi},
    basicstyle=\tt\fontsize{7}{7.5}\selectfont]
sub good(good_result)

X10 := 0x2A
call @malloc with return (*\ref{line:malloc-return-good}*)
mem := mem with [X8 - 0x18, el]:u64 <- X10 (*\label{line:malloc-return-good}*)
X15 := mem[X3 - 0x7C8, el]:u32
when (X15 = 0) goto (*\ref{line:block-2-good}*)
goto (*\ref{line:block-1-good}*)

X10 := mem[X8 - 0x18, el]:u64(*\label{line:block-1-good}*)
call @free with return (*\ref{line:end-bad}*)(*\label{line:call-free-1-good}*)

X10 := mem[X8 - 0x18, el]:u64(*\label{line:block-2-good}*)
call @free with return (*\ref{line:end-good}*)(*\label{line:call-free-2-good}*)

call X1 with noreturn(*\label{line:end-good}*)
\end{lstlisting}
    \end{mdframed}
  \end{minipage}  
\end{figure}


We motivate our analysis using two variations of a double-free
vulnerability (listed as CWE-415 in the MITRE
database\footnote{\label{mitre}\url{https://cwe.mitre.org/data/definitions/415.html}}). A
double-free error occurs when a program calls \code{free} twice with
the same argument. This may cause two later calls to \code{malloc} to
return the same pointer, potentially giving an attacker control over
the data written to memory.

The occurrence of the double-free vulnerability is straightforward to
see in \cref{DF-bad}, but of course, in a real program, the
vulnerability may be much more difficult to identify. \cref{DF-good}
presents an alternative program that contains two occurrences of
\code{free(p)} with no intervening \code{malloc}. However,
\cref{DF-good} does not contain a double-free vulnerability since the
true branch returns from the method call before the second
\code{free(p)} is executed.

The focus of our work is to analyse such programs {\em without} relying
on any high-level language semantics. Instead, we aim to develop a
technique for reasoning about programs (e.g., those in
\cref{DF-bad,DF-good}) at a lower level of abstraction by first
compiling the programs then, lifting the resulting binaries to an
intermediate representation (BIL). Although such analysis contains
more detail, the analysis reflects the behaviour of the {\em actual}
executable, thus is more accurate (e.g., does not require any
additional compiler correctness assumptions). 




\begin{figure}[t]
  \centering
  \begin{minipage}[t]{\columnwidth}
  \centering
  \scalebox{0.8}{
  \begin{tikzpicture}[node distance=1cm, auto]
 \node (dummy) {};
 \node[punkt, dashed, inner sep=5pt,right=0.4cm of dummy] (Compiler) {\sf Compiler};
 \node[punkt, dashed, inner sep=5pt,right=1.2cm of Compiler]
 (BAP) {\sf BAP};
 \node[punkt, inner sep=5pt,right=1.2cm of BAP]
 (Isabelle) {\isabil};
 \node[fill=red!20, punkt, inner sep=2pt,above right=-.2cm and 1cm of Isabelle]
 (Incorrect) {\sf
   \begin{tabular}[t]{@{}c@{}}
     incorrectness proof~\cite{o2019incorrectness}
   \end{tabular}
 };
 \node[fill=green!20, punkt, inner sep=2pt,below right=-0.2cm and 1cm of Isabelle]
 (Correct) {\sf
   \begin{tabular}[t]{@{}c@{}}
correctness proof~\cite{hoare1969axiomatic}
   \end{tabular}
};
 \path[draw, thick, ->] (dummy)-- (Compiler)
 node[above,pos=0.2] {\tt *.c} ; 
 \path[draw, thick, ->] (Compiler)-- (BAP)
 node[above,pos=0.5] {\tt *.exe} ; 
 \path[draw, thick, ->] (BAP)-- (Isabelle)
 node[above,pos=0.5] {\tt *.bil} ; 

 \path[draw, thick, ->] (Isabelle)-- (Incorrect.west); 

 \path[draw, thick, ->] (Isabelle)-- (Correct.west); 
\end{tikzpicture}}
\vspace{-5pt}
 \caption{Translation of C source programs to an Isabelle/HOL proof}
 \label{fig:translation-c-isabelle}
\end{minipage}\medskip

\begin{minipage}[t]{\columnwidth}
  \centering
  \resizebox{\textwidth}{!}{%
\begin{tikzpicture}[yscale=-1]

  \node[spec] (BIL) {
    \large{(\cref{sec:mechanisation})} \\
   \large \BilLocale };
  \node[left = 0ex of BIL] {\Large \textbf{(1)}};
  
  \node[BILInf,below right = 0.4cm and 0.35cm of BIL] (BIL_inf) {
    \large{(\cref{sec:local-locale-1})} \\
    \large \BilInferenceLocale};
  \node[left = 0ex of BIL_inf] {\Large \textbf{(3)}};
  
  \node[specOP,above = 15ex of BIL_inf] (inference) {
    \large \InferenceLocale};

  \node[specOP,above left = 0.7cm and 0.05cm of inference,xshift=5mm] (correctness) {
    \large \CorrectnessLocale\\
    \cite{hoare1969axiomatic}};
  \node[below left = 0ex and 1.1ex of correctness] {\Large \textbf{(2)}};
  
  \node[specOP,above right = 0.7cm and 0cm of inference,xshift=-5mm] (incorrectness)
  { \large \IncorrectnessLocale\\
    \cite{o2019incorrectness}};

  \node[app,below = 1.4cm of BIL_inf] (find_symbol) {
    \large \FindSymbolLocale};

  \node[app,below left = 1.4cm and 1.9cm of BIL_inf,xshift=5mm] (allocation) {
    \phantom{y}\AllocationLocale\phantom{y}};

  \node[app,above right = 0.4cm and 0.1cm of find_symbol,xshift=-7mm] (AV) {
    \large \AvRuleLocale${}^*$};

  \node[app,above left = 0.4cm and 0.1cm of allocation,xshift=7mm] (DF) {
    \large \DoubleFreeBinaryLocale${}^*$};
  \node[above left = 1ex and -2.5ex of DF] {\Large \textbf{(4)}};

  \node[app,below = 0.7cm of find_symbol] (AV_proofs) {
    \large \AvRuleProofLocale};

  \node[app,below = 0.7cm  of allocation] (DF_proofs) {
    \large \DoubleFreeProofLocale};

  \node[curl,below right = 1ex and 16.5ex of BIL_inf] (reallocation) {
    \ReallocLocale};

  \node[curl,below = 2cm of reallocation] (read_data_proofs) {
    \ReadDataBadLocale};

  \node[curl, right = 0.7cm  of reallocation] (LC) {
    \large \LibCurlLocale$^*$};

  \node[curl,right = 3.9cm of read_data_proofs] (SR) {
    \large \SecRecvBadLocale};
  
  \node[curl,above = 0.7cm of SR] (CM) {
    \large \ChooseMechBadLocale};

  \node[curl,above = 0.7cm of CM] (CSL) {
    \large \CurlSecLoginBadLocale};
  \node[above right = 1ex and -2.5ex of CSL] {\Large \textbf{(5)}};

  \draw[refines] (BIL_inf) to node[above,yshift=2mm]{\large sublocale} (reallocation);
  \draw[refines] (reallocation) to node[left]{\large sublocale} (read_data_proofs);

  \draw[faithful] (LC) to node[right]{\large inheritance} (read_data_proofs);

  \draw[refines] (read_data_proofs) to node[below]{\large comp.} (SR);
  \draw[refines] (SR) to node[right]{\large comp.} (CM);
  \draw[refines] (CM) to node[right]{\large comp.} (CSL);

  \draw[refines] (correctness) to node[left]{\large sublocale} (inference);
  \draw[refines] (incorrectness) to node[right]{\large sublocale} (inference);

  \draw[faithful] (BIL) to node[left,pos=0.6]{\large inheritance} (BIL_inf);
  \draw[refines] (inference) to node[right,pos=0.6]{\large sublocale} (BIL_inf);
  \draw[refines] (BIL_inf) to node[left,pos=0.4,xshift=-1mm]{\large sublocale} (allocation);
  \draw[refines] (BIL_inf) to node[left,pos=0.4]{\large sublocale} (find_symbol);

  \draw[refines] (find_symbol) to node[left]{\large sublocale} (AV_proofs);
  \draw[refines] (allocation) to node[left]{\large sublocale} (DF_proofs);

  \draw[faithful] (AV) to[bend right] node[left,rotate=-90,xshift=10mm,yshift=4mm]{\large inheritance} (AV_proofs.east);
  \draw[faithful] (DF) to[bend left] node[right,rotate=-90,xshift=-10mm,yshift=-2.5mm]{\large inheritance} (DF_proofs.west);

  \draw[black,thick,dotted] ($(correctness.north west)+(-0.2,-0.2)$)  rectangle ($(inference.south east)+(2.5,0.2)$) node [pos=0.5,xshift=-0.2cm,yshift=1cm]{\large{(\cref{sec:incorrectness})}};

  \draw[black,thick,dotted] ($(DF.north west)+(-0.2,-0.2)$)  rectangle
  ($(AV_proofs.south east)+(1.4,0.6)$) node [pos=0.5,xshift=8mm,yshift=-1.5cm]{\large{(\cref{sec:example-proofs})}};
  
  \draw[black,thick,dotted] ($(reallocation.north west)+(-0.5,-0.2)$)  rectangle ($(read_data_proofs.south east)+(2.9,0.7)$) node [pos=0.5,yshift=-1.6cm]{\large{(\cref{sec:incorr-readd-subr})}};

  \draw[black,thick,dotted] ($(CSL.north west)+(-0.2,-0.2)$)  rectangle ($(SR.south east)+(0.8,0.7)$) node [pos=0.5,xshift=0cm,yshift=-1.6cm]{\large{(\cref{sec:compositionality})}};


\end{tikzpicture}}
\vspace{-5pt}

\caption{Structure of \isabil locales: theories indicated by ${}^*$ are auto-generated from a given {\tt *.bil} file following the 
         translation in \cref{fig:translation-c-isabelle}.}
\label{fig:proofs}
\vspace{-10pt}
  \end{minipage}
\end{figure}

\subsection{Workflow and Mechanisation}
\label{sec:workfl-mech}
Our overall workflow is given in \cref{fig:translation-c-isabelle},
where the dashed steps represent existing work. The first step to
proving properties about our example program is to use BAP to generate
a BIL representation of a binary executable
(see~\cref{fig:translation-c-isabelle}). For our example program in
\cref{DF-bad,DF-good}, the BIL equivalent (expressed in BIR
format) are given in \cref{DF-bad-bil,DF-good-bil}, respectively.
BIR (Binary Intermediate Representation) is a structured refinement of 
BIL that significantly enhances human readability by presenting 
sequences of BIL as a single instruction. Take for example the BIR 
instruction \lstinline[language=birn]{call @free with return 18} on line 16 in 
\cref{DF-bad-bil,DF-good-bil}. This represents two BIL statements, a 
jump to the address of the symbol ``free'', and the assignment of the 
return value 18 to the return pointer. Note that \isabil takes as 
input BAP's {\em abstract data type} format
(see~\cref{sec:mechanisation}), which we refer to as \BILa, as opposed
to the human-readable forms shown in
\cref{DF-bad-bil,DF-good-bil}. This makes the BIL instructions simpler
to parse when generating the corresponding Isabelle/HOL object
(discussed in more detail below).

Extensions to Isabelle/HOL's logic~\cite{DBLP:conf/cpp/EdmondsP24,DBLP:conf/sp/BauereissG0R17,DBLP:conf/sp/KammullerK16,DBLP:journals/corr/GomesKMB17}
often make use of Isabelle/HOL's {\em locale}
system~\cite{kammuller1999locales,DBLP:conf/types/Ballarin03}, which 
provides convenient modules for building parametric theories. 
Locales are aimed at supporting the use of local 
assumptions and definitions for collections of similar theories,
defined in terms of fixed variables and definitions. 
We refer to the fixed variables as parameters which must satisfy the 
local assumptions. 
For a locale $\mathcal{L}$, we give its definition as $\mathcal{L} = (p_1, p_2, \dots)$
where $p_1, p_2$ are the parameters of the locale.
Each parameter may be any type derived from a set or a relation.
From these variables, assumptions and definitions we can derive 
general theorems within the context of a locale.
These theorems can be exported to the current proof context by \emph{instantiation}, 
in which we provide values to one or all of the parameters and verify the fixed
assumptions of these parameters.
This effectively gives us the general theorems contained within each locale for free 
for a specific instantiation. 
Locales can be extended through \emph{inheritance} or \emph{interpretation}, 
enabling reusability and specialisation by introducing and refining a locale's parameters. 
Inheritance operates hierarchically, directly propagating the context of a parent 
locale to a child locale.
This transfer includes general theorems, assumptions, definitions and fixed variables.
Interpretation is compositional, associating a locale or definition with 
another locale. During interpretation, the assumptions of the 
parent locale must be satisfied as part of the mapping process.
A locale that interprets another locale is referred to as a sublocale,
denoted $\mathcal{L} \sublocale \mathcal{L}'$, where locale $\mathcal{L}'$ 
is the parent of $\mathcal{L}$.
For \isabil, they provide a uniform system for representing
both the program, and the underlying reasoning framework.

The overall structure of the \isabil development is given in
\cref{fig:proofs}, and comprises four main parts (which highlight our
four main contributions). Overall, we have:
\begin{enumerate*}[label=\bfseries(\arabic*)]
\item a locale (see~\cref{sec:mechanisation}), representing a complete
  formalisation of the BIL specification from
  \cref{sec:specification-of-bil},
\item a set of locales (see~\cref{sec:incorrectness}) that encode
  Hoare Logic (to verify correctness) and Incorrectness Logic (as
  developed by O'Hearn),
\item a locale (see~\cref{sec:local-locale-1}) that combines formalisation
  of the BIL specification with (in)correctness logics, 
\item a set of locales (see~\cref{sec:example-proofs}) that apply the
  resulting verification framework to prove correctness of a number of
  examples from the literature and
\item a set of locales
  (see~\cref{sec:incorr-readd-subr,sec:compositionality}) that
  demonstrate scalability of the approach. In particular,
  \cref{sec:incorr-readd-subr} demonstrates the use of subroutines to
  enable one to focus on key functions of interest, even when starting
  with  large BIL files (> 100k LoC). Then
  \cref{sec:compositionality} describes how verified subroutines can
  be used as subcomponents to prove (in)correctness of a larger
  system.
\end{enumerate*}

Our framework is highly flexible and extensible. For instance, the
\BilInferenceLocale locale combines the
\BilLocale and \InferenceLocale locales. In
case one wishes to change the specification (for a different
operational semantics), or use a different set of inference rules (to
perform a type of analysis), one can simply change the specification
and/or inference rule locales. Similarly, a locale can be extended
with additional semantic features: our proofs extend the
\BilInferenceLocale locale with an \AllocationLocale locale (that
models memory allocation) and a \FindSymbolLocale locale (that
tracks a symbol table) to enable reasoning about different program
features.

Each of these contributions contains several innovations. Our encoding
of BIL~(\cref{sec:mechanisation}) is the {\em first} full
mechanisation of BIL in Isabelle/HOL. This mechanisation is a
standalone contribution and can be used by others outside of our
reasoning framework. Our encoding uncovers some inconsistencies in the
official specification, which are addressed by the mechanisation (see
\cref{sec:specification-of-bil}). Moreover, given that BIL is an
intermediate language for many different architectures, future
versions could be further optimised to a particular architecture. As
an example, we present optimisations for RISC-V (see \cref{sec:riscv-optimisations}), which is the
architecture that we focus on.

To maximise flexibility, we provide an encoding of the inference rules
parameterised by an operational semantics, i.e., an {\em instance} of
the correctness and incorrectness rules can be generated for {\em any}
operational semantics. This instance automatically inherits all of the
rules of Hoare and Incorrectness logic that we prove generically in
the \CorrectnessLocale and \IncorrectnessLocale locales (\cref{sec:incorrectness}). 

The BIL semantics from \cref{sec:mechanisation} and the inference
rules from \cref{sec:incorrectness} are combined to form a new locale,
\BilInferenceLocale, which provides facilities to verify BIL
programs. Then, as discussed above, to verify particular types of
examples, we extend the locales with models of allocation and symbols 
to verify particular sets of examples. These proofs represent the {\em
  first} proofs of both correctness and incorrectness for BIL programs
in Isabelle/HOL. As we shall see in \cref{sec:specification-of-bil},
the BIL semantics is highly detailed, with the reductions related to
even single load command potentially splitting into a large number of
{\em small-step} transitions. This has the potential to increase the
verification complexity making the proofs of even small programs
intractable. To address this, we use BIL's big-step semantics
(\cref{sec:big-step-semantics}) and introduce a number of high-level
lemmas that allow one to discharge such proofs
generically. These lemmas are reusable across all of the examples that
we verify.

Our proofs cover key examples from the JSF coding
standards~\cite{JSF2005} and the CWE database\footnoteref{mitre}. This
task is greatly simplified using \isabil's parser, written in
Isabelle/ML~\cite{urban2019isabelle}, that {\em automatically
  generates} a locale corresponding to a BIL program written in \BILa
format. Note that Isabelle isolates Isabelle/ML programming, and hence
our parser cannot interfere with soundness of Isabelle's core logic
engine. The generated locale is then combined with
\BilInferenceLocale to provide a context in which (in)correctness
proofs can be carried out.


\section{\BIL Syntax and Semantics}
\label{sec:specification-of-bil}



Before proceeding with the presentation of our proofs, it is essential
to acquire a basic understanding of \BIL. This section serves as a
succinct introduction to \BIL, emphasizing its type system, memory
model and operational semantics. We encourage interested readers to
explore the \BIL manual~\cite{DBLP:conf/cav/BrumleyJAS11} (see
\cref{appendix:bil-specification}).

\subsection{Syntax}
\label{sec:specification-of-bil-syntax}

\begin{figure}[t] \small\addtolength{\jot}{-2pt}
  \raggedright{\bf \emph{Statements}}, where $str$ denotes a string
  (of type $\String$) and $i \in \mathbb{Z}$\vspace{-2pt}
  \begin{align*}
    stmt \ni s & ::=
                 var := e
                 \mid \StatementJmpE{e}
                 \mid \StatementCpuExnE{i}
                 \mid \StatementSpecialE{str}
                 \mid \StatementWhileE{e}{seq}
                 \mid \StatementIfThen
                 \mid \StatementIfThenElse \\
    bil \ni seq & ::= stmt^*
  \end{align*}
\vspace{-10pt}

\raggedright{\bf \emph{Expressions}}, where $id$ denotes a variable name literal (of type \emph{string})
\vspace{-2pt}
\begin{align*}
exp \ni e ::=& \ \inarr{v \mid var  \mid\ExpressionLoad \mid \ExpressionStore \mid \ExpressionBinOp \mid \ExpressionUnOp \\
         {} \mid \ExpressionCast\mid 
             \ExpressionLet \mid \ExpressionIte \mid \ExpressionConcat \mid\ExpressionExtract } \\
  bop ::=&\ aop\mid lop \quad\ \  var ::= id : type \quad\ \  endian \ni ed ::= \EndianLittle \!\mid\! \EndianBig \quad\ \  
 cast ::=\ \CastLow \!\mid\! \CastHigh \!\mid\! \CastSigned \!\mid\! \CastUnsigned 
\end{align*} \vspace{-10pt}

\raggedright{\bf \emph{Types and Values}}, where $sz, nat \in \nat$ \vspace{-2pt}
\begin{align*}
   type \ni t ::= &\ \TypeImmediate \ | \ \TypeMemory  
   & word \ni w ::= &\ \Word \\ 
  value \ni v ::= &\ \ValueImmediate \ | \ \ValueMemory \ | \ \ValueUnknown
\end{align*}
\vspace{-15pt}
\caption{BIL syntax and types}
\label{BIL:syntax}
\vspace{-10pt}\end{figure}

The basic syntax of BIL programs is given in \cref{BIL:syntax}.

\paragraph{Statements}
A BIL statement may assign an expression $e$ to a variable $var$,
transfer control to a given address $e$, interrupt the CPU with a
given interrupt $num$, be an instruction with unknown semantics, a
while loop, or an if-then-else conditional (with an optional else
clause). Note that each BIL statement (including a compound statement)
corresponds to a single machine instruction, 
thus side effects such as 
setting a status register must be captured in the statement definition
(see \cref{sec:statement-semantics}).

\paragraph{Expressions}
Memory loads and stores are represented at the expression level as $\ExpressionLoad$ for loads, 
and $\ExpressionStore$ for stores. 
In these expressions, the evaluation of $e_1$ represents the target storage being accessed, while the evaluation of $e_2$ specifies the address.
Since a memory operation can span multiple addresses, both the endianess ($ed$) and the size ($sz$) of the operation are provided.
For stores, the evaluated value of $e_3$ is written to memory at the given address.
Binary operations now differentiate arithmetic (\emph{aop})  and logical
(\emph{lop}) operations.  Expressions in
\BIL also include casts, let bindings, if-then-else expressions,
concatenations and extractions.
The extraction operation, denoted as $\ExpressionExtract$, first evaluates the 
expression $e$ to a word $w$ and then extracts a slice from $w$ using BIL's 
existing extraction mechanism. 
Specifically, $\WordExtract$ represents the bits of $w$
from bit $sz_1$ to bit $sz_2$. Thus, for example,
$\WordExtractE{01001011}{5}{2}$ is $0010$. 
Expressions are side-effect free and are evaluated wrt a state
$\BILVariables : var \to val$ mapping variables to values. 

\paragraph{Types and Values}
The highly expressible nature of \BIL is due to its type system, which
defines two distinct types for its values, the irreducible subset of a
BIL expression. These types are  $\TypeImmediate$ (describing an {\em
  immediate type} of size $\DefaultSize$) and $\TypeMemory$
(describing a {\em storage type} with address size $\AddressSize$ and
value size $\ValueSize$). 
A value can be bound to one of the three possibilities.
\begin{enumerate*}[label=\bfseries(\arabic*)]
\item A machine word ($\ValueImmediate$).
\item An abstract storage ($\ValueMemory$), which allows us to define
  a memory as a chain of mappings, where
  $[w \leftarrow v', \ValueSize]$ defines a single mapping from an
  address ($w$) to a value ($v'$) of size $\ValueSize$. Here, $v$
  could either be the root of the chain (in which case it is
  $\ValueUnknown$) or another storage. Technically, $v$ could also be
  a word, but we disallow this by introducing a typing context (see
  below). Similarly, $v'$ could be a storage according to the grammar,
  but this possibility is also eliminated by the typing context.  When
  $v$ or $v'$ is an unknown value, we require that type of $v$ is {\bf
    mem} and the type of $v'$ is {\bf imm}.
\item An ``unknown'' value ($\ValueUnknown$), which is obtained from
  the evaluation of a \BIL expression holding some string information
  {\it str} of some type $t$.
\end{enumerate*}
The type of a value can be obtained using
the function ${\tt type}$, which is define as
follows: 
\begin{align*}
  \TypeFunctionE{\Word} & = \ \TypeImmediate \\
  \TypeFunctionE{\ValueMemoryE{v}{(\WordE{nat}{\AddressSize})}{v'}{\ValueSize}} & = \ \TypeMemory \\
  \TypeFunctionE{\ValueUnknown} & =\ t 
\end{align*}
In \cref{sec:riscv-to-bil} we show the lifted BIL equivalent of a
RISC-V program using the syntax presented in this section.



\subsection{Typing Rules and Typing Context} 
BIL's type system facilitates the need for
typing rules to ensure type correctness. At the lowest
level, the rules ensure that types are correctly defined with the
predicate $\TypeIsOk$. The type of variables must be tracked during symbolic execution to
ensure that they do not change and only values of the correct type can
be assigned. This is achieved via a \emph{typing context}, defined as
$\TypingContext ::= var^{*}$, where type correctness of $\TypingContext$ is expressed 
by overloading the predicate $\TypingContextIsOk$. 
The rules for $\TypeIsOk$ and $\TypingContextIsOk$ are given below:
{\small%
\begin{mathpar} 
\inferrule[(\sc twf\_imm)]{
  sz > 0
}{
  \TypeIsOkE{\TypeImmediate}
}
\and
\inferrule[(\sc twf\_mem)]{\ 
  \AddressSize > 0 \qquad  \ValueSize > 0
}{
  \TypeIsOkE{\TypeMemory}
}
\and
\inferrule[\sc (tg\_nil)]{
}{
  \TypingContextIsOkE{[]}
}
\and
\inferrule[\sc (tg\_cons)]{\ 
  str \notin dom(\TypingContext) \qquad
  \TypeIsOk \qquad
  \TypingContextIsOk
}{
  \TypingContextIsOkE{\TypingContextConcat}
}
\end{mathpar}}
These are extended to expressions, then to the level of statements
(see~\cite{DBLP:conf/cav/BrumleyJAS11} for details). As part of our
formalisation, we discover a missing typing rule for empty sequences,
which would otherwise be needed to ensure the sequencing rules are
type correct.\footnote{We have created a Git pull request in the BAP
  repositories for this issue. }

\subsection{Expression Semantics}
\label{sec:exp-semantics}
Expression evaluation requires the repeated application of small-step
semantics rules until the expression is reduced to a value.
An expression step in \BIL is formally expressed with the
$\BILStep$, where multiple steps can be reduced using the reflexive 
transitive closure $\BILSteps$.
Recall that loads and stores occur at the expression level,
and are sensitive to endian orderings.
We provide the load rules here, and refer the interested
reader to the BIL manual~\cite{DBLP:conf/cav/BrumleyJAS11} for further
details.

A store instruction (semantics not shown) targets a single unit of
addressable memory, typically 8-bits in size. When a storage operation
exceeds this size, it is converted into a sequence of 8-bit stores,
organised in big-endian order. 
The reduction rules for load operations
is given in \cref{fig:load-reduction}. First, all sub-expressions of a
memory object are reduced to values using the rules {\sc
  load\_step\_addr} and {\sc load\_step\_mem}. Then, the resulting
object is recursively deconstructed using the {\sc load\_word\_el} and
{\sc load\_word\_be} rules, depending on the endian. This process
continues until one of {\sc load\_byte}, {\sc load\_un\_mem} or {\sc
  load\_un\_addr} is used. {\sc load\_byte} reduces the expression to
the value, $v'$, when the memory object is a storage of an immediate
(known) value, $w$. {\sc load\_un\_mem} and {\sc load\_un\_addr} both
reduce the expression to an unknown value when the memory or address
being loaded is unknown, respectively. 

\begin{figure}[t]
  \centering \footnotesize
  \begin{mathpar}
\inferrule[(\sc var\_in)]{
  \VarIn
}{\SmallStepExp
  {\Variables}
  {var}
  {v}
}
\and
\inferrule[(\sc load\_step\_addr)]{
  \SmallStepExp{\Variables}{e_2}{e_2'}
}{\SmallStepExp
  {\Variables}
  {\ExpressionLoadE{e_1}{e_2}{ed}{sz}}
  {\ExpressionLoadE{e_1}{e_2'}{ed}{sz}}
}
\and
\inferrule[(\sc load\_step\_mem)]{
  \SmallStepExp{\Variables}{e_1}{e_1'}
}{\SmallStepExp
  {\Variables}
  {\ExpressionLoadE{e_1}{v_2}{ed}{sz}}
  {\ExpressionLoadE{e_1'}{v_2}{ed}{sz}}
}
\and
\inferrule[(\sc load\_byte)]{
}{\SmallStepExp
  {\Variables}
  {\ExpressionLoadE{\ValueMemoryE{v}{w}{v'}{sz}}{w}{ed}{sz}}
  {v'}
}
\and
\inferrule[(\sc load\_byte\_from\_next)]{
  w_1 \neq w_2
}{\SmallStepExp
  {\Variables}
  {\ExpressionLoadE{\ValueMemoryE{v}{w_1}{v'}{sz}}{w_2}{ed}{sz}}
  {\ExpressionLoadE{v}{w_2}{ed}{sz}}
}
\and
\inferrule[(\sc load\_un\_mem)]{
}{\SmallStepExpLB
  {\Variables}
  {\ExpressionLoadE{(\ValueUnknownE{str}{t})}{v}{ed}{sz}}
  {\ValueUnknownE{str}{\TypeImmediateE{sz}}}
}
\and
\inferrule[(\sc load\_un\_addr)]{
}{\SmallStepExpLB
  {\Variables}
  {\ExpressionLoadE{\ValueMemoryE{v}{w_1}{v'}{sz'}}{\ValueUnknownE{str}{t}}{ed}{sz}}
  {\ValueUnknownE{str}{\TypeImmediateE{sz}}}
}
\and
\inferrule[(\sc load\_word\_be)]{
  sz > sz_{mem} \and
  \Succ{w} = w' \\\\
  \TypeFunctionE{v} = \TypeMemoryE{sz_{addr}}{sz_{mem}}
}{\SmallStepExpLB
  {\Variables}
  {\ExpressionLoadE{v}{w}{\EndianBig}{sz}}
  {\ExpressionConcatE
    {\ExpressionLoadE{v}{w}{\EndianBig}{sz_{mem}}}
    {\ExpressionLoadE{v}{w'}{\EndianBig}{(sz - sz_{mem})}}
  }
}
\and 
\inferrule[(\sc load\_word\_el)]{
  sz > sz_{mem} \and
  \Succ{w} = w'\\\\
  \TypeFunctionE{v} = \TypeMemoryE{sz_{addr}}{sz_{mem}}
}{\SmallStepExpLB
  {\Variables}
  {\ExpressionLoadE{v}{w}{\EndianLittle}{sz}}
  {\ExpressionConcatE
    {\ExpressionLoadE{v}{w'}{\EndianLittle}{(sz - sz_{mem})}}
    {\ExpressionLoadE{v}{w}{\EndianLittle}{sz_{mem}}}
  }
}
\end{mathpar}
\vspace{-5pt}
\caption{Reduction rules for Load}
\label{fig:load-reduction}
\end{figure}


\begin{example}
  \label{ex:load}
  Consider  the  instruction, \lstinline[language=birn]{X10 := mem[X8-0x18, el]:u64},
  from line~12 in
  \cref{DF-bad-bil}, which performs a little-endian load of a 64-bit word from address  \lstinline[language=birn]{X8-0x18} in the variable
  \lstinline[language=birn]{mem}.
  We first reduce the address expression,
  \lstinline[language=birn]{X8-0x18}, to a value using {\sc
    load\_step\_addr}, next we apply {\sc load\_step\_mem} to read the
  value stored in \lstinline[language=birn]{mem}. Next, we look up the
  value of \lstinline[language=birn]{mem} in $\Variables$ in {\sc
    var\_in}, which reduces \lstinline[language=birn]{mem} to the
  corresponding storage.  Next, we apply {\sc load\_word\_el} to break
  down the 64-bit load into 8 separate loads in little-endian
  format. Each of the 8 byte-sized loads are individually read from memory
  using {\sc load\_byte} and {\sc load\_byte\_from\_next}. Finally, we
  apply BIL's concatenation rule (not shown) to join each byte into a
  single 64-bit word.
\end{example}

\paragraph{IsaBIL Extensions}
The previous sections present a summary of BIL's semantics from the BIL manual~\cite{DBLP:conf/cav/BrumleyJAS11}
with some minor corrections.
In this section, we present extensions to BIL's big-step
expression evaluation semantics. 
The reduction of expressions, e.g.,
in load and store instructions, results in a significant proof burden.
To make verification tractable, \isabil defines additional big-step
expression evaluation rules which are derived from a combination of
BIL's existing small-step and big-step expression semantics. For
example, the load expression in \cref{ex:load} requires the repeated
application of over 150 small-step rules. Splitting the 64-bit read
into individual reads in \cref{ex:load} (from {\sc load\_word\_el}
onwards) and the concatenation of the result can be combined into a
single big-step rule. We first provide some syntactic sugar for a
storage whose size is a multiple of $8$.

The definition uses the function $\Succ{w}$, which retrieves the
successor of the word $w$ such that $\Succ{w} ::= w +
1$
. Furthermore, we use
$\ReflStorage{en}{v}{w}{v'}{N}$ to refer to a contiguous storage of
the value $v'$ of size $N \in \{8i \mid i \in \nat\}$ (i.e., $N$ is a
multiple of $8$) at address $w$ on the storage $v$:
{\small%
$$ 
\begin{array}[t]{@{}r@{~}l@{}}
  \ReflStorageEL{\InitialStorage}{w}{v'}{N} & \doteq
    \begin{cases}
      \ReflStorageEL
        {\ValueMemoryE{\InitialStorage}{w}{\WordExtractE{v'}{7}{0}}{8}}
        {\\\Succ{w}}
        { \WordExtractE{v'}{N - 1}{8}}
        {N - 8}
      & \textbf{if $N > 8$} \\
      \ValueMemoryE{\InitialStorage}{w}{v'}{8} & \textbf{otherwise}
    \end{cases}
  \\ 
  \ReflStorageBE{\InitialStorage}{w}{v'}{N} & \doteq
    \begin{cases}
      \ReflStorageBE
        {\ValueMemoryE{\InitialStorage}{w}{\WordExtractE{v'}{N - 1}{N - 8}}{8}}
        {\\ \Succ{w}}
        {\WordExtractE{v'}{N - 9}{0}}
        {N - 8}
      & \textbf{if $N > 8$} \\
      \ValueMemoryE{\InitialStorage}{w}{v'}{8} & \textbf{otherwise} 
    \end{cases}
  \\ 
\end{array}
$$}
For example, suppose $B_1$ and $B_2$ are two 8-bit words. We have:
\begin{align*}
\ReflStorageEL{\InitialStorage}{w}{B_1B_2}{16} = v[w \leftarrow B_2, 8][\Succ{w} \leftarrow B_1, 8] \\\ReflStorageBE{\InitialStorage}{w}{B_1B_2}{16} = v[w \leftarrow B_1, 8][\Succ{w} \leftarrow B_2, 8]   
\end{align*}

We give a subset of these rules for little endian 64-bit load and
store operations below.
\begin{small}
\begin{mathpar}
\inferrule[(\sc refl\_load\_el\_word64)]{
}{
  \BigStepExp
    {\Variables}
    {\ExpressionLoadE
        {\ReflStorageEL{v}{w_2}{w_1}{64}}
        {\ReflAddress}
        {64}        
        {\EndianLittle}
    }
    {\ReflValue}
}
\and
\inferrule[(\sc refl\_store\_el\_word64)]{
}{
    {\Variables} \vdash
    {\ExpressionStoreE{v}{\ReflAddress}{\EndianLittle}{64}{\ReflValue}} \leadsto^* \\\\
    {{\ReflStorageEL{v}{w_2}{w_1}{64}}}
}
\end{mathpar}
\end{small}
\begin{example}
  \label{ex:load2}
  Consider the load from \cref{ex:load}. The first few steps proceed
  as before. However, instead of applying {\sc load\_word\_el}, we can
  directly apply {\sc refl\_load\_el\_word64} to obtain the
  corresponding value $\BigStepExp{\Variables}{(\ReflStorageEL{v}{w}{v'}{64})[w, el]:u64}{v'}$. 
\end{example}

All big-step rules from this section, such as {\sc refl\_load\_el\_word64} have been verified in Isabelle wrt to the small step 
semantics.

\subsection{Statement semantics}
\label{sec:statement-semantics}
\newcommand{\BrumleySequenceStepE}[6]{a}
\newcommand{\BrumleySequenceStep}{\BrumleySequenceStepE{\Variables}{w}{seq}{\Variables'}{w'}{seq'}}

In this section, we describe the operational semantics for evaluating
statements. We also provide a correction to the sequencing rule from
the manual~\cite{BAP-TOOLKIT}. Although each statement induces local
control flow, in BIL, they are assumed to execute to completion in a
single step.

The semantics of a statement is defined by
$(\Variables,\ pc) \vdash seq \leadsto (\Variables',\ pc')$, which
executes the $bil$ statement $seq$ from the variable store
$\Variables$ to generate a new variable store $\Variables'$ and
program counter $pc'$. The {\sc move} statement (see below) modifies
$\Delta$ by generating a new variable binding, and {\sc jmp} (not
shown) affects program counter.

Example rules for {\sc move} (aka
assignment) and {\sc if\_true} (for branching on a true guard are
given below). In the move rule, the given expression $e$ is evaluated
to a value using the big-step semantics, and the value for the given
variable $var$ is updated in the variable state. In the {\sc if\_true}
rule, the guard is evaluated (to $true$) then the statement $seq_1$ is
executed. The sequencing rules describe execution of a list of
statements to completion. The sequencing rules in
manual~\cite{BAP-TOOLKIT} do not provide a means to reduce multiple
statements in a single transition\footnote{We have created a Git pull request in the BAP
  repositories for this issue. }.
{\small%
\begin{mathpar}
\inferrule[(\sc move)]{
 \BILStepsE{\Variables}{e}{v}
}
{
\StatementStepE
  {(\Variables,\ pc)}
  {\StatementMoveE{var}{e}}
  {(\Variables(var \mapsto v),\ pc)}
}
\and
\inferrule[(\sc if\_true)]{
 \BILStepsE{\Variables}{e}{true}\and
  \SequenceStepE{\Variables}{pc}{seq_1}{\Variables'}{pc'}
}
{
\StatementStepE
  {(\Variables,\ pc)}
  {\StatementIfThenElse}
  {(\Variables',\ pc')}
}
\and\inferrule[(\sc seq\_rec)]{
\StatementStepE{(\Variables_1,\ pc_1)}{s_1}{(\Variables_2,\ pc_2)} \\ 
\SequenceStepE{\Variables}{pc_2}{s_2 \ldots  s_n}{\Variables_3}{pc_3}
}{
\SequenceStepE{\Variables_1}{pc_1}{s_1 s_2 \ldots  s_n}{\Variables_3}{pc_3}
}
\and
\inferrule[(\sc seq\_nil)]{
}{
\SequenceStepE{\Variables}{pc}{\varepsilon}{\Variables}{pc}
}
{}
\end{mathpar}}

\begin{example}
  \label{ex:move}
  Consider the load expression from \cref{ex:load}, which appears on the right-hand side of the assignment statement given below: 
\[
\SequenceStepE{\Variables}{(\WordE{12}{64}) + (\WordE{1}{64})}{\{\text{\lstinline[language=birn]{X10 := mem[X8-0x18, el]:u64}}\}}{\Variables(X10 \mapsto v')}{(\WordE{13}{64})}
\]
  To apply the statement semantics, we start by deconstructing it using {\sc seq\_rec}, which results in two sub-steps. For the first, we apply {\sc move} and the steps in \cref{ex:load} to  reduce the expression \text{\lstinline[language=birn]{mem[X8-0x18,el]:u64}} to a value, $v'$. Since the original sequence only contained a single statement, our recursive sequence (i.e., the second premise of {\sc seq\_rec}) is empty, which can be trivially reduced using {\sc seq\_nil}. 
\end{example}

\subsection{BIL step relation} 
\label{section:instr-semantics}

A machine instruction in \BIL is represented by a named tuple,
$insn = \InsnTuple$ (implemented later as an Isabelle {\tt
  record}). Here \textbf{addr}, refers to the instruction's address in
memory; \textbf{size}, the size of the instruction; and \textbf{code},
the semantics of the program represented by \BIL
statements. Instructions operate over the \BIL machine state,
$\BILMachineStateE{\Delta}{pc}{mem}$, where $\Delta$ is the variable
store, $pc$ is the program counter and $mem$ is a variable that
denotes the memory. To decode the current instruction from a machine
state, we assume an 
uninterpreted $\BILDecodeTransition$ function, referred to as the
\emph{decode predicate}, which maps a machine state,
$\BILMachineStateE{\Delta}{pc}{mem}$, to the corresponding
instruction, $\InsnTuple$.  The BIL step relation
$\BILMachineStateStep$ defines the execution of a single BIL statement
as shown below:
\begin{gather*}
\inferrule[(\ProgramStepRuleName)]{
\BILDecodeInstructionE{\BILMachineStateE{\Delta}{pc}{mem}}{\InsnExpandedE{pc}{z}{bil}} \\
\SequenceStepE{\Variables}{pc + z}{bil}{\Variables'}{pc'}
}{
\BILMachineStateStepInitialE{\BILMachineStateE{\Variables}{pc}{mem}}{\BILMachineStateE{\Variables'}{pc'}{mem}}
}
\end{gather*}

\begin{example}
  \label{ex:load3}
  Consider the move statement from \cref{ex:move}, which resides at
  the program address 12 and being of one-byte size. This is expressed
  as the BIL machine instruction below:
  \[
  \InsnExpandedE{\WordE{12}{64}}{\WordE{1}{64}}{\{\text{\lstinline[language=birn]{X10 := mem[X8-0x18, el]:u64}}\}}
  \]
  The instruction is obtained by decoding $(\BILDecodeTransition)$ a
  machine state of the form
  $\BILMachineStateE{\Delta}{\WordE{12}{64}}{mem}$ where
  $X8, mem \in {\tt dom}(\Delta)$. By using decode we can structure a
  step proof for the machine state as follows:
  \[
  \BILMachineStateStepInitialE{\BILMachineStateE{\Delta}{\WordE{12}{64}}{mem}}{\BILMachineStateE{\Variables(X10 \mapsto v')}{\WordE{13}{64}}{mem}}
  \]
  By applying the \ProgramStepRuleName rule for programs, we are left with a
  reduction
  $\SequenceStepE{\Variables}{(\WordE{12}{64}) +
    (\WordE{1}{64})}{\{\text{\lstinline[language=birn]{X10 :=
        mem[X8-0x18, el]:u64}}\}}{\Variables(X10 \mapsto v')}{(\WordE{13}{64})}$, which we derived
  in \cref{ex:move}.
 
\end{example}


\section{Mechanisation}
\label{sec:mechanisation}


In this section, we describe the \isabil mechanisation, which utilises
extensible {\em locales} to maximize reusability.  In
\cref{sec:local-locale}, we describe the generic
\BilLocale locale that formalises BIL's syntax and
semantics, and in \cref{sec:program-specification} we describe the
\isabil translation tool that automatically generates locales from BIL
programs.




\subsection{The \BilLocale Locale}
\label{sec:local-locale}

\isabil's \BilLocale locale (see \cref{fig:proofs})
provides a complete mechanisation of BIL's syntax and
semantics\footnote{As outlined in \cref{sec:specification-of-bil} and
  provided in full in \cref{appendix:bil-specification}.}. We encode
BIL's operational rules (\cref{section:instr-semantics}) as Isabelle's
built-in {\em inductive predicates}, which allow one to formalise the
structure of the syntax from \cref{BIL:syntax}.

The \BilLocale = $(\Decode)$ locale provides a single
parameter: the {\em decode} predicate ($\Decode$), which will be
uniquely instantiated for each binary that we wish to verify. 

We provide rules for both {\em introduction} (which are used to introduce compound
statements such as {\bf if-then-else} to the proof goal) as well as
{\em elimination} (which are used to eliminate compound statements
from the proof assumptions) directly within the
\BilLocale locale. 
For example, introduction and elimination rules for the \ProgramStepRuleName rule 
in \cref{section:instr-semantics} which formalises a single step of the
program are given below: 

\noindent\begin{minipage}{0.5\textwidth}
\begin{mdframed}[backgroundcolor=backcolour,hidealllines=true,%
                 innerleftmargin=0,innerrightmargin=0,
                 innertopmargin=-0.2cm,innerbottommargin=-0.2cm]
\begin{lstlisting}[language=isabelle,basicstyle=\tt\scriptsize,numbers=none,escapeinside={(*}{*)}]
lemma step_progI[intro]:
assumes (*\guilsinglleft*)($\Delta$, $pc$, mem) $\mapsto$ 
            $\llparenthesis$addr = $pc$, size = $sz$, code = $seq$$\rrparenthesis$(*\guilsinglright*)
    and (*\guilsinglleft*)($\Delta$, $pc$ + $sz$) $\vdash$ $seq$ $\leadsto$ ($\Delta'$, $pc'$)(*\guilsinglright*)
  shows (*\guilsinglleft*)($\Delta$, $pc$, mem) $\leadsto$ ($\Delta'$, $pc'$, mem)(*\guilsinglright*)
proof ...
\end{lstlisting}
\end{mdframed}
\end{minipage}
\hfill
\begin{minipage}{0.5\textwidth}
\begin{mdframed}[backgroundcolor=backcolour,hidealllines=true,%
                 innerleftmargin=0,innerrightmargin=0,
                 innertopmargin=-0.2cm,innerbottommargin=-0.2cm]
\begin{lstlisting}[language=isabelle,basicstyle=\tt\scriptsize,numbers=none,escapeinside={(*}{*)}]
lemma step_progE[elim]:
assumes (*\guilsinglleft*)($\Delta$, $pc$, mem) $\leadsto$ ($\Delta'$, $pc'$, mem)(*\guilsinglright*)
    and (*\guilsinglleft*)($\Delta$, $pc$, mem) $\mapsto$ 
            $\llparenthesis$addr = $pc$, size = $sz$, code = $seq$$\rrparenthesis$(*\guilsinglright*)
obtains (*\guilsinglleft*)($\Delta$, $pc$ + $sz$) $\vdash$ $seq$ $\leadsto$ ($\Delta'$, $pc'$)(*\guilsinglright*)
proof ...
\end{lstlisting}
\end{mdframed}
\end{minipage}

We prove introduction and elimination lemmas in this way for all of the rules in the official BIL specification (\cref{appendix:bil-specification}) with respect to our changes (\cref{sec:specification-of-bil}). 
In total, the \BilLocale locale consists of over 500 lemmas and rules. 

\subsection{The Program Locale}
\label{sec:program-specification}

Recall that our workflow (\cref{fig:translation-c-isabelle}) proceeds
by
\begin{enumerate*}[label={\it Step \arabic*}:]
\item compiling the given program; 
\item feeding the resulting assembly into BAP to generate the
  corresponding \BILa; and
\item using \isabil to automatically generate an Isabelle/HOL locale
  for the given \BILa input.
\end{enumerate*}
Note that if a binary is distributed without source code, then Step 1
could be skipped and the given assembly could be fed directly into BAP
to generate the \BILa.

The locale for a program {\sf prog}
is generated using \isabil's translation tool, which is written in
Isabelle/ML and is invoked using the custom Isabelle/HOL commands
\CommandBIL or \CommandBILfile for inline or external BIL in \BILa
format, respectively. Both commands take as an input, a name for the
locale and either a BIL string (for \CommandBIL) or a filename (for
\CommandBILfile).
Within this locale, we automatically generate a
decode predicate, $\DecodeProg$, that maps each machine state
$\BILMachineStateE{\Variables}{pc}{mem}$ to an instruction.

Note that, internally, Isabelle maintains the \BILa format,
however, we choose to represent it using {\tt class}
locales~\cite{urban2019isabelle,DBLP:conf/types/Ballarin03,nipkow2002isabelle}
to maintain human-readable syntax.  For instance, without these syntax
classes, expressing a direct jump to the fixed program address 3076
would require writing
\lstinline[language=birn]{RAX := Val(Imm(Word(3076,64)))}. By instantiating the word
syntax ($nat::sz$, see \cref{BIL:syntax}), 
we can simplify the expression to the syntax
\lstinline[language=birn]{RAX := (3076 :: 64)}.

In addition to $\DecodeProg$, a program's locale also defines
\begin{enumerate*}[label={\it Step \arabic*}:]
\item an \emph{address set}, $\ProgramDomain: \powerset(word)$
  (defining the set of addresses that have corresponding
  instructions) and
  
\item a \emph{symbol table}, $\SymbolTable : \String \rightarrow word$
  (mapping the binary's symbols, e.g. main, memcpy, free, as string
  literals to raw addresses). 
\end{enumerate*}
Both $\ProgramDomain$ and $\SymbolTable$ 
capture additional information about a binary that can be used later
in a proof. For example, the $\ProgramDomain$ is useful for
determining whether the program counter points to a valid address and
is particularly important for validating the correctness of indirect
jumps. 
Many binary proofs verify that execution from 
some entry point, such as the {\tt main} function, is either correct or 
incorrect with respect to some property. 
Symbols typically represent entry points to 
functions and although the address of an entrypoint may differ between
binaries, the symbol will remain constant. 
Hence, referring to positions within the binary using symbol names, 
stored in the $\SymbolTable$ rather than program addresses, is often more convenient.
Further details regarding the extraction of the $\ProgramDomain$ and
$\SymbolTable$ from a binary can be found in \cref{sec:parser}.







\begin{example} \label{ex:binary-a} Consider the {\tt Binary\_A} program below, written in
  \BILa, which corresponds to the \BIL instruction
  \lstinline[language=birn]{X8 := X2 + 32}. We assume the program has
  been compiled for RISC-V; for readability, \BILa provides the original
  assembly at \cref{assembly}. The program locale corresponding to
  \cref{exBILadt} starts on \cref{localeBinA}, which fixes the corresponding decode predicate. 

\begin{lstlisting}[language={Isabelle},escapeinside={(*}{*)},label=lst:binary-a]
BIL (*\guilsinglleft*)
105dc: <test>
105dc: addi s0, sp, 32(*\label{assembly}*)
(Move(Var("X8",Imm(64)),PLUS(Var("X2",Imm(64)),Int(32,64))))
(*\guilsinglright*)(*\label{exBILadt}*) defining Binary_A
  
locale Binary_A(*\label{localeBinA}*)
    fixes decode :: (*\guilsinglleft*)(var (*$\Rightarrow$*) val option) (*$\times$*) word (*$\times$*) var (*$\Rightarrow$*) insn (*$\Rightarrow$*) bool(*\guilsinglright*)
    assumes decode_105dc: (*\guilsinglleft*)((*$\Delta$*), 0x105dc :: 64, mem) $\DecodeProg$
               $\llpar$ addr = 0x105dc :: 64, size = 4, bil = [X8 := X2 + (32 :: 64)] $\rrpar$(*\guilsinglright*)
begin
definition $\ProgramDomain$ :: (*\guilsinglleft*)word set(*\guilsinglright*) where (*\guilsinglleft*)$\ProgramDomain$ = {(0x105dc :: 64)}(*\guilsinglright*)
definition $\SymbolTable$ :: (*\guilsinglleft*)str $\rightharpoonup$ word(*\guilsinglright*) where (*\guilsinglleft*)$\SymbolTable$ = ["test" $\mapsto$ (0x105dc :: 64)](*\guilsinglright*)
end
\end{lstlisting}

\end{example}

\paragraph{Externally linked binaries}
Binaries often include links to external code, typically in the form of function calls. 
Accurately modeling these external calls is crucial for understanding the behavior of the original binary. 
\IsaBIL provides two methods for handling externally linked code, depending on the availability of that external code.
If the external binary is available, an Isabelle/HOL locale can be generated for the external binary, 
which is then combined with the existing binary locale through inheritance. 
If the external binary is not available, we can make assumptions about its behavior 
and define an approximate implementation using locale assumptions.
We give an example for both cases in \cref{appendix:external-binary}.

\section{(In)correctness}
\label{sec:incorrectness}








In this section, we present the \isabil encoding of Hoare and
Incorrectness logic (see \cref{fig:proofs}). Our locale-based encoding combines both into a single proof system
\InferenceLocale locale, where we verify the \AgreementRuleName and \DenialRuleName
lemmas (see~\cite{o2019incorrectness}). 

Hoare and O'Hearn triples define (in)correctness of a command $c$
over a pre-state $\sigma$ satisfying the predicate $P$, resulting in a
post-state $\tau$ satisfying the predicate $Q$. In the locale, we assume the
existence of a big-step transition relation $(c,\sigma) \BigStep \tau$
defining the execution of $c$ from state $\sigma$ to termination,
resulting in state $\tau$. Thus, we have the well-known
under- and over-approximating rules:
{\small%
\begin{mathpar}
\inferrule[(\sc hoare)]{
  \HoareProp
}{
  \HoareTriple{P}{c}{Q}
}
\and
\inferrule[(\sc ohearn)]{
  \OHearnProp
}{
  \OHearnTriple{P}{c}{Q}
}  
\end{mathpar}}

\noindent Note that our definition of correctness follows partial correctness, 
meaning termination is not guaranteed.
We encode Hoare and Incorrectness triples as locales
$\CorrectnessLocale = (\BigStep)$ and
$\IncorrectnessLocale = (\BigStep)$, respectively, both of which are
parameterised by $\BigStep$.
Within the $\CorrectnessLocale$ locale,
we verify standard Hoare logic rules, e.g.,
{\small%
\begin{mathpar}
\inferrule[(\sc post\_conj\_corr)]{
  \HoareTriple{P}{c}{Q_1} \\
  \HoareTriple{P}{c}{Q_2}
}{
  \HoareTriple{P}{c}{Q_1 \wedge Q_2}
}
\and
\inferrule[(\sc pre\_disj\_corr)]{
  \HoareTriple{P_1}{c}{Q} \\
  \HoareTriple{P_2}{c}{Q}
}{
  \HoareTriple{P_1 \vee P_2}{c}{Q}
}
\and
\inferrule[(\sc strengthen\_weaken\_corr)]{
  P' \implies P \\
  \HoareTriple{P}{c}{Q} \\
  Q \implies Q'
}{
  \HoareTriple{P'}{c}{Q'}
}
\end{mathpar}}

\noindent Similarly, within the $\IncorrectnessLocale$ locale, we verify rules
for Incorrectness logic, e.g.,
{\small%
\begin{mathpar}
\inferrule[(\sc post\_disj\_incorr)]{
  \OHearnTriple{P}{c}{Q_1} \\
  \OHearnTriple{P}{c}{Q_2}
}{
  \OHearnTriple{P}{c}{Q_1 \vee Q_2}
}
\and
\inferrule[(\sc pre\_disj\_incorr)]{
  \OHearnTriple{P_1}{c}{Q} \\
  \OHearnTriple{P_2}{c}{Q}
}{
  \OHearnTriple{P_1 \vee P_2}{c}{Q}
}
\and
\inferrule[(\sc strengthen\_weaken\_incorr)]{
  P \implies P' \\
  \OHearnTriple{P}{c}{Q} \\
  Q' \implies Q
}{
  \OHearnTriple{P'}{c}{Q'}
}
\end{mathpar}}

\noindent We then combine these to form a locale
$\InferenceLocale = (\BigStep)$, again parameterised by
$\BigStep$. Within this combined locale, we are able to
prove 
the \AgreementRuleName and \DenialRuleName rules~\cite{o2019incorrectness}:
{\small%
\begin{mathpar}
\inferrule[(\AgreementRuleName)]{
  \OHearnTriple{U}{c}{U'} \\
  U \implies O \\
  \HoareTriple{O}{c}{O'} 
}{
  U' \implies O'
}
\and
\inferrule[(\DenialRuleName)]{
  \OHearnTriple{U}{c}{U'} \\
  U \implies O \\
  \neg(U' \implies O')
}{
  \neg(\HoareTriple{O}{c}{O'})
}
\end{mathpar}}

\noindent For proofs of incorrectness, it is typically not enough to simply
state that if $\OHearnTriple{P}{c}{Q}$ holds for an error state $Q$
then the program $c$ is incorrect. A program cannot be incorrect if
there is no valid state that satisfies $Q$ and as such, we must ensure
that $Q$ is {\em non-trivial}, i.e., $\exists \sigma.\
Q(\sigma)$. Proving that $Q$ is non-trivial can be difficult, as $Q$
is often an over-approximation of the actual set of post-states
(defined by some predicate $Q'$) reachable from $P$. 
Therefore, proofs of
incorrectness usually take the form $\IncorrectnessProof{P}{c}{Q'}{Q}$. 
Note that this does not necessarily imply that
$\OHearnTriple{P}{c}{Q} \wedge (\exists \sigma.\ Q(\sigma))$ since the
implication is in the wrong direction to apply the {\sc
  strengthen\_weaken\_incorr} rule.

In \cref{sec:inst-incorr-as}, we provide instantiations for our (in)correctness 
locales to demonstrate their conformance with established encodings of (in)correctness in Isabelle/HOL.


\section{Automation and Examples}
\label{sec:automation}



In this section, we bring together the \BilLocale and
the $\InferenceLocale$ locales to create a new locale,
\BilInferenceLocale (see \cref{sec:local-locale-1}), providing a
proof environment for \BIL programs. This in turn enables us to
develop a large number of highly reusable proof automation procedures
for verifying both correctness and incorrectness of BIL programs (see
\cref{sec:proof-automation}). The final major component of \isabil
automation is the \BILa parser that automatically generates
Isabelle/HOL locales for BIL programs (\cref{sec:parser}), providing a
smooth and seamless pipeline from BAP to Isabelle/HOL proofs. 
We offer an overview of \isabil, discussing 
correctness and incorrectness proofs for a handful of examples.

\subsection{The \BilInferenceLocale locale}
\label{sec:local-locale-1}
\label{sec:big-step-semantics}

As shown in \cref{fig:proofs}, \BilInferenceLocale interfaces with
the \BilLocale and $\InferenceLocale$ locales and is
later extended with instances of state models (e.g.,
\AllocationLocale and \FindSymbolLocale) to enable verification
of generated BIL programs (see \cref{sec:proofs}). Therefore, it is
one of \isabil's most complex components.

Our first task is to define a big-step transition relation,
$\BILBigStep$, required by $\InferenceLocale$, using the BIL step
relation, $(\Delta, pc, mem)\leadsto(\Delta', pc', mem')$, defined in
\cref{section:instr-semantics}. 
One option for defining
$\BILBigStep$ is to simply take the transitive closure of
$\leadsto$. However, this would mean that the pre/postconditions that
we define in our example would be predicates over the full variable
store $\Delta$, which is heavy-handed.

An alternative (which is the approach we take) is to define another
transition relation $\BilSmallStep$ that abstracts
$\leadsto$, and define
$\BILBigStep$ as the transitive closure of
$\BilSmallStep$. As we shall see, this vastly increases the
flexibility of our approach, which becomes
\begin{enumerate*}[label=\bfseries(\arabic*)]
\item extensible, since $\BilSmallStep$ can be used to cover
additional program features such as memory allocation (see
\cref{sec:double-free-paper}), and 
\item more efficient since the components of
  $\Delta$ that are not needed for the proof can be ignored (see
  \cref{sec:forbidden-symbols}).
\end{enumerate*}
Thus, assuming
$\Command = (\Variables, pc, mem)$, we have:
\begin{mathpar} \small
  \scalebox{1}{
  \inferrule[(\sc program\_big\_step)]{
  \Command = (\_, pc, \_) \and pc \in \ProgramDomain \\\\
  \BILMachineStateStepInitialE{\Command}{\Command'}\\
  (\Command, \prestate) \BilSmallStep \prestate' \and
  (\Command', \prestate')\BILBigStep \poststate
}{
  (\Command, \prestate)\BILBigStep \poststate
}{(\ensuremath{\Decode})}
\and
\inferrule[(\sc program\_big\_step\_last)]{
  \Command = (\_, pc, \_) \and \Command' = (\_, pc', \_) \\\\
  pc \in \ProgramDomain \\
  pc' \notin \ProgramDomain
  \\\\
  \BILMachineStateStepInitialE{\Command}{\Command'}\\
  (\Command, \prestate) \BilSmallStep \poststate }{
  (\Command, \prestate)\BILBigStep \poststate
}{(\ensuremath{\Decode})}}
\end{mathpar}
Here, we assume
$\ProgramDomain$ is the set of addresses that contain program
instructions (see \cref{sec:program-specification}), which will be
later instantiated by a program's locale. A program may take a step as
long as the current program counter corresponds to a program
instruction. Note that the precise state model (i.e., type of
$\prestate$, $\prestate'$ and
$\poststate$) that one uses in {\sc program\_big\_step} and {\sc
  program\_big\_step\_last} will depend on the verification task at
hand. If required, one could also take $\sigma =
\Delta$ allowing the proofs to inspect the full variable state (at a low
level of abstraction).

Overall, we obtain a
locale 
$\BilInferenceLocale = (\Decode, \BilSmallStep, \ProgramDomain)$. This
locale extends the $\BilLocale$ (from \cref{sec:mechanisation}) from
which we inherit the decode predicate, $\BILDecodeTransition$.  Then,
we use $\BilSmallStep$ and $\ProgramDomain$ to define $\BILBigStep$ as
above. This allows us to interpret the $\InferenceLocale$ as
sublocale, instantiating the parameter $\BigStep$ to
$\BILBigStep$. 

Within $\BilInferenceLocale$, we can prove several rules of Hoare
and Incorrectness logic in terms of the BIL semantics. These proofs
are trivial at this level, but are nevertheless critical for proof
automation since they can be used by all of the examples in
\cref{sec:proofs}. 

\paragraph{Stuckness and Undefined Behaviour} 
A BIL program is considered stuck if it reaches a state where no transition rules apply. Programs that become stuck are deemed correct, as they do not terminate. 

The most common cause of stuckness in BIL is an attempt to decode an instruction at an address that does not exist within the program (see \cref{section:instr-semantics}). This typically occurs when a jump targets a non-existent address, resulting in a post-state with no valid instruction to execute. However, the big step semantics treat this as valid program termination rather than an explicit error. If the prover wishes to ensure that the program terminates with an expected PC they may assert its value in an (in)correctness triple's post-state $Q$.

Stuckness may also arise due to type errors, such as assigning a variable a value of an incompatible type. These types of stuckness are prevented by ensuring BIL statements are type correct, which can be achieved using IsaBIL's automated type checker (see \cref{sec:eisbach}). In our proofs, we verify type correctness when necessary to prevent stuckness. More details, along with an example, can be found in the BIL manual (\cref{appendix:bil-specification}).

Many traditional sources of stuckness in program semantics, such as dereferencing invalid memory or reading from uninitialized registers, result in undefined behaviour (UB) for BIL. In BIL, UB is represented using the \ValueUnknown\xspace value. For example, if an uninitialized register is used:

\begin{lstlisting}[language=riscv]
mv a6, a5     # UB if a5 was never initialized    
\end{lstlisting}

Since the value of \code{a5} (a 64 bit register) is unspecified, $\ValueUnknownE{str}{\TypeImmediateE{64}}$\xspace will be set to \code{a6}. Evaluating BIL semantics that contain unknown values (such as $\code{a6} + 5$) will propagate further unknowns unless avoided by control flow structures such as if-then-else.
However, this does not necessarily lead to stuckness, except in cases
where the target of a dynamic jump is unknown (i.e. $\StatementJmpE{\ValueUnknownE{str}{\TypeImmediateE{64}}}$). In such cases, the program cannot take a step as it does not know which program address to jump to. 
In (in)correctness logic, unknowns that do not lead to stuckness may still prevent the discharge of the 
post-state ($Q$). 
Unknown values in BIL can be resolved by explicitly specifying constraints on the program and pre-state as required.

\subsection{Proof Automation}
\label{sec:proof-automation}
Instruction set architectures (ISAs), despite their large scale, are inherently
finite, and compilers employ common patterns for
optimization. Additionally, developers leverage reusable components
such as {\em functions} and {\em gadgets} to perform common
tasks. Consequently, binaries often exhibit a significant degree of
repetition. Once identified, this repetition can be verified without
the need for human input. \isabil exploits this repetition to
alleviate the burden associated with proof construction and
verification. In this section, we outline the proof automation
techniques employed by \isabil.

\subsubsection{Eisbach}
\isabil heavily leverages
Eisbach~\cite{DBLP:journals/jar/MatichukMW16}, an Isabelle tactic
framework for proof automation, to discharge proof goals
automatically.  We create Eisbach methods for symbolic execution
(\SymbolicSolver) and type checking (\TypeSolver)
which utilise \isabil's {\em introduction} and {\em
elimination} rules to solve both big- and small-step BIL statements,
expressions and variables as well as proving type correctness. These
methods employ a best-effort approach.
If a case that cannot be solved automatically occurs, the methods 
will backtrack to a safe state and the partially completed 
proof context will be handed back to the human prover, 
whereby corrections or manual inputs can be made before
resuming the automation. 
We describe these methods in more detail in the appendix (\cref{sec:eisbach}).




\subsubsection{Human Readability} 
In assembly languages, registers are typically assigned specific 
names and types. However, in BIL, registers are
represented as generic {\em variables} of type $var$. These variables 
are parameterised with names and types. For example, the RISC-V 
registers {\tt R0-31} are represented by the 
variables $(\text{"R0-31"} : \TypeImmediateE{64})$. 
While this representation offers flexibility, it can be verbose.
It is common knowledge for RISC-V developers that R0-31 are
registers capable of storing 64-bit words.
To enhance program readability, we define the set of registers for \RiscVsf~\cite{Waterman:EECS-2014-54} as the 64bit registers R0-31 : $\TypeImmediateE{64}$.
Additionally, we introduce syntax abbreviations for the 64-bit x86 architecture, presented in detail in \cref{appendix:x86}.

\subsubsection{Architecture-Specific Proof Optimisations (for RISC-V)}
\label{sec:riscv-optimisations}
Whilst \isabil provides sufficient granularity to handle many architectures out of the box, the speed and efficiency of proofs can be improved by tailoring optimizations to specific hardware architectures. This section describes the proof optimizations undertaken for RISC-V.

\paragraph{Instructions}
Modern architectures comprise of large instruction sets. 
Proofs of programs can be optimised
by verifying high-level lemmas corresponding to program steps of the most
commonly used instructions 
directly in the \BilInferenceLocale locale. For example, \isabil
provides step rules for the 32-bit (4-byte) RISC-V instructions
\RiscVauipcNameOnly, \RiscVjalrNameOnly and \RiscVldNameOnly.
\begin{align*}
\RiscVauipc = \InsnExpandedE{pc}{\WordE{4}{64}}{\{\StatementMoveE{rd}{pc + imm}\}}
\end{align*}
The \RiscVauipcNameOnly (Add Upper Immediate to Program Counter) instruction
facilitates the computation of an absolute address. It achieves this by adding 
the immediate value $imm$ to $pc$, storing the result in the destination 
register $rd$. \RiscVauipcNameOnly is commonly used to create jump targets, 
such as those for external functions. 
\begin{align*}
\RiscVjalr = \InsnExpandedE{pc}{\WordE{4}{64}}{\{\StatementMoveE{rd}{pc + 4};\ \StatementJmpE{rs1 + \offset}\}} 
\end{align*}
The \RiscVjalrNameOnly (Jump And Link Register) instruction performs an unconditional jump. 
It sets $pc$ to the address stored in the source register $rs1$ with an optional immediate 
value $\offset$. The address of the subsequent instruction is then stored in the destination
register $rd$. This instruction is commonly used for function calls, where it is important
to preserve the return address.
\begin{align*}
\RiscVld = \InsnExpandedE{pc}{\WordE{4}{64}}{\{\StatementMoveE{rd}{\ExpressionLoadE{mem}{rs1 + \offset}{\EndianLittle}{64}}\}}
\end{align*}
The \RiscVldNameOnly (Load Double) instruction is used to load a 64-bit word, referred to as a double word, from memory. First, \RiscVldNameOnly computes an address by adding the contents of the source register $rs1$ to the immediate $\offset$. The value at this address in memory is then stored in the destination register $rd$. \RiscVldNameOnly is the primary means of retrieving data of this size from memory.

Representing the instructions \RiscVauipcNameOnly, \RiscVjalrNameOnly and \RiscVldNameOnly as the high-level program step lemmas (given below) allows the prover to otherwise skip the lengthy reduction for a program step. The rules for these instructions are provided below:
\vspace{-2pt}
{\small%
\begin{mathpar} 
\scalebox{0.95}{
\inferrule[(\sc auipc\_lemma)]{
  \BILDecodeInstructionE{\BILMachineStateE{\Variables}{pc}{mem}}{\RiscVauipc}
}{
{\BILMachineStateE{\Variables}{pc}{mem}} \BILTransition \\\\
{\BILMachineStateE{\Variables(rd \mapsto pc + imm)}{pc + 4)}{mem}}
}
\quad
\inferrule[(\sc jalr\_lemma)]{
  \BILDecodeInstructionE{\BILMachineStateE{\Variables}{pc}{mem}}{\RiscVjalr} \\\\
  \VarInE{rs1}{addr}{\Variables}
}{
{\BILMachineStateE{\Variables}{pc}{mem}} 
\BILTransition  \\\\
{\BILMachineStateE{\Variables(rd \mapsto pc + 4)}{addr + \offset}{mem}}
}}
\quad
\scalebox{0.9}{
\inferrule[(\sc ld\_lemma)]{
  \BILDecodeInstructionE{\BILMachineStateE{\Variables}{pc}{mem}}{\RiscVld} \\\\\
  \VarInE{mem}{v}{\Variables} \and
  \VarInE{rs1}{addr}{\Variables}  \\\\\
  \BILStepsE{\Variables}{\ExpressionLoadE{v}{addr + \offset}{\EndianLittle}{64}}{w}
}{
{\BILMachineStateE{\Variables}{pc}{mem}} \BILTransition \\\\
{\BILMachineStateE{\Variables(rd \mapsto w)}{pc + 4)}{mem}}
}}
\end{mathpar}}

\medskip
Further semantics for \RiscVaddiNameOnly, \RiscVsdNameOnly and \RiscVretNameOnly RISC-V instructions are provided in \cref{sec:furth-arch-spec}.

\paragraph{Execution} We also prove high-level big-step semantics
rules for common gadgets in binary executables. For example, the {\em
  procedure linkage table} (PLT) acts as an intermediary between the
current program and shared libraries, including C's standard library.
The PLT facilitates indirect calls to external functions, whose
locations are unknown until runtime when they are resolved by the
dynamic loader. In the dump of the RISC-V binary for \cref{DF-bad}
given in \cref{pltdump}, we can observe the presence of stubs within
the PLT, representing entries for each external call. Specifically, in
\cref{pltdump}, the stubs correspond to the functions {\tt free} and
{\tt malloc} that are called by the program.

\begin{figure}[t]
  \centering
  \begin{mdframed}[backgroundcolor=backcolour,hidealllines=true,%
                   innerleftmargin=0,innerrightmargin=0,
                   innertopmargin=-0.2cm,innerbottommargin=-0.2cm]
\begin{lstlisting}[linerange={17-27}, label={pltdump},language=riscv]
00000000000104a0 <malloc@plt>:
   104a0: 00002e17            auipc  t3,0x2
   104a4: b78e3e03            ld     t3,-1160(t3) # 12018 <malloc@GLIBC_2.27>
   104a8: 000e0367            jalr   t1,t3
   104ac: 00000013            nop

00000000000104b0 <free@plt>:
   104b0: 00002e17            auipc  t3,0x2
   104b4: b70e3e03            ld     t3,-1168(t3) # 12020 <free@GLIBC_2.27>
   104b8: 000e0367            jalr   t1,t3
   104bc: 00000013            nop
\end{lstlisting}
  \end{mdframed}
  \vspace{-12pt}
  \caption{PLT stubs for free and malloc}
  \Description{Dump of a RISC-V binary, showing the free and malloc PLT stubs}
  \vspace{-15pt}
\end{figure}
By leveraging the consistent pattern observed in PLT entries, we can construct universal big-step rules for PLT stubs, as demonstrated below.
{\small%
\begin{mathpar} 
\inferrule[(\PltStubLemmaName)]{
  \BILDecodeInstructionE{\BILMachineStateE{\Variables_1}{pc}{mem}}{\RiscVauipcE{t3}{0x2}} \and
  \BILDecodeInstructionE{\BILMachineStateE{\Variables_2}{pc + 4}{mem}}{\RiscVldE{t3}{t3}{\offset}} \\\\
  \BILDecodeInstructionE{\BILMachineStateE{\Variables_3}{pc + 8}{mem}}{\RiscVjalrE{t1}{t3}{0}} \and
  \{pc, pc + 4, pc + 8\} \subseteq \ProgramDomain \\\\
  \Variables_2 = \Variables_1(t3 \mapsto pc + 0x2) \and
  \Variables_3 = \Variables_2(t3 \mapsto w) \and
  \Variables_4 = \Variables_3(t1 \mapsto pc + 12) \\\\
  (\BILMachineStateE{\Variables_1}{pc}{mem}, \sigma_1) \BilSmallStep \sigma_2 \and
  (\BILMachineStateE{\Variables_2}{pc + 4}{mem}, \sigma_2) \BilSmallStep \sigma_3 \and
  (\BILMachineStateE{\Variables_3}{pc + 8}{mem}, \sigma_3) \BilSmallStep \sigma_4 \\\\
  (\BILMachineStateE{\Variables_4}{w}{mem}, \sigma_4)\BILBigStep \tau \and
  \VarInE{mem}{v}{\Variables_1} \and
  \BILStepsE{\Variables}{\ExpressionLoadE{v}{pc + {\tt 0x2} + \offset}{\EndianLittle}{64}}{w}
}{
  (\BILMachineStateE{\Variables_1}{pc}{mem}, \sigma_1)\BILBigStep \tau
}    
\end{mathpar}}

Now, if we encounter a PLT stub in a proof at a lower level, we can discharge the proof obligation efficiently using \PltStubLemmaName. This lemma significantly reduces the proof effort required. Without \PltStubLemmaName, one must apply over 150 rules to achieve the same result. Furthermore, by applying this approach to stack allocations and deallocations, we eliminate the need for approximately 50 rules in the case of allocation and 250 rules in the case of deallocation. Further details on this are provided in \cref{sec:furth-arch-spec}.



\subsection{BIL to \isabil transpiler}
\label{sec:parser}
\newcommand{\BILml}{\BIL_{\text{ML}}}

A key component of our \isabil framework is a verified BIL parser that
transpiles \BILa programs 
into an Isabelle/HOL locale to enable verification.

Parsing occurs in two distinct phases. The first phase involves the translation of \BILa into $\BILml$, a format suitable for manipulation within Isabelle/ML. 
This intermediate step enables analysis tasks in Isabelle/ML, such as calculating the size of instructions.
The second phase involves translating $\BILml$ into an Isabelle/HOL representation, which is the starting point for verification. Parsing can be invoked directly on a \BILa string 
with \CommandBIL or alternately on a file containing \BILa with \CommandBILfile. An overview of the parsing process is given in \cref{fig:parse-process}.

During the first phase, the parser iterates over the instructions in a \BILa input. It captures and stores all program addresses in $\ProgramDomain$. Moreover, the parser associates symbols with a program address in $\SymbolTable$. 
Since the syntax of \BILa resembles a tree structure, it is intuitive to convert it into an Abstract Syntax Tree (AST) using a lexer.
This AST is processed by a Recursive Descent Parser (RDP), which traverses each node in the tree depth-first, matching the node's value to a corresponding parsing function. For example, if the parser encounters a {\tt Var} node, it will attempt to parse the first child as a {\tt string} and the second child as a \BIL type. This process transforms the input into $\BILml$, a structured data type closely resembling \BILa which can be manipulated within Isabelle/ML.

The second phase defines a locale within the current proof context with a fixed decode predicate. For each $\BILml$ instruction, an assumption is added to the locale, stating how a program with any variable state and memory, but with the instructions program address, decodes to an Isabelle/HOL representation of said instruction. 
To obtain the Isabelle/HOL representation, an RDP translator recursively converts each $\BILml$ instruction to Isabelle/HOL terms defined in the \BilLocale\ locale (see \cref{sec:mechanisation}). 

The first phase of the parser harnesses Isabelle/HOL's \emph{code generation}, where specifications 
for both the lexer and parser were defined within Isabelle/HOL and subsequently translated into ML.
$\BILml$ for example, is a direct ML representation of the BIL specification from the 
\IsaBIL framework.
Using \emph{code reflection}, we import the generated ML code back into Isabelle/HOL as a plugin.
Consequently, the lexer and parser are formally verified under the Isabelle/HOL framework.
In contrast, the second phase employs non-code-generated ML. 
This decision stems from its necessity to interface with Isabelle/HOL's core ML library,
which cannot be achieved via code-generated ML. 
To bridge this gap, we introduce a minimal layer of unverified ML code that lifts the BIL representation 
obtained from the verified parser into Isabelle/HOL’s proof system. 
Below, as an example, we present a snippet of the translation code responsible for converting $\BILml$'s 
binary operations to IsaBIL’s binary operations:
\begin{lstlisting}[language=ML,mathescape = false, basicstyle = \footnotesize\ttfamily]
mk_exp (AstParser.BinOp (e1,bop,e2)) = @{term "BinOp"} $ mk_exp e1 $ mk_bop bop $ mk_exp e2     
\end{lstlisting}
Note that Isabelle isolates Isabelle/ML programming, and hence
our parser cannot interfere with the soundness of Isabelle's core logic
engine.

\definecolor{colorHOL}{HTML}{dbccff}
\definecolor{colorML}{HTML}{c7d9fc}

\definecolor{colorAstStatement}{HTML}{c9e7ff}
\definecolor{colorAstChild}{HTML}{98ff9c}
\definecolor{colorAstExpression}{HTML}{fffec9}
\definecolor{colorAstType}{HTML}{fad0d0}

\begin{figure}[t]
  \begin{minipage}[t]{0.47\columnwidth}
    \centering
    \scalebox{0.75}{
      \begin{tikzpicture}[node distance=1cm, auto]
  \node[] (Dummy) {};
  \node[punkt, fill=colorHOL, dashed, inner sep=5pt, below right=0.5cm and 1.4cm of Dummy] (IO) {\sf IO};
  \node[punkt, fill=colorML, below left=0.5cm and 1.4cm of Dummy] (bil-file-cmd) {\tt BIL\_file};
  \node[punkt, fill=colorHOL, inner sep=5pt, inner sep=5pt,below=0.5cm of IO]
  (Lexer) {\sf Lexer};
  \node[punkt, fill=colorML,inner sep=5pt,below=0.5cm of bil-file-cmd] (bil-cmd) {\tt BIL};
  \node[punkt, fill=colorHOL,inner sep=5pt,below=0.5cm of Lexer]
  (Parser) {Parser};
  \node[punkt, fill=colorHOL, inner sep=5pt,below=0.5cm of Parser]
  (Translator) {Translator};
  \node[punkt, fill=colorML, inner sep=5pt,below=1.7cm of bil-cmd]
  (IsaBIL) {IsaBIL Locale};
  \path[draw, thick, ->] (bil-file-cmd)-- (IO)
  node[below,pos=0.5] {\tt filename} ; 
  \path[draw, thick, ->] (bil-cmd)-- (Lexer)
  node[below,pos=0.5] {\sc \BILa} ; 
  \path[draw, thick, ->] (IO)-- (Lexer)
  node[right,pos=0.5] {\sc \BILa} ; 
  \path[draw, thick, ->] (Lexer)-- (Parser)
  node[right,pos=0.5] {\sc AST} ; 
  \path[draw, thick, ->] (Parser)-- (Translator)
  node[right,pos=0.5] {\sc $\BILml$} ; 

  \path[draw, thick, ->] (Translator)-- (IsaBIL)
  node[below,pos=0.5] {\tt thy} ; 

  \draw[black,thick,dotted] ($(IO)+(-1.2,0.6)$)  rectangle ($(Translator)+(1.2,-0.6)$) node [pos=0.5,xshift=-1cm,yshift=-2.9cm]{Isabelle/ML};

  \draw[black,thick,dotted] ($(bil-file-cmd)+(-1.4,0.6)$)  rectangle ($(IsaBIL)+(1.4,-0.6)$) node [pos=0.5,xshift=-1cm,yshift=-2.9cm]{Isabelle/HOL};
 
    \end{tikzpicture}}
    \vspace{-10pt}
    \caption{Overview of the BIL parser, entrypoints\\ via the \CommandBIL or  \CommandBILfile commands}
       \vspace{-10pt}
    \label{fig:parse-process}
  \end{minipage}
  \begin{minipage}[t]{0.50\columnwidth}
    \centering
    \scalebox{0.7}{
    \begin{tikzpicture}[node distance=0.4cm, auto]
      \node[punkt, fill=colorAstStatement, inner sep=5pt] (Move) {\tt Move};
      \node[punkt, fill=colorAstExpression, inner sep=5pt,below left=0.4cm and 1cm of Move] (Var) {\sf Var};
      \node[punkt, fill=colorAstChild, inner sep=5pt,below left=0.4cm and -0.2cm of Var] (Var-Name) {\sf "X8"};
      \node[punkt, fill=colorAstType, inner sep=5pt,below right=0.4cm and -0.2cm of Var] (Var-Type) {\sf Imm};
      \node[punkt, fill=colorAstChild, inner sep=5pt,below=0.4cm of Var-Type] (Var-Type-sz) {\sf 64};
      \node[punkt, fill=colorAstExpression, inner sep=5pt,below right=0.4cm and 1cm of Move] (Plus) {\sf PLUS};
      \node[punkt, fill=colorAstExpression, inner sep=5pt,below left=0.4cm and 0.3cm of Plus] (Plus-Lhs) {\sf Var};
      \node[punkt, fill=colorAstChild, inner sep=5pt,below left=0.4cm and -0.2cm of Plus-Lhs] (Plus-Lhs-Name) {\sf "X2"};
      \node[punkt, fill=colorAstType, inner sep=5pt,below right=0.4cm and -0.2cm of Plus-Lhs] (Plus-Lhs-Type) {\sf Imm};
      \node[punkt, fill=colorAstChild, inner sep=5pt,below=0.4cm of Plus-Lhs-Type] (Plus-Lhs-Type-sz) {\sf 64};
      \node[punkt, fill=colorAstExpression, inner sep=5pt,below right=0.4cm and 0.3cm of Plus] (Plus-Rhs) {\sf Int};
      \node[punkt, fill=colorAstChild, inner sep=5pt,below left=0.4cm and -0.2cm of Plus-Rhs] (Plus-Rhs-Val) {\sf 32};
      \node[punkt, fill=colorAstChild, inner sep=5pt,below right=0.4cm and -0.2cm of Plus-Rhs] (Plus-Rhs-sz) {\sf 64};
 
      \path[draw, thick, -] (Move)-- (Var); 
      \path[draw, thick, -] (Var)-- (Var-Name); 
      \path[draw, thick, -] (Var)-- (Var-Type); 
      \path[draw, thick, -] (Var-Type)-- (Var-Type-sz); 
      \path[draw, thick, -] (Move)-- (Plus); 
      \path[draw, thick, -] (Plus)-- (Plus-Lhs); 
      \path[draw, thick, -] (Plus-Lhs)-- (Plus-Lhs-Name); 
      \path[draw, thick, -] (Plus-Lhs)-- (Plus-Lhs-Type); 
      \path[draw, thick, -] (Plus-Lhs-Type)-- (Plus-Lhs-Type-sz); 
      \path[draw, thick, -] (Plus)-- (Plus-Rhs); 
      \path[draw, thick, -] (Plus-Rhs)-- (Plus-Rhs-Val); 
      \path[draw, thick, -] (Plus-Rhs)-- (Plus-Rhs-sz); 
    \end{tikzpicture}}
        \vspace{-10pt}

    \caption{AST representation of the \BILa statement {\tt Move(Var("X8", Imm(64)), PLUS(Var("X2", Imm(64)), Int(32, 64)))}}
    \vspace{-10pt}
    \label{fig:ast-tree}
  \end{minipage}
\end{figure}

%
%
%

\subsection{Example Proofs}
\label{sec:example-proofs}
The majority of the development consists of general tactics and lemmas which can be reused across multiple developments. We give the number of lemmas for each component in \cref{tab:development-overview}.

\begin{table}[t] \footnotesize
\begin{minipage}[b]{0.44\columnwidth}
    \begin{tabular}{|@{~}l|l@{\ }|}
\hline
\textbf{Component}                      & \textbf{Lemmas} \\ \hline
Correctness                             & 13              \\ \hline
Incorrectness                           & 3               \\ \hline
Inference Rules                         & 2               \\ \hline
BIL Specification (including syntax)                      & 594             \\ \hline
BIL Inference                           & 13              \\ \hline
RISC-V optimisations                    & 74              \\ \hline
Alloc Model                             & 6               \\ \hline
Find Symbol                             & 0               \\ \hline
Double Free (total for both examples)   & 77              \\ \hline
AV Rules (total for all seven examples) & 187             \\ \hline
\end{tabular}
\caption{Overview of theories and lemma for each of \isabil's components.}
\label{tab:development-overview}
\vspace{-10pt}
\end{minipage}
\hfill
\begin{minipage}[b]{0.54\columnwidth}
\begin{tabular}{|@{~}l|p{0.85\columnwidth}@{~}|}
  \hline
  \textbf{AVR} & \textbf{Description (the following shall not be used)}                                                                                               \\ \hline
  17               & The error indicator errno 
  \\ \hline
  19               & {\tt \textless{}locale.h\textgreater} and the setlocale function\footnotemark 
  \\ \hline
  20               & The {\tt setjmp} macro and the {\tt longjmp} functions
  \\ \hline
  21               & The signal handling facilities of {\tt \textless{}signal.h\textgreater} 
  \\ \hline
  23               & The library functions {\tt atof}, {\tt atoi} and {\tt atol} from library {\tt \textless{}stdlib.h\textgreater} 
  \\ \hline
  24               & The library functions {\tt abort}, {\tt exit}, {\tt getenv} and system from library {\tt \textless{}stdlib.h\textgreater} 
  \\ \hline
  25               & The time handling functions of library {\tt \textless{}time.h\textgreater} 
  \\ \hline
\end{tabular} \smallskip

\caption{AV rules with description}
\label{tab:av-rules}
\end{minipage}
\vspace{-10pt}
\end{table}

All of the sublocales and inherited locales can reuse the proofs of the higher-level (more abstract) locales. Thus, for instance, the 594 lemmas of the \BilLocale locale are applicable to \BilInferenceLocale and all of the examples, which improves re-usability. The RISC-V optimisations help improve performance and are reusable across all examples. The theories for \DoubleFreeLocale and \FindSymbolLocale are general and apply to all corresponding examples. Finally, each example comprises a step lemma for each line of code - but using the earlier optimisations, solving them is mostly trivial and only requires running the automated tactics.

To motivate \isabil, we employ a combination of correctness and incorrectness 
proofs, focusing on illustrative examples frequently referenced in BAP 
literature \cite{DBLP:conf/cav/BrumleyJAS11}. These examples demonstrate 
representations of patterns commonly found within much larger programs/libraries.

\subsubsection{Double Free}
\label{sec:double-free-paper}
\label{sec:double-free}
We detail the (in)correctness proofs for CWE-415: Double Free outlined
in \cref{sec:overview-and-motivation}. To classify this vulnerability,
we require a memory allocation model, which is not part of the BIL
semantics (\cref{sec:specification-of-bil}). Memory allocation in C is
reflected as a PLT stub in the binary. 
In particular, we are only required to track the pointer (memory
address) that is allocated (and later freed). 
To this end, we develop a locale, \AllocationLocale,
that extends the BIL semantics (see \cref{fig:proofs}) with an abstract allocation model inspired by earlier works~\cite{DBLP:journals/pacmpl/MemarianGDKRWS19,DBLP:journals/jar/LeroyB08}.


\paragraph{The \AllocationLocale locale} 
We define \AllocationLocale by instantiating the small-step
transition relation $\BILSmallStep$ of \BilLocale (see
\cref{sec:local-locale-1}). We start by defining two memory
operations:
\[
  \memop ::= \AllocOp \mid \FreeOp
\]
where $w$ is the memory being allocated or freed and $sz$ is the size. Additionally, we
assume an abstract allocation function
$\NextAddrAllocatorName : \memop^* \rightarrow word$ that serves the
purpose of selecting suitable memory addresses for allocation and
retrieves the next address from the allocator based on the given
history (sequence) of allocations and deallocations. For each
allocation, we use $\NextAddrAllocatorName$ to obtain the next
available memory address. Consequently, in the state of
\AllocationLocale, we must track the sequence of $\memop$
operations, which is provided to $\NextAddrAllocatorName$ as an input
whenever an allocation occurs. For the Double Free example, no other
variables need to be tracked, hence we simply assume the state to be
this sequence, which we denote 
$\MemoryTrace : \memop^*$.

Recall that by using the $\Decode$ predicate
(\cref{section:instr-semantics}), the BIL instruction corresponding to
$\Command = \BILMachineStateE{\Variables}{pc}{mem}$ can be uniquely
identified. We use predicates
$\FreePredicateName : prog \rightarrow bool$ and
$\AllocPredicateName : prog \rightarrow bool$, which hold when the
next instruction to be executed in the given $\Command \in prog$
corresponds to a free and allocation instruction, respectively.
We assume that an instruction that frees memory cannot simultaneously
perform an allocation and vice versa, i.e.,
$\neg \FreePredicate{\Command} \lor \neg
\AllocPredicate{\Command}$. 
When the instruction for $\Command$ is an allocation, 
we utilise the function
$\GetSzName: prog \rightarrow sz$ to obtain the size allocated and
when the instruction for $\Command$ is a
deallocation, we utilise the function
$\GetFreedAddrName: prog \rightarrow word$ to obtain the address that
is freed. The size and location can be determined using the $\Decode$ function
on the given $\Command$.
Thus, 
we define a small-step transition relation, $\AllocSmallStep$, 
over
allocations in \cref{fig:allocation-steps}. 

\begin{figure}
    \centering
    {\footnotesize%
    \begin{mathpar} 
  \inferrule[(\AllocRule)]{
    \AllocPredicate{\Command} \and
    w = \NextAddrAllocator \\\\ sz = \GetSz{\Command} \\\\
    \MemoryTrace' = \MemoryTrace.[\AllocOp] }{ (\Command, \MopState)
    \AllocSmallStep \MopStateE{\MemoryTrace'} } 
\and 
\inferrule[(\FreeRule)]{
    \FreePredicate{\Command} \and
    w = \GetFreedAddr{\Command} \\\\
    \MemoryTrace' = \MemoryTrace.[\FreeOp] }{ (\Command, \MopState)
    \AllocSmallStep \MopStateE{\MemoryTrace'} } \and \inferrule[(\SkipRule)]{
    \neg\FreePredicate{\Command} \\\\
    \neg\AllocPredicate{\Command} }{ (\Command, \MopState)
    \AllocSmallStep \MopState }
\end{mathpar}}
\vspace{-10pt}
    \caption{Allocation semantics}
    \Description{Inference rules for the small-step allocation semantics}
    \label{fig:allocation-steps}
\vspace{-10pt}
\end{figure}

We encode this {\em allocation model} as the locale
$\AllocatorLocale$, fixing the functions \NextAddrAllocatorName,
\FreePredicateName, \AllocPredicateName, \GetSzName and \GetFreedAddrName. This
locale is derived as an \emph{interpretation} of the locale $\BilInferenceLocale$
retaining the decode predicate $\BILDecodeTransition$ and address set
$\ProgramDomain$ as abstract, but overriding
$\BilSmallStep\ =\ \AllocSmallStep$. Thus, we have the
signature $\AllocatorLocale = (\BILDecodeTransition, \ProgramDomain,
  \NextAddrAllocatorName, $ $ \FreePredicateName, \AllocPredicateName,
  \GetSzName, \GetFreedAddrName)$, 
where the parameters $\NextAddrAllocatorName$, $\FreePredicateName$,
$\AllocPredicateName$, $\GetSzName$ and $\GetFreedAddrName$ are used to derive
$\AllocSmallStep$ within the locale.


\paragraph{The \DoubleFreeBadLocale and \DoubleFreeGoodLocale
  locales} These locales correspond to the double free program 
and are {\em auto-generated} from the corresponding BIL programs
(shown in \cref{DF-bad-bil,DF-good-bil}). This step is trivial when
using \isabil's \lstinline[language=isabellen]{BIL_file} command,
which we have developed using Isabelle/ML. In particular, we simply
invoke the Isabelle/HOL commands:

\medskip
\begin{lstlisting}[language=isabellen,escapeinside={(*}{*)},numbers=none,backgroundcolor=\color{white}]
  BIL_file (*\guilsinglleft*)double-free-bad.bil.adt(*\guilsinglright*)  defining double_free_bad
  BIL_file (*\guilsinglleft*)double-free-good.bil.adt(*\guilsinglright*) defining double_free_good
\end{lstlisting}
\medskip 
\noindent where {\tt double-free-bad.bil.adt} is a file containing the
BIL for \autoref{DF-bad} in \BILa format and {\tt double\_free\_bad}
is the name of the locale to be generated. The good version is
similar.


\paragraph{The \DoubleFreeProofLocale locales}
We combine the two locales generated from the BIL programs with our
allocator locale ($\AllocationLocale$) to produce a two new locales
\DoubleFreeBadProofLocale and
\DoubleFreeGoodProofLocale. Recall from
\cref{sec:program-specification} that for each locale generated from a
\BILa program, we have access to a symbol table {\tt sym\_table}
mapping strings (i.e., external function calls) to the program address
at which the symbol appears. Note that when a BIL program calls a
function, it typically stores the return address, then jumps to the
function. The symbol corresponding to the program counter after the
jump contains the name of the external function, which is stored in
the symbol table. In RISC-V convention, the first argument of a
function call is stored in the register {\tt X10}. To allocate memory, 
function {\tt malloc} is called with the size as the first argument 
whereas to deallocate a pointer, function {\tt free} is called with 
the pointer as the first argument. Therefore, we can obtain the size 
allocated by a call to {\tt malloc} and pointer freed by a call to 
{\tt free} by retrieving the value stored in register~{\tt X10}.


We therefore instantiate the predicates $\FreePredicateName$,
$\AllocPredicateName$, $\GetSzName$ and $\GetFreedAddrName$ from the
\AllocationLocale locale as follows:
\begin{align*}
  \FreePredicate{(\_,pc,\_)}         &= (pc = \SymbolTable(``{\tt free}")) &
  \GetFreedAddr{(\Variables,\_,\_)}  &= \Variables({\tt X10})
  \\
  \AllocPredicate{(\_,pc,\_)}        &= (pc = \SymbolTable(``{\tt malloc}")) 
   &
  \GetSz{(\Variables,\_,\_)}         &= \Variables({\tt X10})
\end{align*}

The exact implementation of {\tt next\_free} occurs at a lower level
of abstraction, and depends on the allocation model. At this level, we
provide the main requirement axiomatically, i.e., we require
\vspace{2pt}

\hfill
$w = \NextAddrAllocator \implies (\forall i.\ \MemoryTrace(i) = \AllocOp \implies \exists j.\ j > i \wedge \MemoryTrace(i) = \FreeOp)$\hfill {}

\vspace{2pt}\noindent
This presents us with all the components necessary to prove
(in)correctness of our examples. The following predicate captures the
double-free over the memory state defined above:
\begin{align*}
    \IsDF 
   & = \lambda \MemoryTrace.\ 
      \exists i,j,w.\ \inarr{
      i < j \wedge \MemoryTrace[i] = \FreeOp \wedge \MemoryTrace[j] = \FreeOp \wedge {}   \\
  (\forall k.\  i <  k < j \implies \MemoryTrace[k] \neq \AllocOp)}
\end{align*}
Note that the predicate captures {\em only} the double-free
vulnerability and does not formalise other pointer misuse
vulnerabilities, e.g., use-after-free or free-before-alloc.

\begin{theorem} \label{thy:double-free} Let $\Command_{bad}$ and $\Command_{good}$ be the
  programs corresponding to \DoubleFreeBadLocale and
  \DoubleFreeGoodLocale, respectively.  Both of the following
  hold
  \begin{enumerate}[leftmargin=15pt]
  \item $\IncorrectnessProof{\neg \IsDF}{\ \Command_{bad}\ }{Q}{\IsDF}$

  \item $\HoareTriple{\neg \IsDF}{
\ \Command_{good}\ }{\neg \IsDF}$    
\end{enumerate}
\end{theorem}

\subsubsection{AV rules} 
\label{sec:av-rules-paper}
In a similar manner, we prove incorrectness of the AV rules from \cref{tab:av-rules}. Details of these proofs are provided in \cref{sec:forbidden-symbols}. 






\section{Case Study: Double-Free Vulnerability in cURL (CVE-2016-8619)}
\label{sec:double-free-curl}

\begin{figure}
  \centering
  \begin{mdframed}[backgroundcolor=backcolour,hidealllines=true,%
                   innerleftmargin=0,innerrightmargin=0,
                   innertopmargin=-0.2cm,innerbottommargin=-0.2cm]
\begin{lstlisting}[language=c,escapeinside={(*}{*)},lastline=11,basicstyle=\tt\fontsize{7}{7.5}\selectfont]
static __attribute__ ((noinline)) CURLcode read_data(struct connectdata *conn,
                                                     curl_socket_t fd,
                                                     struct krb5buffer *buf)
{
  int len;
  void* tmp = NULL;
  CURLcode result;

  result = socket_read(fd, &len, sizeof(len));(*\label{line:read-data-read-len}*) // result = 0 iff socket_read successful
  if(result)(*\label{line:read-data-result}*) 
    return result;
\end{lstlisting}
  \end{mdframed}
  \vspace{-1.7ex}
  \begin{mdframed}[backgroundcolor=red!10,hidealllines=true,%
                     innerleftmargin=0,innerrightmargin=0,
                     innertopmargin=-0.15cm,innerbottommargin=-0.12cm]
  \begin{minipage}{0.5\textwidth}
    \begin{mdframed}[backgroundcolor=red!10,hidealllines=true,%
                     innerleftmargin=0,innerrightmargin=0,
                     innertopmargin=-0.15cm,innerbottommargin=-0.12cm]
\begin{lstlisting}[language=c,escapeinside={(*}{*)},backgroundcolor=\color{red!10},   
                       firstline=12,lastline=15,showlines=true,firstnumber=12,basicstyle=\tt\fontsize{7}{7.5}\selectfont]

  len = ntohl(len);(*\label{line:read-data-ntohl}*)
  tmp = realloc(buf->data, len);(*\label{line:read-data-realloc}*)
 
\end{lstlisting}
    \end{mdframed}
  \end{minipage}
  \hfill
  \begin{minipage}{0.494\textwidth}
    \begin{mdframed}[backgroundcolor=green!10,hidealllines=true,%
                     innerleftmargin=0,innerrightmargin=0,
                     innertopmargin=-0.15cm,innerbottommargin=-0.12cm]
\begin{lstlisting}[language=c,escapeinside={(*}{*)},backgroundcolor=\color{green!10},
                       firstline=12,firstnumber=12,lastline=15,numbers=none,basicstyle=\tt\fontsize{7}{7.5}\selectfont]
  if (len) {(*\label{line:read-data-guard}*)
    len = ntohl(len);
    tmp = realloc(buf->data, len);
  }    
\end{lstlisting}
    \end{mdframed}
  \end{minipage}
  \end{mdframed}
  \vspace{-1.7ex}
  \begin{mdframed}[backgroundcolor=backcolour,hidealllines=true,%
                   innerleftmargin=0,innerrightmargin=0,
                   innertopmargin=-0.2cm,innerbottommargin=-0.2cm]
\begin{lstlisting}[language=c,escapeinside={(*}{*)},firstline=16,firstnumber=16,basicstyle=\tt\fontsize{7}{7.5}\selectfont]
  if(tmp == NULL)
    return CURLE_OUT_OF_MEMORY;

  buf->data = tmp;
  result = socket_read(fd, buf->data, len);
  if(result)
    return result;
  buf->size = conn->mech->decode(conn->app_data, buf->data, len,
                                 conn->data_prot, conn);
  buf->index = 0;
  return CURLE_OK;
}
\end{lstlisting}
  \end{mdframed}
  \vspace{-2ex}
  \caption{The \ReadData function of the cURL library in \code{security.c}.
           The snippet highlighted in \colorbox{red!10}{red} shows the vulnerable code in version 7.50.3, and that in \colorbox{green!10}{green} shows the corrected vulnerability in version 7.51.0.}
  \label{fig:curl-double-free}
\end{figure}

cURL is an open-source library primarily used for transferring data over the internet 
using various protocols, including TLS.
It has a large user base, supporting multiple operating systems and architectures (including \RiscV), and thus ensuring the security of cURL is critically important.

Unfortunately, however, cURL has been known to contain vulnerabilities.
Consider the function \ReadData in \cref{fig:curl-double-free}, which underpins the Kerberos authentication protocol in cURL (prior to version 7.51.0), where 
\cref{line:read-data-read-len} reads the length of an incoming 
data packet from a socket.
If this succeeds, \code{socket\_read} will return a 0, causing the if statement on 
\cref{line:read-data-result} to fall through.
Then, on \cref{line:read-data-ntohl} the call to \code{ntohl} converts 
the data length (\code{len}) from network byte order (big-endian) to host byte order (big- or little-endian). 
Note that this method does not sanitize its input.
Next, \Realloc on \cref{line:read-data-realloc} resizes the \code{data} buffer in \KrbBuffer, 
and returns 
either a new pointer to the resized memory, or \code{NULL} 
if an error occurs.
The rest of the method then reads and decodes the remaining data from the socket.

The caller is responsible for managing the \KrbBuffer pointer by allocating it before and freeing it after calling \ReadData.
Here we assume \Malloc is called directly before \ReadData
and free is called directly after, as follows:
\lstinline{malloc(buf, sz); read_data(conn, fd, buf); free(buf);}

If receiving data from the socket on \cref{line:read-data-read-len} yields a length of 0 ($\code{len}=0$), the call to \Realloc with $\code{len} = 0$ results in \emph{zero reallocation}, leading to \emph{undefined behaviour}\footnote{\url{https://www.open-std.org/jtc1/sc22/wg14/www/docs/n2464.pdf}}. 
Most implementations of \Realloc 
will return NULL and free the pointer following a \emph{zero reallocation}.
As such, since the \KrbBuffer buffer is already freed before \Realloc returns and the caller is expected to free it after the function returns, this will lead to a \emph{double-free vulnerability}, identified by the vulnerability enumeration \href{https://cve.mitre.org/cgi-bin/cvename.cgi?name=2016-8619}{CVE-2016-8619}. 


This vulnerability is resolved in version 7.51.0 of cURL by preempting the zero-allocation that leads to it: as highlighted in \colorbox{green!10}{green} in \cref{fig:curl-double-free}, placing a guard around the call to \Realloc \footnote{\url{https://github.com/curl/curl/commit/3d6460edeee21d7d790ec570d0887bed1f4366dd}} ensures that it is not called with a zero length. As such, if $\code{len}=0$, then the call to \Realloc is skipped,
\code{TMP} remains \code{NULL} and returns 0.
This ensures consistent behaviour across all C implementations.

\subsection{Scalability of the Parser}

\begin{figure}
  \centering
  \begin{mathpar} \footnotesize
  \inferrule[(\ReallocRule)]{
    \ReallocPredicate(\Command) \and
    w = \GetFreedAddr{\Command} \\\\ 
    sz = \GetSz{\Command} \\\\
    \MemoryTrace' = \MemoryTrace.[\AllocOp] }{ (\Command, \MopState)
    \ReallocSmallStep \MopStateE{\MemoryTrace'} } 
  \and 
  \inferrule[(\ZeroReallocRule)]{
    \ReallocPredicate(\Command) \and
    w = \GetFreedAddr{\Command} \\\\
    sz = \GetSz{\Command} \and
    sz = 0 \\\\
    \MemoryTrace' = \MemoryTrace.[\FreeOp] }{ (\Command, \MopState)
    \ReallocSmallStep \MopStateE{\MemoryTrace'} } 
    \and 
  \inferrule[(\SkipRule)]{
    \neg\FreePredicate{\Command} \\\\
    \neg\AllocPredicate{\Command} \\\\
    \neg\ReallocPredicate(\Command)
  }{ (\Command, \MopState)
    \ReallocSmallStep \MopState 
  }
  \end{mathpar}
  \vspace{-10pt}
  \caption{Reallocation semantics}
  \label{fig:reallocation-steps}
  \vspace{-10pt}
\end{figure}

We cross compile cURL \curlold for \RiscV on Linux using the GNU \RiscV compiler 
with the default options.
We modify the source code for \ReadData by adding the \lstinline{noinline} attribute, which ensures the function is not inlined.
Note that \lstinline{noinline} is not a requirement for verification, but facilitates compositional reasoning (see \cref{sec:curl-comp}).

Using the generated BIL directly creates a scalability challenge. 
The assembly corresponding to version 7.50.3 contains {\em 63,592 \RiscV instructions}, which equates to {\em 127,023 LoC in BIL} with {\em 63,592 BIL instructions}. 
%
Naively generating the corresponding \IsaBIL locales 
(see \cref{sec:parser}) takes approximately 14 minutes, which is impractical for verification. Of this, 
the lexer takes 4s, the parser takes 2s and the translator takes approximately 13 minutes.
The wait time in the translator arises from instantiating the Isabelle/HOL locale
in the proof context, an operation which cannot be avoided.

To address this, we extend the \CommandBILfile command with a \emph{new option}, \CommandwithSR, to focus only on the subroutines of interest (and their dependencies). For example, to generate the \IsaBIL locale for \code{read\_data} in version 7.50.3, we use the following command: 
\begin{lstlisting}[language=isabelle,escapeinside={(*}{*)},numbers=none,backgroundcolor=\color{white}]
    BIL_file (*\guilsinglleft*)libcurl.7.50.3.bil.adt(*\guilsinglright*) defining read_data_7_50_3
      with_subroutines read_data and ... 
\end{lstlisting}
where ``\dots'' contains functions that \ReadData calls (e.g., \code{ntohl}). 
Running \CommandBILfile now takes only 6s, with 131 instructions to verify. 

We next show how we detect this vulnerability in version 7.50.3 using an incorrectness proof. 



\subsection{Incorrectness of the \ReadData Subroutine}
\label{sec:incorr-readd-subr}





As described above, \ReadData contains a double-free vulnerability. However, unlike our prior example (\cref{sec:double-free-paper}), the vulnerability arises from a reallocation of memory. 
Thus, we first describe the construction of a \ReallocLocale locale, which reuses components from the \AllocationLocale locale. 
Then, we prove the incorrectness of the \ReadData function in version \curlold.

\paragraph{The \ReallocLocale locale}
%
We define \ReallocLocale by instantiating the small-step transition 
relation $\BilSmallStep$ of \BilLocale (see
\cref{sec:local-locale-1}).
Whilst we do not extend \AllocationLocale from \cref{sec:double-free-paper} directly, as it already defines small step 
semantics ($\AllocSmallStep$) that are incompatible with reallocation, we 
can borrow:
\begin{enumerate*}
  \item the definition of $\memop$,
  \item our abstract allocator $\NextAddrAllocatorName$,
  \item our memory trace $\MemoryTrace$,
  \item the getter for the size $\GetSzName$,
  \item the getter for the pointer to free $\GetFreedAddrName$,
  \item the predicate that denotes a program is freeing memory $\FreePredicateName$ and
  \item the predicate that denotes a program is allocating memory $\AllocPredicateName$.
\end{enumerate*}
These concepts are explained in detail in \cref{sec:double-free-paper}.

We use the predicate
$\ReallocPredicate$ to determine whether the 
next instruction to be executed in the given $\Command$
corresponds to a \Realloc instruction. We assume that an instruction that frees or allocates memory cannot simultaneously
perform an reallocation and vice versa, i.e.,
$\neg \FreePredicate{\Command} \lor 
 \neg \AllocPredicate{\Command} \lor 
 \neg \ReallocPredicate(\Command)$. 

When a program reallocates memory it has two parameters, 
a pointer to the memory to be resized, and the new size of that memory.
We utilise the existing functions $\GetSzName$ to obtain the 
size reallocated and 
$\GetFreedAddrName$ to obtain the address that
is reallocated. 
The size and location can be determined using the $\Decode$ function
on the given $\Command$.
The size and location are then used to add an $\AllocOp$ to the 
$\MemoryTrace$.
However, if the size of a reallocation is 0, this represents a zero-reallocation, and hence a $\FreeOp$ is added to the $\MemoryTrace$.
This results in a small-step transition relation, $\ReallocSmallStep$, 
over reallocations as given in \cref{fig:reallocation-steps}.     
Note that we retain the \AllocRule and \FreeRule rules 
from \cref{fig:allocation-steps}, and adapt the 
\SkipRule to encompass is reallocations.
Finally, we obtain the signature 
\[\ReallocLocale = (\BILDecodeTransition, \ProgramDomain,
  \NextAddrAllocatorName, \FreePredicateName, \AllocPredicateName, \ReallocPredicate,
  \GetSzName, \GetFreedAddrName)\] 
where the parameters $\NextAddrAllocatorName$, $\FreePredicateName$, $\ReallocPredicate$,
$\AllocPredicateName$, $\GetSzName$ and $\GetFreedAddrName$ are used to derive
$\ReallocSmallStep$ within the locale.

\paragraph{The {\tt read\_data} proof locale}
We combine the  locale generated for cURL binaries to produce
a new locale \CurlBadProofLocale. 
Reallocations are handled by calls to a PLT stub for \Realloc
so we can proceed in the same way as \code{malloc}/\code{free}.
In \cref{sec:double-free-paper}, we discussed how the first argument
of a subroutine call is stored in the register {\tt X10}.
Reallocations take two arguments:
\begin{enumerate*}
  \item a pointer to memory, and 
  \item a size.
\end{enumerate*}
The second argument (2) is stored in the return register ({\tt X11}),
thus 
we instantiate the parameters of \ReallocLocale as follows
for both locales:
\begin{align*}
  \ReallocPredicate{(\_,pc,\_)} &= (pc = \SymbolTable(``{\tt realloc}")) \\
  \GetSz{\Variables,pc,\_} &= 
    \begin{cases}
      \Variables({\tt X11}) & \textbf{if $pc = \SymbolTable(``{\tt realloc}")$} \\
      \Variables({\tt X10}) & \textbf{otherwise} 
    \end{cases}
\end{align*}
The remaining definitions of \GetFreedAddrName, \FreePredicateName
and \AllocPredicateName match \AllocatorLocale (see \cref{sec:double-free-paper}).

\begin{theorem} 
\label{thm:curl-read-data-locale} 
Let $\ProgBad$ 
be the program corresponding to \CurlBadLocale. Then 

\noindent\hfill$\IncorrectnessProof{\neg \IsDF}{\ \ProgBad\ }{Q}{\IsDF}$.\hfill{}

\end{theorem}



\subsection{Compositionality}
\label{sec:compositionality}
\label{sec:double-free-curl-compositional}
\label{sec:curl-comp}
\begin{figure}
  \centering
  \scalebox{0.9}{%
  \begin{tikzpicture}
    \node[rect] (ReadData) {\ReadData};
    \node[rect, right=2em of ReadData] (SecRecv) {\SecRecv};
    \node[rect, right=2em of SecRecv] (ChooseMech) {\code{choose\_mech}};
    \node[rect, fill=red!30, right=2em of ChooseMech] (CurlSecLogin) {\code{Curl\_sec\_login}};


    \path[draw, thick, ->] (SecRecv) -- (ReadData); 
    \path[draw, thick, ->] (ChooseMech) -- (SecRecv); 
    \path[draw, thick, ->] (CurlSecLogin) -- (ChooseMech); 

  \end{tikzpicture}}%
  \caption{A partial call graph in cURL demonstrating a path from an external function (\colorbox{red!30}{highlighted}), \code{Curl\_sec\_login}, to the internal function \ReadData.
           }
  \label{fig:curl-callgraph}
  \vspace{-10pt}
\end{figure}
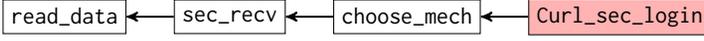

The proof in \cref{fig:curl-double-free} covers the \ReadData subroutine in the cURL library. 
However, this subroutine is an \emph{internal} function (i.e. not defined in a header file), making the vulnerability difficult to exploit directly in practice.
Nevertheless, as shown in the call graph in \cref{fig:curl-callgraph}, there is a path from an \emph{external} function (defined in a header file), namely \code{Curl\_sec\_login}, to \ReadData.
That is, a call to \code{Curl\_sec\_login} includes a call to \code{choose\_mech}, which in turn includes a call to \code{sec\_recv}, which itself calls \ReadData.
As such, the vulnerability in \ReadData \emph{could potentially} be exploited by a call to \code{Curl\_sec\_login}.  

Note that this may not always be the case, as freeing the data buffer pointer is left up to the caller, not the callee, and thus further incorrect behaviour to by caller (in this case a function above \ReadData in the call graph) could
rectify this vulnerability by forgetting to call free, or 
perhaps reallocating the pointer again before freeing it. 
Thus, we should aim to establish that each of these methods are (in)correct formally.

We do this in a \emph{compositional} fashion through \emph{proof reuse}: we reuse the (in)correctness specifications we prove at lower levels of the call graph in higher levels.
For instance, when proving that \code{sec\_recv} is (in)correct, we reuse the specification of \ReadData in \cref{thm:curl-read-data-locale} as a lemma. 
Indeed, this is precisely why we used the \code{noinline} attribute when compiling \ReadData: this allows us to reuse its specification; otherwise each call to \ReadData would be inlined, preventing us from proof reuse.

\newcommand{\SecRecvBad}{{\it SR}_{7.50.3}}
\newcommand{\ChooseMechBad}{{\it CM}_{7.50.3}}
\newcommand{\CurlSecLoginBad}{{\it CSL}_{7.50.3}}

In \cref{thm:sec-rev-vuln} below, we prove incorrectness specifications for the $\SecRecvBad$, $\ChooseMechBad$ and $\CurlSecLoginBad$ functions, showing that each of these functions do indeed exhibit a double-free vulnerability. 
Our proof is compositional in that the proof for each function reuses the specification of the function below it, as discussed above (e.g. the proof of $\SecRecvBad$ reuses the incorrectness specification of \ReadData in \cref{thm:curl-read-data-locale} as a lemma). 

\begin{theorem}
\label{thm:sec-rev-vuln}
Let $C \in \{\SecRecvBad, \ChooseMechBad, \CurlSecLoginBad\}$, be a program corresponding to the \curlold versions of \code{sec\_recv}, \code{choose\_mech} and \code{Curl\_sec\_login}. Then: 

$\IncorrectnessProof{\neg \IsDF}{\ C\ }{Q}{\IsDF}$. 



\end{theorem}

\cref{thm:sec-rev-vuln} demonstrates a scalable proof for the incorrectness specifications of the functions in \cref{fig:curl-callgraph}. 
We can apply similar techniques to verify their corresponding correctness specifications for version \curlfixed; we have elided these proofs as they are similar. 






\section{Related Work}
\label{sec:related-work}

\newcommand{\picinae}{{\sc picin\ae}\xspace}

The importance of binary-level analysis and the growing power of
verification tools has meant that there is increasing work on theorem
provers 
being applied to verify program binaries.

\vspace{-2pt}
\paragraph{Verified machine code}
Closest to our work is the \picinae
framework~\cite{10.1145/3338502.3359759}, which has been developed to
reason about BIL in the Coq theorem
prover~\cite{bertot2013interactive}. 
Like \IsaBIL, they lift BIL outputs to Coq, in turn allowing proofs of generic
properties ranging from termination to full functional verification
within a Coq proof environment. 
However, the Picinæ lifter is unverified and introduces an additional step by 
lifting BIL to a bespoke intermediate representation (IR) with unclear semantics.
The authors of Picinæ describe their work as preliminary, with examples mainly 
covering smaller subroutines such as memset and string comparison.

Early works on using verification of machine-code include works by
\citet{DBLP:conf/fsen/MyreenFG07} and
\citet{DBLP:conf/tacas/MyreenG07}, who developed high-level models of
Arm hardware followed by Hoare logic proofs against these
models. 
Similarly, \citet{DBLP:conf/popl/JensenBK13} develop high-level
separation logics for verifying low-level machine code applied to an
{\em abstraction} of x86 instructions. The main focus of the work is
the design of a separation logic and appropriate frame rules for
unstructured programs, with the framework being encoded in Coq.

Later works by \citet{DBLP:conf/fmcad/MyreenGS08,DBLP:conf/fmcad/MyreenGS12} 
develop \emph{decompilation into logic}, capable of targeting x86, ARM and PowerPC.
Instead of using a lifter to convert to an established IR such as BIL, then
performing verification according to the IR's operational semantics, 
this approach decompiles binaries into tail-recursive functions 
in the logic of the HOL4~\cite{DBLP:journals/jfp/Hutton94} theorem prover.
Verification then occurs over Hoare triples for this logic. 
Here, the trust is shifted away from conventional lifters like BAP, and onto
mechanised models of x86 \cite{DBLP:conf/cade/CraryS03}, PowerPC \cite{DBLP:conf/popl/Leroy06}
and ARM \cite{DBLP:conf/tphol/Fox03,DBLP:conf/itp/FoxM10}, 
theoretically achieving a smaller trusted base.
\citeauthor{DBLP:conf/fsen/MyreenFG07} acknowledge that most \cite{DBLP:conf/tphol/Fox03,DBLP:conf/cade/CraryS03,DBLP:conf/popl/Leroy06} of these models were not created for the purpose of decompilation into logic. 
Moreover, hardware has changed significantly since these early works,
so it is unclear if these models are an accurate reflection of
current-day ISAs.

More recently, the Islaris
framework~\cite{DBLP:conf/pldi/SammlerHL0PD0S22} has been developed to
leverage the extensive set of works on Sail-based Armv8-A and RISC-V
ISA
semantics~\cite{DBLP:conf/esop/SimnerAPPGS22,DBLP:conf/esop/SimnerFPAPMS20,DBLP:journals/pacmpl/ArmstrongBCRGNM19}. Islaris
is based on higher-order separation logic implemented in Coq and
provides automatic translations from Sail to Coq models. Their
end-goals are however slightly different to ours: while they focus on
high-fidelity reasoning directly over ISA models, we focus on the
lifted versions of assembly to provide a unifying intermediate-layer
proof environment that applies to multiple architectures. As such, our
automation techniques (described in \cref{sec:automation}) are
reusable across any architecture and compilation toolchain that is
supported by BAP. 
Moreover, we have used \isabil to analyse a larger body of examples: 
Islaris has been used to verify 173 assembly LoC (across nine examples), while we have used IsaBIL to verify 612 assembly LoC (across eleven examples) -- our largest example alone comprises 131 assembly LoC

MiniSail~\cite{wassellmechanised} encodes a subset of the Sail
language in various proof assistants, including Isabelle/HOL, 
providing the method to reason over lifted Sail models at a level
similar to \IsaBIL.
Sail has a thorough type system with 
type checking operational semantics, which can be used to verify
if a Sail model is typed correctly.
At the time of writing, the MiniSail development was developed predominantly  
to ensure that the Sail language is sound. 
Thus it lacks the mature automation techniques and higher levels of reasoning 
present in \IsaBIL.

\vspace{-2pt}
\paragraph{Verified compilation} Perhaps the two most famous examples
of compilation correctness are the
CakeML~\cite{DBLP:conf/popl/KumarMNO14} and
CompCert~\cite{DBLP:conf/fm/BlazyDL06} projects, aimed at the
correctness of compilations to (subsets of) ML and C,
respectively. \citet{DBLP:conf/pldi/SewellMK13} check compilation
correctness of high-level C code for the seL4 microkernel down to
binaries generated by gcc 4.5.1. The Bedrock Structured Programming
System~\cite{DBLP:conf/icfp/Chlipala13} provides an intermediate layer
representation combined with strongest postcondition calculations to
verify low-level C macros within the Coq theorem prover. The system
has been designed to optimise reasoning about C macros unlike \isabil,
which allows verification of binaries that may not be generated by a
compiler.

Unlike \isabil, these works leverage proofs of functional correctness
at the source level, i.e., are performed at a higher level of
abstraction. However, the proofs must take into account the compiler
models and optimisation levels that are ultimately to be applied, with
extra checks to ensure that the implemented architectures do not
interfere with any high-level assumptions. Such approaches would not
be suitable for situations where the source code is not available,
whereas BAP and consequently \isabil does not require the source code.


\vspace{-2pt}
\paragraph{Verified lifting}
\citet{DBLP:conf/pldi/VerbeekBFR22} have formalised and integrated a
subset of the x86-64 instruction set (i.e., the 64-bit subset of x86)
within Isabelle/HOL and used it to verify so called {\em sanity
  properties} (e.g., the integrity of a return address) for a large
codebase. However, the focus of this verification is correctness of
the lifting, as opposed to properties about the program itself (e.g.,
termination, functionality).
\IsaBIL employs type verification, however, this method only verifies that the lifted BIL is correct 
according to its specification, and not the specification of the assembly 
it was lifted from.
\citet{DBLP:conf/fmcad/LamC23} developed a verified BAP lifter for ARMv8, 
though other ISAs rely on external validation for correctness of their respective 
BAP lifters. 
However, we note that validation is already
accompanied by an extensive set of validation tools that uses trace
files to verify the lifter's
accuracy~\footnote{\url{https://github.com/BinaryAnalysisPlatform/bap-veri}},
that are checked ``per instruction''. We consider formal verification
of BAP lifters to be future work.

\vspace{-2pt}
\paragraph{Verified instrumentation}
Armor~\cite{DBLP:conf/emsoft/ZhaoLSR11} presents a checker for memory
safety and control flow integrity of Arm binaries, which instruments a
check before particular operations, then uses HOL to prove that the
modified binary is correct. Similarly,
RockSalt~\cite{DBLP:conf/pldi/MorrisettTTTG12} is a Coq-based checker
for a subset of x86 that is used to verify software-based fault
isolation implementations. RockSalt is designed to check specific
sandbox policies, e.g., that the code will only read/write data from
specified contiguous segments in memory. Unlike \isabil, which has the
full flexibility of Hoare and O'Hearn's logics for proving
(in)correctness, both Armor and RockSalt focus on a subset of
properties. 

\vspace{-2pt}
\paragraph{Security analysis}
An overview of security properties, exploits and tools for binary
analysis has been given by \citet{shoshitaishvili2016state}.  Above,
we have focussed mainly on safety, though also describe how BAP can be
extended to reason about transient execution vulnerabilities (e.g., as
exploited by Spectre and
Meltdown)~\cite{DBLP:journals/tse/WangCGMR21,DBLP:conf/csfw/CheangRSS19,DBLP:conf/fm/GriffinD21}. \citet{DBLP:conf/csfw/CheangRSS19}
cover model checking (using UCLID5), while
\citet{DBLP:journals/tse/WangCGMR21} develop a taint tracking tool
based on BAP to check correctness of Spectre mitigations.
\citet{DBLP:conf/fm/GriffinD21} consider proofs of speculative
execution over BAP outputs, but only cover a subset of the BIL
language called AIR. They apply Hoare-style proofs to verify a
hyperproperty~\cite{DBLP:journals/jcs/ClarksonS10} known as
TPOD~\cite{DBLP:conf/csfw/CheangRSS19} directly over AIR without any
of the modularity offered by \isabil.  Nevertheless, it would be
interesting to extend our work to cover proofs of (in)correctness of
speculative execution~\cite{DBLP:conf/pldi/CauligiDGTSRB20,DBLP:conf/csfw/DongolGPW24}.

\paragraph{(Incorrectness) separation logic}
In \isabil we do not currently have support for separation logic (SL) or its incorrectness analogue, ISL \cite{isl}.
While the absence of (I)SL support does not impede the scalability of our approach, it is an interesting direction of future work to pursue. 
Doing so, however, is non-trivial as the memory representation in BIL is incompatible with that of (I)SL.
More concretely, in (I)SL the memory (heap) is represented as a \emph{partial commutative monoid} (typically as a map from locations to values) that can be broken down to smaller pieces (heaplets), thus affording seamless ``separation'' of memory resources. This also makes it straightforward to track \emph{knowledge} and \emph{ownership} of the existence of locations in memory and their associated values. 
For instance, the (I)SL assertion $x \mapsto 1 * y \mapsto 2$ describes a memory comprising exactly two locations, $x$ and $y$, respectively holding values $1$ and $2$. 

By contrast, as described in \cref{sec:specification-of-bil-syntax}, memory in BIL is represented as a ``history'' of updates (mutations) that must be ``replayed'' to ascertain its contents. 
For instance, the BIL memory $[x \leftarrow 1][y \leftarrow 2][x \leftarrow 3]$ describes \emph{any} memory obtainable after applying (replaying) the listed updates in order, namely by first updating location $x$ to hold $1$, then updating $y$ to $2$ and finally re-updating $x$ to $3$.
While this confers the existence and the final values of locations $x$ and $y$, it has no bearing on the existence or the values of other memory locations.

Representing the memory in BIL as described above is a design choice we \emph{inherited} from BIL, and it is not immediately conducive to ``separation'' as in (I)SL. 
As such, to add support for (I)SL in BIL we would have to fundamentally change the memory representation in BIL, an undertaking which is far from trivial and will have far-reaching ramifications. While we believe this a worthy direction of work in the future, doing so is beyond the scope of our work here. 

\paragraph{Modular Language Semantics.}
By structuring our mechanisation using locales, we enable a modular formalization of (in)correctness, allowing each component to be parameterised by different program semantics. This approach aligns with existing work on modular program semantics \cite{DBLP:journals/pacmpl/ManskyD24}, such as Iris~\cite{DBLP:conf/popl/JungSSSTBD15}, where Hoare logic is defined parametrically with respect to the underlying language. IsaBIL presents the first encoding of parameterised semantics for Hoare and O’Hearn logic within Isabelle/HOL. Prior research~\cite{DBLP:conf/csfw/CheangRSS19,DBLP:conf/fm/GriffinD21} has shown that extending BIL's semantics is both straightforward and beneficial. By structuring our development using locales, we afford a degree of modularity, presenting the first abstract formalisation of (in)correctness triples in Isabelle/HOL.






%

%



\section{Conclusions}

Program analysis in the absence of source code techniques typically focus on a combination of hardware (e.g., x86, ARM), intermediary representations, e.g., LLVM-IR, high-level language models (e.g., C/C++). Our work adds to the growing literature on low-level verification using Hoare logic, and is the first to also apply incorrectness logic to verify the absence and presence of bugs in binaries. While many works in the literature focus on particular architectures, e.g., RISC-V, x86 or ARM, our work is on the generic BAP platform, and hence inherits its generality.
BAP is a production ready system with 1000s of users but lack a verification component that IsaBIL provides. We have developed a complete formalisation of BIL, BAP's intermediary language in Isabelle/HOL, and developed the first (in)correctness logics for BIL. Our formalisation of the BIL specification uncovered bugs in that specification, which have been reported to the developers and corrected. We provide automation to enable better proof scalability, and demonstration compositionality on case studies larger than our counterparts (Islaris).

\bibliography{refs}

\newpage
\appendix

\newcommand{\ImpImmediateE}[1]{\textbf{N}(#1)}
\newcommand{\ImpImmediate}{\ImpImmediateE{i}}
\newcommand{\ImpVarE}[1]{\textbf{V}(#1)}
\newcommand{\ImpVar}{\ImpVarE{str}}
\newcommand{\ImpBoolConditionE}[1]{\textbf{Bc}(#1)}
\newcommand{\ImpBoolCondition}{\ImpBoolConditionE{bool}}

\section{Instantiating (In)correctness in Isabelle/HOL}
\label{sec:inst-incorr-as}

The Isabelle/HOL standard library~\cite{nipkow2002isabelle} and the Archive of Formal
Proofs contain many formulations of Hoare and O'Hearn
logic built for different purposes.
In particular, Isabelle/HOL has the following theories for 
Hoare logic: IMP, IMPP, Owicki-Gries, Rely Guarantee, Abstract Hoare Logics~\cite{DBLP:journals/afp/Nipkow06} 
and A Hoare Logic for Diverging Programs~\cite{DBLP:journals/afp/PohjolaMT23}.
For incorrectness logic, this includes IMP~\cite{DBLP:journals/afp/Murray20}.

\begin{figure}[t]
  \begin{minipage}[t]{0.48\columnwidth}
  \vspace{-2ex}
  \resizebox{\textwidth}{!}{%
  \begin{tikzpicture}[yscale=-1]

  \node[specOP] (inference) {
    \large \Locale{inference}};
  
  \node[specOP,above = 0.7cm of inference] (incorrectness)
  { \large \Locale{incorrectness}};

  \node[specOP, left = 0.7cm of incorrectness] (correctness) {
    \large \Locale{correctness}};

  \node[BILss, below = 9ex of inference] (IMP) { \large \Locale{IMP}};

  \node[BILss, left = of IMP] (IMPP) { \large \Locale{IMPP}};
  \node[BILss, left = of IMPP] (OG) { \large \Locale{OG}};
  \node[BILss, left = of OG] (RG) { \large \Locale{RG}};
  \node[BILss, left = of RG] (FD) { \large \Locale{For\_Divergence}};

  \node[BILss, left = 16.5ex of correctness] (While) { \large \Locale{Hoare\_While}};
  \node[BILss, above = of FD] (AH) { \large \Locale{Abstract\_Hoare}};

  \draw[refines] (correctness) to node[left,pos=0.6,yshift=-1mm]{\large sublocale} (inference);
  \draw[refines] (incorrectness) to node[right]{\large sublocale} (inference);

  \draw[extends] (inference)   to node[left,pos=0.6]{\large instantiate} (IMP);

  \draw[extendsl] (correctness) to node[left,pos=0.85]{\large instantiate} ($(correctness.south)+(0,2.2)$);
  \draw[extends] ($(correctness.south)+(0,2.2)$) to node[right,pos=0.6]{} (IMPP);
  \draw[extends] ($(correctness.south)+(0,2.2)$) to node[right,pos=0.6]{} (FD);
  \draw[extends] ($(correctness.south)+(0,2.2)$) to node[right,pos=0.6]{} (OG);
  \draw[extends] ($(correctness.south)+(0,2.2)$) to node[right,pos=0.6]{} (RG);
  \draw[extends] (correctness) to node[left,yshift=1mm]{\large instantiate} (AH);
  \draw[extends] (correctness) to node[above,pos=0.52]{\large instantiate} (While);
  
  \draw[black,thick,dotted] ($(correctness.north west)+(-0.2,-0.2)$)  rectangle ($(inference.south east)+(0.7,0.2)$) node [pos=0,xshift=4mm,yshift=-1.8cm]{\large{(\cref{sec:incorrectness})}};
 
  \end{tikzpicture}}
  \end{minipage}
  \hfill
  \begin{minipage}[t]{0.51\columnwidth}
  \vspace{-2ex}
  \resizebox{\textwidth}{!}{%
  \begin{tabular}{|l|c|c|}
\hline
\textbf{Library}                     & \CorrectnessLocale & \IncorrectnessLocale \\ \hline
Hoare While Language 
& {\color[HTML]{32CB00} $\checked$}         & -             \\ \hline
Imperative Language (IMP) 
& {\color[HTML]{32CB00} $\checked$}         & {\color[HTML]{32CB00} $\checked$}           \\ \hline
Parallel IMP (IMPP) 
& {\color[HTML]{32CB00} $\checked$}         & -             \\ \hline
Owicki-Gries 
& {\color[HTML]{32CB00} $\checked$}         & -             \\ \hline
Rely Guarantee 
& {\color[HTML]{32CB00} $\checked$}         & -             \\ \hline
Abstract Hoare Logic 
& {\color[HTML]{32CB00} $\checked$}         & -             \\ \hline
A Hoare Logic for Diverging Programs 
& {\color[HTML]{32CB00} $\checked$}         & -             \\ \hline
  \end{tabular}}
  \end{minipage}
  \caption{Interpretation of the \CorrectnessLocale and \IncorrectnessLocale locales for existing libraries in Isabelle/HOL. Most libraries do not provide semantics for incorrectness, this is indicated with a dash (-).}
  \label{fig:incorrectness-imp}
  \label{tab:incorrectness-imp}
\end{figure}
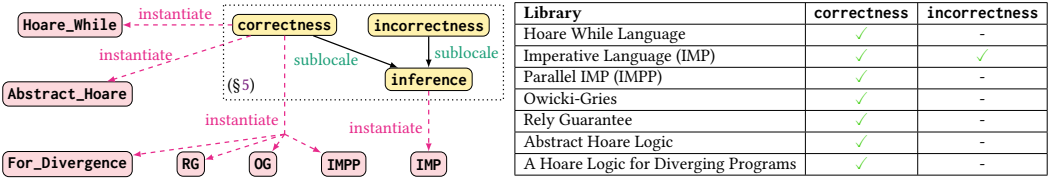

It is straightforward to show that these semantics are consistent
with our abstract theories of (in)correctness from \cref{sec:incorrectness}
by proving that they are an interpretation of our \InferenceLocale locale.
This 
\begin{enumerate*}[label=(\alph*)]
  \item shows that our locales are more general, and
  \item validates the \CorrectnessLocale and \IncorrectnessLocale locales.
\end{enumerate*}
An overview of these proofs can be found in \cref{tab:incorrectness-imp} and
as an illustrative example, we walk through the process for a basic imperative language, IMP.

To validate our $\InferenceLocale$ locale and demonstrate the
flexibility of the locale-based approach, we show that prior encodings
of Hoare and Incorrectness logic in Isabelle/HOL for the simple $IMP$
language (see below) can be derived as instances of
$\InferenceLocale$. $IMP$ is defined by the following well-known
syntax:
\begin{align*}
  IMP \ni h ::=&\ {\bf skip} \mid var := val \mid h_1 ; h_2 \mid \textbf{if}\ b\ \textbf{then}\ h_1\ \textbf{else}\ h_2 \mid  \textbf{while}\ b\ \textbf{do}\ h
\end{align*}
The HOL-IMP library provides a big-step transition relation
($\HolBigStep$) for $IMP$ as well as a Hoare logic semantics and
decomposition rules rules. \citet{DBLP:journals/afp/Murray20} extends
this to additionally cover Incorrectness logic (for $\HolBigStep$).
We instantiate the big-step transition relation ($\BigStep$) in
$\InferenceLocale$ as the transition relation for $IMP$ (i.e.,
$\HolBigStep$), allowing us to derive the inference rules described in
\cref{sec:incorrectness} for the $IMP$ language. In particular, we
prove that our triples $\HoareTriple{P}{c}{Q}$ and
$\OHearnTriple{P}{c}{Q}$ are equivalent to existing triples in 
HOL-IMP~\cite{DBLP:journals/afp/Murray20}.

\section{Further Architecture-Specific Proof Optimisations}
\label{sec:furth-arch-spec}

In addition to the step rules for the RISC-V instructions covered in \cref{sec:riscv-optimisations}, \isabil also provides rules for \RiscVsdNameOnly, \RiscVaddiNameOnly and \RiscVretNameOnly.
\begin{align*}
\RiscVsd = \InsnExpandedE{pc}{\WordE{2}{64}}{\{\StatementMoveE{rd}{\ExpressionLoadE{mem}{rs1 + \offset}{\EndianLittle}{64}}\}}
\end{align*}
The \RiscVsdNameOnly (Store Double) instruction stores a 64-bit word in memory. Similar to \RiscVldNameOnly, \RiscVsdNameOnly computes an address by adding the contents of the first source register $rs1$ to the immediate $\offset$. The value in the second source register $rs2$ is then stored at this address in memory. \RiscVsdNameOnly is the primary means of storing data of this size in memory.
\begin{align*}
\RiscVaddi = \InsnExpandedE{pc}{\WordE{2}{64}}{\{\StatementMoveE{rd}{rs1 + imm\}}}
\end{align*}
The \RiscVaddiNameOnly (Add Immediate) instruction is used to perform arithmetic addition. It adds the immediate value $imm$ to the source register $rs1$ and saves the result to the destination register $rd$. \RiscVaddiNameOnly is frequently used to calculate address offsets, as well as performing conventional addition.
\begin{align*}
\RiscVret = \InsnExpandedE{pc}{\WordE{2}{64}}{\{\StatementJmpE{ra}\}}
\end{align*}
The \RiscVretNameOnly (Return) instruction transfers control to the calling function. It achieves this by jumping to the program address stored in the return address register $ra$.

\begin{figure}[!t]
  \centering
  \begin{lstlisting}[language=riscv, caption={Stack allocation and deallocation for the {\tt good} function}, label={lst:stackdump}]
0000000000010554 <good>:
   10554:	1101                addi  sp,sp,-32
   10556:	ec06                sd    ra,24(sp)
   10558:	e822                sd    s0,16(sp)
   1055a:	1000                addi  s0,sp,32
   ...
   10582:	60e2                ld    ra,24(sp)
   10584:	6442                ld    s0,16(sp)
   10586:	6105                addi  sp,sp,32
   10588:	8082                ret
\end{lstlisting}
\end{figure}

Functions use stack memory to store arguments and local variables with allocation and deallocation of stack memory generally occurring at the start and end of the function respectively. In \cref{lst:stackdump}, we provide an example for the {\tt good} function in \cref{DF-good}. By using \isabil's definitions for the RISC-V instructions \RiscVaddiNameOnly, \RiscVsdNameOnly, \RiscVldNameOnly and \RiscVretNameOnly we can formulate comprehensive big-step lemmas for both allocation and deallocation, as shown below: 

\begin{mathpar}
\inferrule[(\StackAllocLemmaName)]{
  \BILDecodeInstructionE{\BILMachineStateE{\Variables_1}{pc}{mem}}{\RiscVaddiE{sp}{sp}{space}} \and
  \BILDecodeInstructionE{\BILMachineStateE{\Variables_2}{pc + 2}{mem}}{\RiscVsdE{ra}{space - 8}{sp}} \\\\
  \BILDecodeInstructionE{\BILMachineStateE{\Variables_3}{pc + 4}{mem}}{\RiscVsdE{s0}{space - 16}{sp}} \and
  \BILDecodeInstructionE{\BILMachineStateE{\Variables_4}{pc + 6}{mem}}{\RiscVaddiE{s0}{sp}{space}} \\\\
  \{pc, pc + 2, pc + 4, pc + 6\} \subseteq \ProgramDomain \\\\
  \Variables_2 = \Variables_1(sp \mapsto stack - space) \and
  \Variables_3 = \Variables_2(mem \mapsto v') \and
  \Variables_4 = \Variables_3(mem \mapsto v'') \\\\
  \Variables_5 = \Variables_4(s0 \mapsto stack) \\\\
  (\BILMachineStateE{\Variables_1}{pc}{mem}, \sigma_1) \BilSmallStep \sigma_2 \and
  (\BILMachineStateE{\Variables_2}{pc + 2}{mem}, \sigma_2) \BilSmallStep \sigma_3 \and
  (\BILMachineStateE{\Variables_3}{pc + 4}{mem}, \sigma_3) \BilSmallStep \sigma_4 \\\\
  (\BILMachineStateE{\Variables_4}{pc + 6}{mem}, \sigma_4) \BilSmallStep \sigma_5 \and
  \BILMachineStateE{\Variables_5}{pc + 8}{mem}, \sigma_5)\BILBigStep \tau \\\\
  \VarInE{mem}{v}{\Variables_1} \and \VarInE{sp}{stack}{\Variables_1} \and \VarInE{ra}{ret}{\Variables_1} \and \VarInE{r0}{frame}{\Variables_1}  \\\\
  v' = \ReflStorageEL{v}{stack - 8}{ret}{64} \and v'' = \ReflStorageEL{v'}{stack - 16}{frame}{64}
}{
  (\BILMachineStateE{\Variables_1}{pc}{mem}, \sigma_1)\BILBigStep \tau
}
\and
\inferrule[(\StackDeallocLemmaName)]{
  \BILDecodeInstructionE{\BILMachineStateE{\Variables_1}{pc}{mem}}{\RiscVldE{ra}{sp}{space - 8}} \and
  \BILDecodeInstructionE{\BILMachineStateE{\Variables_2}{pc + 2}{mem}}{\RiscVldE{s0}{sp}{space - 16}} \\\\
  \BILDecodeInstructionE{\BILMachineStateE{\Variables_3}{pc + 4}{mem}}{\RiscVaddiE{sp}{sp}{space}} \and
  \BILDecodeInstructionE{\BILMachineStateE{\Variables_4}{pc + 6}{mem}}{\RiscVret} \\\\
  \{pc, pc + 2, pc + 4, pc + 6\} \subseteq \ProgramDomain \\\\
  \Variables_2 = \Variables_1(ra \mapsto ret) \and
  \Variables_3 = \Variables_2(s0 \mapsto frame) \and
  \Variables_4 = \Variables_3(sp \mapsto stack + space) \\\\
  (\BILMachineStateE{\Variables_1}{pc}{mem}, \sigma_1) \BilSmallStep \sigma_2 \and
  (\BILMachineStateE{\Variables_2}{pc + 2}{mem}, \sigma_2) \BilSmallStep \sigma_3 \and
  (\BILMachineStateE{\Variables_3}{pc + 4}{mem}, \sigma_3) \BilSmallStep \sigma_4 \\\\
  (\BILMachineStateE{\Variables_4}{pc + 6}{mem}, \sigma_4) \BilSmallStep \sigma_5 \and
  \BILMachineStateE{\Variables_4}{ret}{mem}, \sigma_5)\BILBigStep \tau \\\\
  \VarInE{mem}{v}{\Variables_1} \and \VarInE{sp}{stack}{\Variables_1} \\\\
  \BILStepsE{\Variables}{\ExpressionLoadE{v}{stack + (space - 8))}{\EndianLittle}{64}}{ret} \and
  \BILStepsE{\Variables}{\ExpressionLoadE{v}{stack + (space - 16))}{\EndianLittle}{64}}{frame}
}{
  (\BILMachineStateE{\Variables_1}{pc}{mem}, \sigma_1)\BILBigStep \tau
}    
\end{mathpar}

Similar to the \PltStubLemmaName, we can also handle stack allocations and deallocations by utilising \StackAllocLemmaName and \StackDeallocLemmaName, respectively. This significantly reduces the proof burden, eliminating the need for approximately 50 rules in the case of allocation and 250 rules in the case of deallocation.

\section{Further details of the example proofs}
\label{sec:furth-deta-example}
\label{sec:proofs}

To motivate \isabil, we employ a combination of correctness and incorrectness 
proofs, focusing on illustrative examples frequently referenced in BAP 
literature \cite{DBLP:conf/cav/BrumleyJAS11}. These examples demonstrate 
representations of patterns commonly found within much larger programs/libraries.
We fix the target operating system to Ubuntu
22.04.2 LTS, the architecture to RISC-V and the compiler to the RISC-V
GNU toolchain 12.2.0 (g2ee5e430018) cross-compiler. To generate the
BIL files we use a pre-release of BAP (511b64c). 
The proofs for our first example, CWE-415: Double Free outlined
in \cref{sec:overview-and-motivation} are given in \cref{sec:example-proofs}.
Additionally, we verify a set of examples from the Joint Strike Fighter Air 
Vehicle (JSF AV) coding standards, described in \cref{tab:av-rules-appendix}. Details 
on the proofs are given in \cref{sec:forbidden-symbols}.

\subsection{Joint Strike Fighter (JSF) Air Vehicle (AV) Examples}
\label{sec:forbidden-symbols}

Our next set of examples are a subset of the JSF AV rules, which
prohibit the utilisation of specific external functions, collectively
referred to as {\em forbidden symbols}. We summarise these in
\autoref{tab:av-rules-appendix}, together with the corresponding forbidden
symbols. 
One could postulate that analysis of a binary's symbol table using tools
such as {\tt objdump} would be sufficient for identifying forbidden symbols.
However, the existence of a symbol in this table does not guarantee its usage, 
nor does it ensure it is invoked from a location reachable 
from a specific entry point. 
We have the full generality of Hoare and Incorrectness Logic providing 
stronger guarantees than allowed by static checking. 
Since these examples are about forbidden behaviours, we use
incorrectness logic in their verification. Proving incorrectness for
each of these examples requires the identification of such symbols in
the execution of a program. To this end, we define a new locale (see
\cref{fig:proofs}) that simplifies symbol tracking.

\paragraph{The \FindSymbolLocale locale}
We must first define the finding of symbols by instantiating the
small-step transition relation $\BILSmallStep$ of
\BilLocale (see \cref{sec:local-locale-1}). To identify
the presence of symbols in a program execution we require a history of
the accessed program addresses, which we denote
$\omega : \powerset(word)$.  This history is sufficient enough to
determine the presence of symbols in an execution. Consequently, we
can simply assume the state $\sigma$ to be this history $\omega$.  In
this context, a transition of the verification state $\omega$ can
occur with the step predicate $\FindSymbolSmallStep$ with the
following rule:
{\small%
\begin{mathpar}
\inferrule[(\sc next\_symbol)]{
  \Command = \BILMachineStateE{\_}{pc}{\_} \and
  \omega' = \omega \cup \{pc\}
}{
  (\Command , \omega)\FindSymbolSmallStep \omega'
}    
\end{mathpar}}

We use the predicate $\IsSymbol(str, \omega)$ to denote the occurrence
of symbol $str$ in the program address trace $\omega$. Recall from
\cref{sec:mechanisation} that the symbols in a binary are provided by
the $\SymbolTable$ of the locale that is auto-generated from for each
BIL program. We define
$\IsSymbol(str, \omega) = \SymbolTable(str) \in \omega$.  We encode
{\em finding symbols} in the locale $\FindSymbolLocale$, fixing the
parameter $\SymbolTable$, where we assume that $\SymbolTable$ will
later be instantiated by a program’s locale. This locale is derived as
an {\em interpretation} of $\BilInferenceLocale$, overriding
$\BilSmallStep\ =\ \FindSymbolSmallStep$ and leaving
$\BILDecodeTransition$ and $\ProgramDomain$ abstract. Therefore, we
define this new locale with the following signature:
\[
\FindSymbolLocale = (\BILDecodeTransition, \ProgramDomain, \SymbolTable)
\]

\begin{table}[t]
\begin{tabular}{|p{0.1\columnwidth}|p{0.8\columnwidth}|}
  \hline
  \textbf{AV Rule} & \textbf{Description}                                                                                                \\ \hline
  17               & The error indicator errno shall not be used                                                                        \\ \hline
  19               & {\tt \textless{}locale.h\textgreater} and the setlocale function shall not be used$^{*}$                                        \\ \hline
  20               & The {\tt setjmp} macro and the {\tt longjmp} functions shall not be used                                                         \\ \hline
  21               & The signal handling facilities of {\tt \textless{}signal.h\textgreater} shall not be used                                 \\ \hline
  23               & The library functions {\tt atof}, {\tt atoi} and {\tt atol} from library {\tt \textless{}stdlib.h\textgreater} shall not be used            \\ \hline
  24               & The library functions {\tt abort}, {\tt exit}, {\tt getenv} and system from library {\tt \textless{}stdlib.h\textgreater} shall not be used \\ \hline
  25               & The time handling functions of library {\tt \textless{}time.h\textgreater} shall not be used                              \\ \hline
\end{tabular} \smallskip

\raggedright {\small * Of course {\tt \textless{}locale.h\textgreater}
  refers to the {\tt C} locale library and not the Isabelle locales
  used by \isabil.}
\caption{AV rules with description}
\label{tab:av-rules-appendix}
\vspace{-10pt}
\end{table}

\paragraph{The \AvRuleLocale locale}
We refer to the AV Rule locales collectively using
\AvRuleLocale, where \Locale{*} represents the rule number.  For each
of the examples in \autoref{tab:av-rules-appendix} we auto-generate a
corresponding locale using the Isabelle command:
\begin{center}
\lstinline[language=isabellen]{BIL_file Av_Rule_* "av-rule-*.bil.adt"}
\end{center}
where {\tt av-rule-*.bil.adt} refers to a file 
for
each example in \autoref{tab:av-rules-appendix} in \BILa format.

\paragraph{The \AvRuleProofLocale locale}
To join each $\AvRuleLocale$ locales with $\FindSymbolLocale$, we
create new {\em proof} locales ($\AvRuleProofLocale$), which inherit
their respective $\AvRuleLocale$ locale. We derive this locale as an
{\em interpretation} of $\FindSymbolLocale$, setting its parameters
$\BILDecodeTransition$, $\ProgramDomain$ and $\SymbolTable$ to those
provided by the auto-generated $\AvRuleLocale$.

It is straightforward to define predicates that capture rule
violations in the AV rule examples (\autoref{tab:av-rules-appendix}). For
example, Rule 23 prohibits the use of {\tt atof}, {\tt atoi} and {\tt
  atol}. As such, we must derive predicates specific to a given AV
Rule from \IsSymbol. An example of this predicate for Rule 23 is:
\[
{\tt violation\_av\_23}(\omega) = \IsSymbol(\text{``{\tt atof}''}, \omega) \vee \IsSymbol(\text{``{\tt atoi}''}, \omega) \vee \IsSymbol(\text{``{\tt atol}''}, \omega)
\]

Recall from \cref{sec:incorrectness} that proofs of incorrectness 
require a non-triviality condition on the post-state. Proving 
non-triviality is often only possible using intermediate post-state 
predicates. Using this fact, we can formulate a proof of incorrectness
for forbidden symbols given in \cref{thy:forbidden-symbol}.

\begin{theorem} \label{thy:forbidden-symbol}
  For $i \in \{17,19,20,21,23,24,25\}$, given that $\Command_i$ is
  program corresponding to\\ {\tt av\_rule\_i}, we have:
  \begin{mathpar}
    \IncorrectnessProof{\neg{\tt violation\_av\_i}}{\ \Command_{i}\ }{Q}{{\tt
        violation\_av\_i}}
  \end{mathpar}
\end{theorem}
The proof for \cref{thy:forbidden-symbol} necessitates the 
repeated application of O'Hearn step lemmas described in \cref{sec:local-locale-1} 
until the program reaches termination. 
The application of O'Hearn step lemmas is partly automated, 
although human intervention is often necessary to ascertain
the post-state of a program step. Once this post state has 
been determined, the deconstruction of the BIL program step 
leading to this post state is automated for the majority of the
AV Rules. The optimizations described in 
\cref{sec:riscv-optimisations}, which enable efficient 
handling of RISC-V instructions and external symbol calls, 
require less intervention. The identification of 
these cases is manual, though proof discharge is automatic. 
Upon termination, we examine the post-state to determine if 
a forbidden symbol is present and that the post-state is 
non-trivial. Both of these conditions can be automatically 
verified for the AV Rule examples using the built-in solvers 
of Isabelle/HOL. The size of each proof, given in \cref{table:proof-stats},
is determined by the length of the execution trace, not the size of the binary.
The largest proof, associated with AV Rule 25, consists of 
approximately 950 lines and takes 4 minutes to verify in Isabelle/HOL, 
while the average size for the examples is around 363 lines.
The proofs show that for all AV Rules in \cref{tab:av-rules-appendix}, the 
corresponding forbidden symbols occur, rendering the examples incorrect.
Even without source code, identifying the external symbols present in a binary is a straightforward task using tools available in GNU \cite{stallman2003using, Hemel:2021}. However, it should be noted that the presence of these symbols does not necessarily imply their actual usage. The symbols can reside in impossible control paths or functions that are never called. This proof guarantees the utilization of such forbidden symbols in a valid execution, not just their presence within the binary.

\section{Syntax abbreviations for x86 registers}
\label{appendix:x86}
The representation of x86 registers in BIL is necessarily verbose. 
For example, the x86 register {\tt RAX} is represented by the 
variable $(\text{"rax"} : \TypeImmediateE{64})$. 
It is common knowledge for x86 developers that RAX is a
register capable of storing a 64-bit word.
Consequently, we define the set of
registers for \XESSF~\cite{intel-x86-manual} 
in \cref{tab:x86-registers}.
This approach encompasses the general purpose registers for 64-bit, 128-bit and 256-bit 
as \BIRRegisterSixFourE{8-15}, \BIRRegisterOneTwoEightE{0-7}, \BIRRegisterTwoFiveSixE{0-7}, and \BIRRegisterFiveOneTwoE{0-7} respectively,
as well as the state of the processor, usually held in the EFLAGS
register, represented by single-bit (boolean) variables 
\BIRCarryFlag, \BIROverflowFlag, \BIRAdjustFlag, \BIRParryFlag,
\BIRSizeFlag, and \BIRZeroFlag.

\begin{table}
  \begin{minipage}{0.49\textwidth}
  \centering
\begin{tabular}{|c|c|c|}
\hline
  Definition & <name>  & <type>  \\ \hline
  \BIRAccumulatorRegisterSixFour  & rax & $\TypeImmediateE{64}$ \\ \hline
  \BIRBaseRegisterSixFour         & rbx & $\TypeImmediateE{64}$ \\ \hline
  \BIRCounterRegisterSixFour      & rcx & $\TypeImmediateE{64}$ \\ \hline
  \BIRDataRegisterSixFour         & rdx & $\TypeImmediateE{64}$ \\ \hline
  \BIRStackPointerRegisterSixFour & rsp & $\TypeImmediateE{64}$ \\ \hline
  \BIRStackBaseRegisterSixFour    & rbp & $\TypeImmediateE{64}$ \\ \hline
  \BIRSourceRegisterSixFour       & rsi & $\TypeImmediateE{64}$ \\ \hline
  \BIRDestinationRegisterSixFour  & rdi & $\TypeImmediateE{64}$ \\ \hline
  \BIRRegisterSixFourE{8-15}      & r8-15 & $\TypeImmediateE{64}$ \\ \hline
\end{tabular}
  \end{minipage}
  \hfill
  \begin{minipage}{0.49\textwidth}
  \centering
\begin{tabular}{|c|c|c|}
\hline
  Definition                    & <name> & <type>  \\ \hline
  \BIRRegisterOneTwoEightE{0-7} & xmm0-7 & $\TypeImmediateE{128}$ \\ \hline
  \BIRRegisterTwoFiveSixE{0-7}  & ymm0-7 & $\TypeImmediateE{256}$ \\ \hline
  \BIRRegisterFiveOneTwoE{0-7}  & zmm0-7 & $\TypeImmediateE{512}$ \\ \hline
  \BIRCarryFlag    & cf  & $\TypeImmediateE{1}$ \\ \hline 
  \BIROverflowFlag & of  & $\TypeImmediateE{1}$ \\ \hline
  \BIRAdjustFlag   & af  & $\TypeImmediateE{1}$ \\ \hline
  \BIRParryFlag    & pf  & $\TypeImmediateE{1}$ \\ \hline
  \BIRSizeFlag     & sf  & $\TypeImmediateE{1}$ \\ \hline
  \BIRZeroFlag     & zf  & $\TypeImmediateE{1}$ \\ \hline
\end{tabular}
  \end{minipage}

  \caption{Definitions for the \XESSF\xspace registers as BIL variables where $(\text{"<name>"} : \text{<type>})$. }
  \label{tab:x86-registers}
\end{table}

\section{Further detail on the Eisbach Automated Solvers}
\label{sec:eisbach}

The Eisbach methods for symbolic execution (\SymbolicSolver)
and type checking (\TypeSolver) in \IsaBIL both facilitate extensions 
through the use of parameterised proof methods,
allowing a new Eisbach method to be \emph{installed} in the solver.
The parameterised proof method is placed at the highest precedence in the solver (such that they run first).
If the parameterised proof method fails, then the standard solver will run.
It is through these proof methods that optimisations for 16, 32 and 64-bit memory 
(see~\cref{sec:big-step-semantics}) 
as well as x86 and \RiscV (see~\cref{sec:automation}) are installed
in the \SymbolicSolver solver.

\paragraph{Symbolic Execution}
Symbolic execution refers to the repeated application of 
rules in BIL's operational semantics to advance the program
state.
We provide an abridged example for introduction rules of sequences and statements 
of BIL below: 
\begin{lstlisting}[language=isabelle,escapechar=']
method sexc_bilI methods extensions = (
  extensions | '\label{line:sexc-exts}'
  (rule step_stmt_cpuexnI) | (rule step_stmt_specialI) | '\label{line:sexc-cpuexn}'

  (rule step_stmt_moveI.word, (sexc_expsI; fail)) | '\label{line:sexc-move-word}'
  (* other syntax-specific rules... *)
  (rule step_stmt_moveI, sexc_expsI?) | '\label{line:sexc-move}'

  (* syntax specific rules... *)
  (rule step_stmt_jmpI, type_contextI?) | '\label{line:sexc-jmp}'

  (rule step_stmt_if_trueI, sexc_expsI, sexc_bilI?) | '\label{line:sexc-if-true}'
  (rule step_stmt_if_falseI, sexc_expsI, sexc_bilI?) | '\label{line:sexc-if-false}'

  (rule step_stmt_while_falseI, sexc_expsI) | '\label{line:while-false}'
  (rule step_stmt_whileI, sexc_expsI, sexc_bilI?) | '\label{line:sexc-while-true}'

  (rule step_bil_emptyI) | '\label{line:sexc-empty}'
  (rule step_bil_empty_eqI; (simp; fail)) | '\label{line:sexc-empty-eq}'
  (rule step_bil_singleI, sexc_bilI?) | '\label{line:sexc-single}'
  (rule step_bil_seqI, sexc_bilI?, sexc_bilI?)'\label{line:sexc-seq}'
)
\end{lstlisting}
The Eisbach method \SymbolicSolverBILI attempts to deconstruct a
sequence of BIL statements using introduction rules for statements and 
sequences. 
Note that BIL statements and sequences are mutually recursive, therefore
a method that deconstructs statements ($stmt$) must also deconstruct
sequences ($bil$).
The deconstruction process involves several steps. 
First, \SymbolicSolverBILI attempts to apply any extension methods on \cref{line:sexc-exts} 
using the parameterised proof method \code{extensions}.
If this fails the method proceeds to the next case, denoted by the pipe ``$\mid$'' operator.

If the statement is a \StatementCpuExnName or \StatementSpecialName then it is trivially discharged on \cref{line:sexc-cpuexn}
by applying the \code{step\_stmt\_cpuexnI} or \code{step\_stmt\_specialI} rules.
Those familiar with Isabelle/HOL may question why the \code{intros} method does not 
discharge both goals in one.
Unfortunately, this method is unable to retain \emph{schematic variables} of the form
\code{?x} which serve as a placeholder for terms that are yet to be determined, and are 
an important facet of exploratory proofs.
the lack of support for \emph{schematic variables} also governs the choice not to use 
Eisbach's expressive pattern-matching features.

The next case on \cref{line:sexc-move-word} tackles the move statement ($\StatementMove$), 
specifically in the case where $exp$ evaluates to a word $(\Word)$ by attempting to apply 
\code{step\_stmt\_moveI.word}, if successful, this generates a further
deconstruction obligation for the expression $exp$, which is discharged 
with \code{sexc\_expsI} using the ``,'' operator.
Note that if the application of \code{sexc\_expsI} generates  
any unsolved goals, indicated by the ``;'' operator, then the \code{fail} proof method
will fire, causing the method to proceed to the next case.
This is a form of backtracking that prevents the automated solver from reaching an 
invalid proof state where $exp$ does not evaluate to a word.
What would usually follow in the next lines are other syntax-specific rules,
these have been omitted in this example to maintain clarity.
\cref{line:sexc-move} defines a \emph{last resort} case for the move statement in 
which $v$ is interpreted as a \emph{catch-all} value and not any particular syntax (i.e., word, storage). 
Applying this case will never lead to an invalid proof state, 
though may require manual intervention to interpret the values type correctly. 
If this is the case then the unsolved goals generated by \code{sexc\_expsI} will be handed 
back to the human prover using the \code{succeed} method.
Cases for the \StatementJmpName statement proceed the same as the move statement.

If statements of the form $\StatementIf$ introduce different control flows
depending upon the evaluation of $e$, if true then the rule 
\code{step\_stmt\_if\_trueI} should be applied, otherwise
\code{step\_stmt\_if\_falseI}.
Both rules generate a subgoal requiring the evaluation of $e$
to be proven as either true or false.
Whilst the method can infer that the statement is an $\StatementIfName$
statically, this cannot be inferred for the evaluation of $e$, and thus must be 
discharged with the \code{sexc\_expsI}.
First, \cref{line:sexc-if-true} attempts this for true case, if the first 
subgoal cannot be discharged then we know that either the evaluation is not 
true or the rule \code{step\_stmt\_if\_trueI} cannot safely be applied. 
Thus, we backtrack and attempt the next case for false on \cref{line:sexc-if-false}.
If this also fails then the remaining proof methods will be attempted before 
the context is handed back to the human prover.
If either case's $e$ is provably true or false then this obligation is discharged
and $seq_1$ or $seq_2$ is discharged with \SymbolicSolverBILI.
Control flow is also present in $\StatementWhileName$ statements, albeit with a simpler
false case.

If we are executing a sequence instead of a statement,
then the first cases on \cref{line:sexc-empty} will attempt to
match an empty sequence by applying \code{step\_bil\_emptyI}.  
Occasionally, the sequence may be empty, but some additional
simplification must occur on the variable state and program
counter to match the post state. 
In this case the method on line \cref{line:sexc-empty-eq} will
match empty sequences with the \code{step\_bil\_empty\_eqI} rule, 
then attempt to discharge any goals it spits out with the \code{simp}
method.
The simplifier is destructive and may unintentionally leave the proof 
in an unsafe state. As such, if the simplifier cannot discharge its 
goal, then we backtrack to a state before the simplifier was applied
using \code{fail} and force the human prover to solve a safe goal. 

If the sequence is not empty, and instead contains a single element 
then on \cref{line:sexc-single} the rule \code{step\_bil\_singleI} 
is applied, 
followed by \code{sexc\_bilI}, keeping the solvers progress.

If the sequence contains more than one element then the next case
on \cref{line:sexc-seq} attempts to apply \code{step\_bil\_seqI}, 
if successful this generates two
further deconstruction obligations, one for the statement and one for 
the recursive sequence, for both of which we recursively apply \SymbolicSolverBILI. 
  
Lastly, if none of these cases match, the proof method fails without
making any progress, returning control to the user (i.e., human prover).
The user can then assess if the proof is impossible with the current 
set of assumptions, or if manual intervention is required.

Most failures or backtracking occur due to issues with Isabelle's math solvers
being intractable (especially the case for modulo operations common in bitwise
arithmetic),
or are the result of introduced syntax such as definitions which must be unfolded.
A manual fix for one of these problems may be fed back into the automated solver
using the \code{extensions} proof method so that this problem does not cause issues 
in future.

Additionally, the inherent non-determinism in the operational semantics 
may cause the automated solver to fail.
Failures like this are harder to automate, but generally indicate errors in the 
program's specification or initial variables, which would warrant manual 
intervention.

\paragraph{Type Checking}
BIL is a typed language, in our proofs, we are often
required to verify type correctness of our instructions 
in order to execute the operational semantics.
Type checking is a static process and does not require symbolic execution to infer type correctness. 
Typing rules (see~\cref{sec:specification-of-bil}) ensure that BIL code is well-formed, which guarantees that it is executable under the operational semantics.
We apply Eisbach proof tactics to perform type checking,
without this, we would otherwise be left with repetitive proof goals we must prove manually. 
Therefore, \TypeSolver is generally applied to all the instructions 
before any symbolic execution in a 
given program locale to ensure that they are executable.
This, in turn, allows us to verify that the BIL generated by BAP is in fact 
correct according to its operational semantics.
Through this method, we found a bug
for the \code{C\_LUI} \RiscV instruction\footnote{\url{https://github.com/BinaryAnalysisPlatform/bap/pull/1588}}.
The equivalent \TypeSolver method for \TypeSolverBILI is given below: 
\begin{lstlisting}[language=isabelle]
method typec_bilI methods extensions = (
  extensions |
  (rule type_stmt_cpuexnI, typec_contextI?) | 
  (rule type_stmt_specialI, typec_contextI?) |
  (rule type_stmt_moveI, (typec_varI, typec_expI)?) |
  (rule type_stmt_jmpI, typec_expI?) |
  (rule type_stmt_ifI, (typec_expI, typec_bilI, typec_bilI)?) |
  (rule type_stmt_whileI, (typec_expI, typec_bilI)?) |

  (rule type_bil_emptyI) | 
  (rule type_bil_singleI, typec_bilI?) | 
  (rule type_bil_seqI, (typec_bilI, typec_bilI)?)
)
\end{lstlisting}
The \TypeSolverBILI method is structured similarly to \SymbolicSolverBILI, 
albeit with fewer cases.
Typing covers both true/false cases for \StatementWhileName and \StatementIfName 
statements, and does not need to consider additional syntax at this level.
Type checking has a much faster proof time than 
symbolic execution owing to its much smaller ruleset and the
fact that it is deterministic.
Therefore, many of the backtracking and failed proof attempts present in 
\SymbolicSolver are not present in \TypeSolver.
However, it may still run into issues with unknown syntax/definitions
or intractable math problems.
These are solved by the same means as \SymbolicSolverBILI.

\section{Additional detail on the structure of the Locales}
\label{sec:locale-summary}

\begin{figure}
\centering
\begin{minipage}{0.39\textwidth}
  \vspace{17ex}
  \resizebox{\textwidth}{!}{\begin{tikzpicture}
 
  \node[parambox] (inference) {\InferenceLocale(};
  \node[parambox, right = 0ex of inference] (inference1) {$\BigStep$};
  \node[parambox, right = 0ex of inference1] (inference_End) {)};
  
  \node[parambox,above left = 7ex and 0ex of inference] (correctness) {\CorrectnessLocale(};
  \node[parambox,right = 0ex of correctness] (correctness1) {$\BigStep$};
  \node[parambox,right = 0ex of correctness1] (correctness_End) {)};
  
  \node[parambox,right = 2ex of correctness_End] (incorrectness) {
     \IncorrectnessLocale(};
  \node[parambox,right = 0ex of incorrectness] (incorrectness1) {
     $\BigStep$};
  \node[parambox,right = 0ex of incorrectness1] (incorrectness_End) {
     )};

  \node[parambox,below left = 4ex and -4ex of inference] (BIL) {
     \BilLocale(};
  \node[parambox,right = 0ex of BIL] (BIL1) {
     $\Decode$};
  \node[parambox,right = 0ex of BIL1] (BIL_End) {
     )};

  \node[parambox,right = 4ex of BIL_End] (BIL_BigStep) {
     $\BILBigStep$ };
  \node[parambox,right = 1ex of BIL_BigStep]  {
     (see \cref{sec:big-step-semantics}) };

  \node[parambox,below left = 8ex and 10ex of BIL_BigStep] (BIL_inf) {
     \BilInferenceLocale(};
  \node[parambox,right = 0ex of BIL_inf] (BIL_inf1) {
     $\Decode$,};
  \node[parambox,right = 1ex of BIL_inf1] (BIL_inf2) {
     \ProgramDomain,};
  \node[parambox,right = 1ex of BIL_inf2] (BIL_inf3) {
     $\BILSmallStep$};
  \node[parambox,right = 0ex of BIL_inf3] (BIL_inf_End) {
     )};

  \path[draw=gray,text=gray,->] (inference1) to node[below,sloped]{\small inherit} (correctness1);
  \path[draw=gray,text=gray,->] (inference1) to node[below,sloped]{\small inherit} (incorrectness1);
  \path[draw=gray,text=gray,->] (BIL_inf1) to node[above,sloped]{\small inherit} (BIL1);
  \path[draw=gray,text=gray,dashed,->] (BIL_inf1)
    to ($(BIL_BigStep)-(0ex,6ex)$)
    to node[right]{uses} (BIL_BigStep);
  \path[draw=gray,text=gray,dashed,-] (BIL_inf2) to ($(BIL_BigStep)-(0ex,6ex)$);
  \path[draw=gray,text=gray,dashed,-] (BIL_inf3) to ($(BIL_BigStep)-(0ex,6ex)$);  
  \path[draw=gray,text=gray,->] (BIL_BigStep) to node[right]{\small override} (inference1); 
      
\end{tikzpicture}}
  \caption{Instantiation of \BilInferenceLocale.}
  \label{fig:bil-inference-params}
\end{minipage}
\hfill
\begin{minipage}{0.6\textwidth}
  \resizebox{\textwidth}{!}{\begin{tikzpicture}

  \node (BIL_inf) {
     \BilInferenceLocale(};
  \node[parambox, right = 0ex of BIL_inf] (BIL_inf1) {
     $\Decode$,};
  \node[parambox, right = 1ex of BIL_inf1] (BIL_inf2) {
     \ProgramDomain,};
  \node[parambox, right = 1ex of BIL_inf2] (BIL_inf3) {
     $\BILSmallStep$};
  \node[parambox, right = 0ex of BIL_inf3] (BIL_inf_End) {
     )};

  \node[parambox,below = 4ex of BIL_inf3] (AllocStep) {
     $\ReallocSmallStep$};
  \node[parambox,right = 1ex of AllocStep]  {
     (see \cref{fig:reallocation-steps}) };

  \node[parambox,below left = 12ex and 8ex of BIL_inf] (allocation) {
     \ReallocLocale(};
  \node[parambox,right = 0ex of allocation] (allocation1) {
     $\Decode$,};
  \node[parambox,right = 1ex of allocation1] (allocation2) {
     \ProgramDomain,};
  \node[parambox,right = 1ex of allocation2] (allocation3) {
     \NextAddrAllocatorName,};
  \node[parambox,right = 1ex of allocation3] (allocation4) {
     \GetFreedAddrName,};
  \node[parambox,right = 1ex of allocation4] (allocation5) {
     \GetSzName,};
  \node[parambox,right = 1ex of allocation5] (allocation6) {
     \FreePredicateName,};
  \node[parambox,right = 1ex of allocation6] (allocation7) {
     \AllocPredicateName,};
  \node[parambox,right = 1ex of allocation7] (allocation8) {
     \ReallocPredicate};
  \node[parambox,right = 0ex of allocation8] (allocation_End) {
     )};

  \node[parambox, below = 3ex of allocation6, xshift=-3ex] (Free) {\SymbolTable[free]};
  \node[parambox, below = 7ex of allocation7, xshift=-1ex] (Alloc) {\SymbolTable[malloc]};
  \node[parambox, below = 11ex of allocation8, xshift=-3ex] (Realloc) {\SymbolTable[realloc]};

  \node[parambox, below = 3ex of allocation4] (X10) {$\Variables$({\tt X10})};

  \node[parambox,below right = 20ex and 8ex of allocation] (DF_proofs) {\ReadDataBadLocale(};
  \node[parambox,right = 0ex of DF_proofs] (DF_proofs1) {$\Decode$,};
  \node[parambox,right = 1ex of DF_proofs1] (DF_proofs2) {\NextAddrAllocatorName,};
  \node[parambox,right = 1ex of DF_proofs2] (DF_proofs3) {...};
  \node[parambox,right = 0ex of DF_proofs3] (DF_proofsEnd) {)};
  \node[text=gray,parambox,right = 1ex of DF_proofsEnd] (DF_proofs_Defs) {
    [};
  \node[text=gray,parambox,right = 0ex of DF_proofs_Defs] (DF_proofs_Defs1) {
    $\ProgramDomain$,};
  \node[text=gray,parambox,right = 1ex of DF_proofs_Defs1] (DF_proofs_Defs2) {
    $\SymbolTable$};
  \node[text=gray,parambox,right = 0ex of DF_proofs_Defs2] (DF_proofs_Defs_End) {
    ]};

  \node[parambox,below left = 6ex and 12ex of DF_proofs] (DF) {
     \LibCurlLocale(};
  \node[parambox,right = 0ex of DF] (DF1) {
     $\Decode$};
  \node[parambox,right = 0ex of DF1] (DFEnd) {)};
  \node[text=gray,parambox,right = 1ex of DFEnd] (DF_Defs) {
    [};
  \node[text=gray,parambox,right = 0ex of DF_Defs] (DF_Defs1) {
    $\ProgramDomain$,};
  \node[text=gray,parambox,right = 1ex of DF_Defs1] (DF_Defs2) {
    $\SymbolTable$};
  \node[text=gray,parambox,right = 0ex of DF_Defs2] (DF_Defs_End) {
    ]};

  \node[parambox,below = 12ex of DF_proofs1] (SR1) {$\Decode$,};
  \node[parambox,left = 0ex of SR1] (SR) {
     \SecRecvBadLocale(};
  \node[parambox,right = 1ex of SR1] (SR2) {\NextAddrAllocatorName,};
  \node[parambox,right = 1ex of SR2] (SR3) {...};
  \node[parambox,right = 0ex of SR3] (SR_End) {)};
  \node[text=gray,parambox,right = 1ex of SR_End] (SR_Defs) {
    [}; 
  \node[text=gray,parambox,right = 0ex of SR_Defs] (SR_Defs1) {
    $\ProgramDomain$,};
  \node[text=gray,parambox,right = 1ex of SR_Defs1] (SR_Defs2) {
    $\SymbolTable$};
  \node[text=gray,parambox,right = 0ex of SR_Defs2] (SR_Defs_End) {
    ]};

  \node[parambox,below = 5ex of SR1] (CM1) {$\Decode$,};
  \node[parambox,left = 0ex of CM1] (CM) {
     \ChooseMechBadLocale(};
  \node[parambox,right = 1ex of CM1] (CM2) {\NextAddrAllocatorName,};
  \node[parambox,right = 1ex of CM2] (CM3) {...};
  \node[parambox,right = 0ex of CM3] (CM_End) {)};     
  \node[text=gray,parambox,right = 1ex of CM_End] (CM_Defs) {
    [}; 
  \node[text=gray,parambox,right = 0ex of CM_Defs] (CM_Defs1) {
    $\ProgramDomain$,};
  \node[text=gray,parambox,right = 1ex of CM_Defs1] (CM_Defs2) {
    $\SymbolTable$};
  \node[text=gray,parambox,right = 0ex of CM_Defs2] (CM_Defs_End) {
    ]};

  \node[parambox,below = 5ex of CM1] (CSL1) {$\Decode$,};
  \node[parambox,left = 0ex of CSL1] (CSL) {
     \CurlSecLoginBadLocale(};
  \node[parambox,right = 1ex of CSL1] (CSL2) {\NextAddrAllocatorName,};
  \node[parambox,right = 1ex of CSL2] (CSL3) {...};
  \node[parambox,right = 0ex of CSL3] (CSL_End) {)};     
  \node[text=gray,parambox,right = 1ex of CSL_End] (CSL_Defs) {
    [}; 
  \node[text=gray,parambox,right = 0ex of CSL_Defs] (CSL_Defs1) {
    $\ProgramDomain$,};
  \node[text=gray,parambox,right = 1ex of CSL_Defs1] (CSL_Defs2) {
    $\SymbolTable$};
  \node[text=gray,parambox,right = 0ex of CSL_Defs2] (CSL_Defs_End) {
    ]};

  \path[draw=gray,text=gray,->] (allocation1) to node[sloped,above]{\small inherit} (BIL_inf1);
  \path[draw=gray,text=gray,->] (allocation2) to node[sloped,above]{\small inherit} (BIL_inf2);
  \path[draw=gray,text=gray,dashed,->] (allocation3) 
    to ($(AllocStep)-(0ex,4ex)$)
    to node[right]{\small uses} (AllocStep);
  \path[draw=gray,text=gray,dashed,-] (allocation4) to ($(AllocStep)-(0ex,4ex)$);
  \path[draw=gray,text=gray,dashed,-] (allocation5) to ($(AllocStep)-(0ex,4ex)$);
  \path[draw=gray,text=gray,dashed,-] (allocation6) to ($(AllocStep)-(0ex,4ex)$);
  \path[draw=gray,text=gray,dashed,-] (allocation7) to ($(AllocStep)-(0ex,4ex)$);
  \path[draw=gray,text=gray,dashed,-] (allocation8) to ($(AllocStep)-(0ex,4ex)$);
  \path[draw=gray,text=gray,->] (AllocStep) to node[right]{\small override} (BIL_inf3);
  \path[draw=gray,text=gray,->] (DF_proofs1) to node[sloped,below]{\small inherit} (allocation1);
  \path[draw=gray,text=gray,->] (DF_proofs1.south) to node[above, sloped]{\small inherit} (DF1.north);

  \path[draw=gray,text=gray,->] (DF_proofs2) to node[below, near start, sloped]{\small inherit} (allocation3.south);
  
  \path[draw=gray,text=gray,dashed,->] (DF_Defs1) to node {} ($(DF_proofs_Defs1.south)-(1ex,0)$);
  \path[draw=gray,text=gray,dashed,->] (DF_Defs2) to node[left,xshift=-1ex]{} ($(DF_proofs_Defs2.south)-(1ex,0)$);

  \path[draw=gray,text=gray,near start,->] (DF_proofs_Defs1) to node[below, sloped]{\small override} (allocation2.south);
  \path[draw=gray,text=gray,dashed,->] (DF_proofs_Defs2) to node[below, sloped]{\small uses} (Alloc.south);
  \path[draw=gray,text=gray,dashed,->] ($(DF_proofs_Defs2.north west)+(2ex,0)$) to node[below, sloped]{\small uses} (Free.south);
  \path[draw=gray,text=gray,dashed,->] (DF_proofs_Defs2) to node[below, sloped]{\small uses} (Realloc);

  \path[draw=gray,text=gray,->] (Free) to node[left]{\small override} (allocation6);
  \path[draw=gray,text=gray,->] (Alloc) to node[right]{\small override} (allocation7);
  \path[draw=gray,text=gray,->] (Realloc) to node[right]{\small override} (allocation8);

  \path[draw=gray,text=gray,->] (X10) to node[left]{\small override} (allocation4);
  \path[draw=gray,text=gray,->] (X10) to node{} (allocation5);

  \path[draw=gray,text=gray,<-] (SR1) to node[near start,right] {inherit} (DF_proofs1);
  \path[draw=gray,text=gray,<-] (SR2) to node {} (DF_proofs2);
  \path[draw=gray,text=gray,<-] (SR3) to node {} (DF_proofs3);
  \path[draw=gray,text=gray,dashed,<-] (SR_Defs1) to node {} (DF_proofs_Defs1);
  \path[draw=gray,text=gray,dashed,<-] (SR_Defs2) to node[near start,left,xshift=-1.5ex]{\small provide} (DF_proofs_Defs2);
  
  \path[draw=gray,text=gray,<-] (CSL1) to node[right] {inherit} (CM1);
  \path[draw=gray,text=gray,<-] (CSL2) to node {} (CM2);
  \path[draw=gray,text=gray,<-] (CSL3) to node {} (CM3);
  \path[draw=gray,text=gray,dashed,<-] (CSL_Defs1) to node {} (CM_Defs1);
  \path[draw=gray,text=gray,dashed,<-] (CSL_Defs2) to node[left,xshift=-1.5ex]{\small provide} (CM_Defs2);

  \path[draw=gray,text=gray,<-] (CM1) to node[right] {inherit} (SR1);
  \path[draw=gray,text=gray,<-] (CM2) to node {} (SR2);
  \path[draw=gray,text=gray,<-] (CM3) to node {} (SR3);  
  \path[draw=gray,text=gray,dashed,<-] (CM_Defs1) to node {} (SR_Defs1);
  \path[draw=gray,text=gray,dashed,<-] (CM_Defs2) to node[left,xshift=-1.5ex]{\small provide} (SR_Defs2);
      
\end{tikzpicture}}
  \caption{Locale instantiation of the \ReallocLocale examples.}
  \label{fig:realloc-params}
\end{minipage}

\begin{minipage}{0.44\textwidth}
  \vspace{0.2ex}
  \resizebox{\textwidth}{!}{\begin{tikzpicture}

  \node[parambox] (BIL_inf) {
     \BilInferenceLocale(};
  \node[parambox,right = 0ex of BIL_inf] (BIL_inf1) {
     $\Decode$,};
  \node[parambox,right = 1ex of BIL_inf1] (BIL_inf2) {
     \ProgramDomain,};
  \node[parambox,right = 1ex of BIL_inf2] (BIL_inf3) {
     $\BILSmallStep$};
  \node[parambox,right = 0ex of BIL_inf3] (BIL_inf_End) {
     )};

  \node[parambox,below = 3ex of BIL_inf3] (FindSymbolStep) {
     $\FindSymbolSmallStep$ };
  \node[parambox,right = 1ex of FindSymbolStep]  {
     (see \cref{sec:forbidden-symbols}) };

  \node[parambox,below left = 3ex and 20ex of FindSymbolStep] (find_symbol) {
    \FindSymbolLocale(};
  \node[parambox,right = 0ex of find_symbol] (find_symbol1) {
     $\Decode$,};
  \node[parambox,right = 1ex of find_symbol1] (find_symbol2) {
     $\ProgramDomain$,};
  \node[parambox,right = 1ex of find_symbol2] (find_symbol3) {
     $\SymbolTable$};
  \node[parambox,right = 0ex of find_symbol3] (find_symbol_End) {
    )};

  \node[parambox,below left = 8ex and -19ex of find_symbol] (AV_proofs) {
     \AvRuleProofLocale(};
  \node[parambox,right = 0ex of AV_proofs] (AV_proofs1) {
     $\Decode$,};
  \node[parambox,right = 1ex of AV_proofs1] (AV_proofs2) {
     ...};
  \node[parambox,right = 0ex of AV_proofs2] (AV_proofs_End) {
    )};
  \node[text=gray,parambox,right = 1ex of AV_proofs_End] (AV_proofs_Defs) {
    [};
  \node[text=gray,parambox,right = 0ex of AV_proofs_Defs] (AV_proofs_Defs1) {
    $\ProgramDomain$,};
  \node[text=gray,parambox,right = 1ex of AV_proofs_Defs1] (AV_proofs_Defs2) {
    $\SymbolTable$};
  \node[text=gray,parambox,right = 0ex of AV_proofs_Defs2] (AV_proofs_Defs_End) {
    ]};

  \node[parambox,below right = 7ex and -10ex of AV_proofs] (AV) {
     \AvRuleLocale(};
  \node[parambox,right = 0ex of AV] (AV1) {
     $\Decode$};
  \node[parambox,right = 0ex of AV1] (AV_End) {
    )};
  \node[text=gray,parambox,right = 1ex of AV_End] (AV_Defs) {
    [};
  \node[text=gray,parambox,right = 0ex of AV_Defs] (AV_Defs1) {
    $\ProgramDomain$,};
  \node[text=gray,parambox,right = 1ex of AV_Defs1] (AV_Defs2) {
    $\SymbolTable$};
  \node[text=gray,parambox,right = 0ex of AV_Defs2] (AV_Defs_End) {
    ]};
    
  \path[draw=gray,text=gray,->] (find_symbol1) to node[above,sloped]{\small inherit} (BIL_inf1);  
  \path[draw=gray,text=gray,->] (find_symbol2) to node[above,sloped]{\small inherit} (BIL_inf2); 
  \path[draw=gray,text=gray,dashed,->] (find_symbol3) to node[right]{\small uses} (FindSymbolStep);
  \path[draw=gray,text=gray,->] (FindSymbolStep) to node[right]{\small override} (BIL_inf3);  

  \path[draw=gray,text=gray,->] (AV_proofs1.north west) to node[below,sloped]{\small inherit} (find_symbol1.south);  
  \path[draw=gray,text=gray,->] (AV_proofs1) to node[below,sloped]{\small inherit} (AV1);  

  \path[draw=gray,text=gray,->,dashed] (AV_Defs1)
        to node[above,sloped]{\small provide} (AV_proofs_Defs1);

  \path[draw=gray,text=gray,->,dashed] (AV_Defs2)
        to node[above,sloped]{\small provide} (AV_proofs_Defs2);

  \path[draw=gray,text=gray,->] (AV_proofs_Defs1) 
        to node[above,sloped] {\small override} (find_symbol2);

  \path[draw=gray,text=gray,->] (AV_proofs_Defs2) 
        to node[above,sloped] {\small override} (find_symbol3);
 
\end{tikzpicture}}
  \caption{Instantiation of the \FindSymbolLocale.}
  \label{fig:find-symbol-params}
\end{minipage}
\hfill
\begin{minipage}{0.55\textwidth}
  \resizebox{\textwidth}{!}{\begin{tikzpicture}

  \node (BIL_inf) {
     \BilInferenceLocale(};
  \node[parambox, right = 0ex of BIL_inf] (BIL_inf1) {
     $\Decode$,};
  \node[parambox, right = 1ex of BIL_inf1] (BIL_inf2) {
     \ProgramDomain,};
  \node[parambox, right = 1ex of BIL_inf2] (BIL_inf3) {
     $\BILSmallStep$};
  \node[parambox, right = 0ex of BIL_inf3] (BIL_inf_End) {
     )};

  \node[parambox,below = 4ex of BIL_inf3] (AllocStep) {
     $\AllocSmallStep$};
  \node[parambox,right = 1ex of AllocStep]  {
     (see \cref{fig:allocation-steps}) };

  \node[parambox,below left = 12ex and 4ex of BIL_inf] (allocation) {
     \AllocationLocale(};
  \node[parambox,right = 0ex of allocation] (allocation1) {
     $\Decode$,};
  \node[parambox,right = 1ex of allocation1] (allocation2) {
     \ProgramDomain,};
  \node[parambox,right = 1ex of allocation2] (allocation3) {
     \NextAddrAllocatorName,};
  \node[parambox,right = 1ex of allocation3] (allocation4) {
     \GetFreedAddrName,};
  \node[parambox,right = 1ex of allocation4] (allocation5) {
     \GetSzName,};
  \node[parambox,right = 1ex of allocation5] (allocation6) {
     \FreePredicateName,};
  \node[parambox,right = 1ex of allocation6] (allocation7) {
     \AllocPredicateName};
  \node[parambox,right = 0ex of allocation7] (allocation_End) {
     )};

  \node[parambox, below = 3ex of allocation6, xshift=-3ex] (Free) {\SymbolTable[free]};
  \node[parambox, below = 7ex of allocation7, xshift=-4ex] (Alloc) {\SymbolTable[malloc]};

  \node[parambox, below = 3ex of allocation4] (X10) {$\Variables$({\tt X10})};

  \node[parambox,below right = 15ex and -12ex of allocation] (DF_proofs) {\DoubleFreeProofLocale(};
  \node[parambox,right = 0ex of DF_proofs] (DF_proofs1) {$\Decode$,};
  \node[parambox,right = 1ex of DF_proofs1] (DF_proofs2) {\NextAddrAllocatorName,};
  \node[parambox,right = 1ex of DF_proofs2] (DF_proofs3) {...};
  \node[parambox,right = 0ex of DF_proofs3] (DF_proofsEnd) {)};
  \node[text=gray,parambox,right = 1ex of DF_proofsEnd] (DF_proofs_Defs) {
    [};
  \node[text=gray,parambox,right = 0ex of DF_proofs_Defs] (DF_proofs_Defs1) {
    $\ProgramDomain$,};
  \node[text=gray,parambox,right = 1ex of DF_proofs_Defs1] (DF_proofs_Defs2) {
    $\SymbolTable$};
  \node[text=gray,parambox,right = 0ex of DF_proofs_Defs2] (DF_proofs_Defs_End) {
    ]};

  \node[parambox,below = 7ex of DF_proofs,xshift=8ex] (DF) {
     \DoubleFreeBinaryLocale(};
  \node[parambox,right = 0ex of DF] (DF1) {
     $\Decode$};
  \node[parambox,right = 0ex of DF1] (DFEnd) {)};
  \node[text=gray,parambox,right = 1ex of DFEnd] (DF_Defs) {
    [};
  \node[text=gray,parambox,right = 0ex of DF_Defs] (DF_Defs1) {
    $\ProgramDomain$,};
  \node[text=gray,parambox,right = 1ex of DF_Defs1] (DF_Defs2) {
    $\SymbolTable$};
  \node[text=gray,parambox,right = 0ex of DF_Defs2] (DF_Defs_End) {
    ]};

  \path[draw=gray,text=gray,->] (allocation1) to node[sloped,above]{\small inherit} (BIL_inf1);
  \path[draw=gray,text=gray,->] (allocation2) to node[sloped,above]{\small inherit} (BIL_inf2);
  \path[draw=gray,text=gray,dashed,->] (allocation3) 
    to ($(AllocStep)-(0ex,4ex)$)
    to node[right]{\small uses} (AllocStep);
  \path[draw=gray,text=gray,dashed,-] (allocation4) to ($(AllocStep)-(0ex,4ex)$);
  \path[draw=gray,text=gray,dashed,-] (allocation5) to ($(AllocStep)-(0ex,4ex)$);
  \path[draw=gray,text=gray,dashed,-] (allocation6) to ($(AllocStep)-(0ex,4ex)$);
  \path[draw=gray,text=gray,dashed,-] (allocation7) to ($(AllocStep)-(0ex,4ex)$);
  \path[draw=gray,text=gray,->] (AllocStep) to node[right]{\small override} (BIL_inf3);
  \path[draw=gray,text=gray,->] (DF_proofs1) to node[sloped,below]{\small inherit} (allocation1);
  \path[draw=gray,text=gray,->] (DF_proofs1) to node[above, sloped]{\small inherit} (DF1);

  \path[draw=gray,text=gray,->] (DF_proofs2) to node[below, near start, sloped]{\small inherit} (allocation3);


  \path[draw=gray,text=gray,dashed,->] (DF_Defs1) to node[sloped,above]{\small provide} (DF_proofs_Defs1);
  \path[draw=gray,text=gray,dashed,->] (DF_Defs2) to node[sloped,above]{\small provide} (DF_proofs_Defs2);

  \path[draw=gray,text=gray,near start,->] (DF_proofs_Defs1) to node[below, sloped]{\small override} (allocation2);
  \path[draw=gray,text=gray,dashed,->] (DF_proofs_Defs2) to node[above, sloped]{\small uses} (Alloc);
  \path[draw=gray,text=gray,dashed,->] ($(DF_proofs_Defs2.north west)+(2ex,0)$) to node[above, sloped]{\small uses} (Free);

  \path[draw=gray,text=gray,->] (Alloc) to node[right]{\small override} (allocation7);
  \path[draw=gray,text=gray,->] (Free) to node[left]{\small override} (allocation6);

  \path[draw=gray,text=gray,->] (X10) to node[left]{\small override} (allocation4);
  \path[draw=gray,text=gray,->] (X10) to node{} (allocation5);
      
\end{tikzpicture}}
  \caption{Locale instantiation of the \AllocationLocale examples.}
  \label{fig:alloc-params}
\end{minipage}

\end{figure}

In \cref{sec:overview-and-motivation}, we provided a high-level overview of the 
locale structure. 
Throughout this paper, we have progressively introduced parameters and assumptions 
to these locales, thereby increasing the complexity of our locale hierarchy. 
\cref{fig:bil-inference-params,fig:find-symbol-params,fig:alloc-params,fig:realloc-params} offer a more detailed view of how the locale parameters are instantiated. 
We discuss these instantiations in the text, including the process of discharging 
the assumptions placed upon parameters of these locales.

We begin with \IsaBIL's core framework, which encompasses all the locales given 
in \cref{fig:bil-inference-params}. 
The \CorrectnessLocale, \IncorrectnessLocale and \InferenceLocale locales 
(see \cref{sec:incorrectness}) are all parameterised by big-step semantics ($\BigStep$).
The \InferenceLocale locale extends both \CorrectnessLocale and \IncorrectnessLocale,
ensuring that $\BigStep$ remains consistent. 
Notably, none of these locales impose any assumptions on $\BigStep$.

The \BilLocale locale is parameterised by the decode predicate ($\Decode$)
and introduces a key assumption: that ($\Decode$) is deterministic. 
Formally, we state:
\begin{assumption}[$\Decode$ is deterministic]
\label{lem:bil-deterministic}
For all BIL programs $prog$, any two instructions ($insn_1$, $insn_2$) decoded from the same program must be equal:
\begin{align*}
\forall prog\ insn_1\ insn_2.\
prog \Decode insn_1 \wedge
prog \Decode insn_2 \implies
 insn_1 = insn_2
\end{align*}
\end{assumption}
\cref{lem:bil-deterministic} ensures that decode behaves like a real binary, 
where a given program state yields a unique decoded instruction.
However, this does not preclude self-modifying code, which must modify $prog$ to 
produce a new instruction $insn$.

The \InferenceLocale and \BilLocale locales are unified in \BilInferenceLocale,
which inherits \BilLocale, including $\Decode$.
The \BilInferenceLocale locale is parameterised by an 
auxiliary small-step relation ($\BilSmallStep$) and a set of valid program address 
($\ProgramDomain$). 
Additionally, \BilInferenceLocale defines big-step semantics for BIL (see \cref{sec:big-step-semantics}) using its parameters, and sublocales \InferenceLocale using these big-step semantics.
\BilInferenceLocale adds no additional assumptions beyond \cref{lem:bil-deterministic} 
from \BilLocale.

Next, we explore the remaining locales which refine \BilInferenceLocale for specific proof tasks.

\paragraph{AV Rule Examples}
The AV Rule examples outlined in \cref{sec:forbidden-symbols} belong 
to the same class of proof, employing operational semantics defined in the 
\FindSymbolLocale locale. 
The parameter instantiations for these locales are shown in \cref{fig:find-symbol-params}.
The \FindSymbolLocale locale extends \BilInferenceLocale, inheriting
$\Decode$ and \ProgramDomain while introducing the parameter, \SymbolTable, 
which represents a table of symbols in the binary.
This symbol table is used to define a concrete small-step semantics for 
symbol lookup ($\FindSymbolSmallStep$), as outlined in \cref{sec:forbidden-symbols}.
The relation ($\FindSymbolSmallStep$) is then used to instantiate $\BilSmallStep$, the remaining parameter of \BilInferenceLocale.
Notably, \FindSymbolLocale introduces no additional assumptions.

The program locales (\AvRuleLocale) are auto-generated for each example.
As described in \cref{sec:program-specification}, this process:
\begin{itemize}
  \item fixes the $\Decode$ parameter to the target locale.
  \item adds each instruction as an assumption on $\Decode$.
  \item defines both a valid address set ($\ProgramDomain$) and a symbol table ($\SymbolTable$) for the binary.
\end{itemize} 
The proof locales (\AvRuleProofLocale) extend individual program locales,
inheriting their $\Decode$ parameter and assumptions.
This provides access to $\ProgramDomain$ and $\SymbolTable$,
which are then used to instantiate the corresponding parameters in \FindSymbolLocale, 
including its $\Decode$ parameter.
Additional parameters and assumptions may be introduced within the proof locales, depending on the specific example. These are represented as (...) in \cref{sec:forbidden-symbols} and discussed more generally in \cref{{sec:runtime-assumptions}}.

\paragraph{Double Free Examples}
\cref{fig:alloc-params} details the instantiation of the double free 
locale parameters.
The double free examples from \cref{sec:double-free} use the
allocation model defined in the \AllocationLocale locale.

The \AllocationLocale locale inherits $\Decode$ and \ProgramDomain\xspace from 
the \BilInferenceLocale locale directly, while introducing various parameters
used in the allocation semantics, namely
\NextAddrAllocatorName, \GetFreedAddrName, \GetSzName, \FreePredicateName and
\AllocPredicateName, 
the purpose of these parameters are explained in \cref{sec:double-free}.
From these parameters, \AllocationLocale assumes that a program state cannot 
both free and allocated a pointer:

\begin{assumption}[Free and Alloc operations are mutually exclusive]
\begin{align*}
\neg\AllocPredicate{x} \vee \neg \FreePredicate{x}      
\end{align*}
\end{assumption}

These parameters are used to define the auxiliary small-step $\AllocSmallStep$
for tracking pointer allocations, which then overrides $\BilSmallStep$.
The operational semantics of $\AllocSmallStep$ are given 
in~\cref{fig:allocation-steps}.

Program locales (\DoubleFreeBinaryLocale) are auto-generated for 
the good and bad examples, this process is the same as the AV Rules
explained in the section above.

The proof locales (\DoubleFreeProofLocale) inherit the decode 
predicate and assumptions from both \AllocationLocale and \DoubleFreeBinaryLocale,
adding additional assumptions related to the decoding of \code{malloc} and \code{free} PLT calls:

\begin{assumption}[Decode predicate for \code{malloc} and \code{free} calls]
\label{asm:free-malloc-decode}
Instead of providing a specific implementation for \code{malloc} and \code{free}, we 
provide a minimal abstraction useful for our proofs.

For a given call to \code{malloc}, the program will simply assign a new pointer to the
functions return value stored in \code{X10}, before jumping to the return address stored in \code{X1}.
{\small%
\begin{align*}
\BILDecodeInstructionE{
  \BILMachineStateE{\Variables}{\SymbolTable[\text{``malloc''}]}{mem}}{
  \llparenthesis~ &\textbf{addr} = \SymbolTable[\text{``malloc''}], 
    \textbf{size} = \WordE{4}{64},\\
    &\textbf{code} = [\code{X10} := \NextAddrAllocator,~\StatementJmpE{\code{X1}}]~\rrparenthesis
}
\end{align*}}

For a given call to \code{free}, the program will simply jump to the return address stored
in \code{X1}.
{\small%
\begin{align*}
\BILDecodeInstructionE{
  \BILMachineStateE{\Variables}{\SymbolTable[\text{``free''}]}{mem}}{
  \ &\InsnExpandedE{\SymbolTable[\text{``free''}]}{\WordE{4}{64}}{[\StatementJmpE{\code{X1}}]]}
}
\end{align*}}
\end{assumption}

\cref{asm:free-malloc-decode} provides an approximate abstraction of \code{malloc} and \code{free}, while additional external calls (e.g., \code{printf}) are handled similarly.

To instantiate the \AllocPredicateName and \FreePredicateName predicates, we retrieve the program addresses for \code{malloc} and \code{free} from the \SymbolTable.
The \GetFreedAddrName and \GetSzName retrieve the value in register \code{X10}, which holds the first parameter of each function call (for \code{malloc}, this is the requested size; for \code{free}, it is the pointer to be freed).
The \NextAddrAllocatorName function remains as a parameter without additional assumptions.
For further details on the parameters instantiations, see \cref{sec:double-free}.
Similar to the \FindSymbolLocale examples, the proof locales may introduce additional parameters and assumptions, as discussed in \cref{{sec:runtime-assumptions}}.

\paragraph{cURL Examples}
\cref{fig:realloc-params} outlines the locale structure of the cURL case studies discussed in \cref{sec:double-free-curl}. 
These proofs introduce a reallocation model, which extends the allocation model from the double-free examples by adding an additional predicate: \ReallocPredicate\xspace.

This predicate identifies when a pointer is reallocated and is instantiated similarly to \AllocPredicateName and \FreePredicateName, using the symbol table and the \code{realloc} function provided by the auto-generated program locale (\LibCurlLocale).
We then construct proof locales for each function outlined in \cref{fig:curl-callgraph} and verify their incorrectness compositionally.

\subsection{Discharging Runtime Parameters}
\label{sec:runtime-assumptions}

IsaBIL's symbolic execution reveals values that are only determined at runtime,
indicated as (...) parameters in \cref{fig:find-symbol-params,fig:alloc-params,fig:realloc-params}. In RISC-V, these include the stack pointer (\code{stack}), frame pointer (\code{frame}), return address (\code{ret}), and the global pointer (\code{global}) for global variables stored in memory. We assume all of these are valid 64-bit words, meaning they fall within the range $[0,264)$. 
Additionally, to ensure termination, we assume that the return address does not belong to the program's address set (\code{ret} $\notin$ \ProgramDomain).

Each PLT stub contains a dynamically resolved jump address, which points to the target function in memory at runtime for external calls. To handle these external calls symbolically, we retain the decode predicate, allowing us to introduce additional decode rules that mock these calls. For our proofs, this is sufficient; however, if the prover needed to directly call an external function, the process is detailed in \cref{sec:program-specification}.

Lastly, recall that storage in BIL ($\ValueMemory$) is a recursive concatenation of writes on $\InitialStorage$ with the last write being $\ValueUnknownE{str}{\TypeMemory}$. Since we do not assume a specific value for $str$ in our proofs, we fix it as a locale parameter.

For example, the \AvRuleProofLocaleE{19} locale is defined as:
\begin{align*}
\AvRuleProofLocaleE{19} = (\Decode, stack, frame, ret, global, setlocale\_jmp, str)    
\end{align*}
Where $setlocale\_jmp$ is the dynamically loaded address of the \code{setlocale} function.
Since there are no restrictions on the stack pointer, we introduce reasonable assumptions to ensure a sound proof. 
Namely, we assume that the memory reserved for the stack is distinct from the PLT, global variables, and other memory segments, ensuring no unintended overlaps. 
In the case of 64-bit words, two addresses $w_1$ and $w_2$ do not overlap if:
\begin{align*}
(w_1 + 8) < w_2 \vee (w_2 + 8) < w_1    
\end{align*}
For \AvRuleProofLocaleE{19}, this means that the frame and return pointers — stored at addresses $stack - 16$ and $stack - 8$, respectively — must not overlap with the address of the PLT table entry for $setlocale\_jmp$ (65562). Although these assumptions are not generated automatically, their impact on the automated solver makes them easy to identify in the proof. That is, if an overlap exists, the solver will generate a proof obligation, requiring the human prover to verify whether the given addresses conflict with 65562.

Introducing such assumptions poses the risk of inconsistency — where an instantiation of a locale cannot satisfy all constraints (e.g., assuming both $x=5$ and $x=7$). To ensure soundness, we fully instantiate all proof locales.
For example, a valid instantiation of \AvRuleProofLocaleE{19} is:
\begin{align*}
\AvRuleProofLocaleE{19} = (\Decode_{19},128,256,512,1024,2048, ``hello")
\end{align*}
Where $\Decode_{19}$ is an inductive predicate that satisfies all the required decode instruction assumptions,
including \cref{lem:bil-deterministic}. 128 is the stack pointer, 256 is the frame pointer, 512 is the return address, 1024 is the global pointer, 2048 is $setlocale\_jmp$, and \emph{``hello''} is the arbitrary string for the base program memory ($\ValueUnknownE{str}{\TypeMemory}$).

Finally, we must discharge the locale's assumptions using these values. Since all provided values are valid 64-bit words (i.e., less than $2^{64}$), and $setlocale\_jmp = 2048$ does not overlap with $stack = 128$, the instantiation is valid.

\section{Handling External Binaries}
\label{appendix:external-binary}

Binaries often include links to external code, typically in the form of function calls. 
Accurately modeling these external calls is crucial for understanding the behavior of the original binary. 
\IsaBIL provides two methods for handling externally linked code, depending on the availability of that external code.
If the external binary is available, an Isabelle/HOL locale can be generated for the external binary, 
which is then combined with the existing binary locale through inheritance. 
If the external binary is not available, we can make assumptions about its behavior 
and define an approximate implementation using locale assumptions.
We give an example for both cases in \cref{appendix:external-binary}.

\begin{example} \label{ex:linked}
    
Consider \code{Binary\_A} from \cref{ex:binary-a}, which now calls external functions \code{Fun\_B} and 
\code{Fun\_C} from \code{Binary\_B} and \code{Binary\_C}, respectively.
We have access to \code{Binary\_B}, but \code{Binary\_C} has been lost.
To resolve these external dependencies, we create a new locale, \code{Binary\_ABC}, as follows:

\begin{lstlisting}[language={Isabelle},escapeinside={(*}{*)},basicstyle=\tt\scriptsize,label=lst:binary-concat]
locale Binary_ABC = Binary_A decode + Binary_B decode (*\label{line:localeAandB}*)
  for decode :: (*\guilsinglleft*)(var (*$\Rightarrow$*) val option) (*$\times$*) word (*$\times$*) var (*$\Rightarrow$*) insn (*$\Rightarrow$*) bool(*\guilsinglright*) 
+
  assumes decode_Fun_C: (*\guilsinglleft*)((*$\Delta$*), 0x42, mem) $\DecodeProg$ $\llpar$addr = 0x42 :: 64, size = 4, (*\label{line:assume-1}*)
                 bil = [X8 := X8 * (5 :: 64), jmp X1]$\rrpar$(*\guilsinglright*)  (*\label{line:assume-2}*)
\end{lstlisting}
Since \code{Binary\_B} is available, we generate a program locale for it.
In \cref{line:localeAandB}, our new locale \code{Binary\_ABC} inherits both \code{Binary\_A} and \code{Binary\_B} program locales.
As \code{Binary\_C} is unavailable, we cannot generate a program locale for it. 
However, we know that \code{Fun\_C} multiplies register \code{X8} by five and then jumps to the return address stored in \code{X1}. 
We also know that this function resides at address \code{0x42}. 
Thus, we approximate the behavior of \code{Fun\_C} by fixing an assumption to \code{Binary\_ABC} that decodes address \code{0x42} 
to an BIL sequence that describes the behaviour above, as shown in \cref{line:assume-1,line:assume-2}
\end{example}
While this is a simple example, additional \code{decode} assumptions can be added to the locale to model more complex behaviors. 
Both techniques are further demonstrated in \cref{sec:compositionality,sec:example-proofs,sec:furth-deta-example}.

\section{Translation of RISC-V program to BIL}
\label{sec:riscv-to-bil}

Translation of a RISC-V program (left) to BIL (right) with BAP, using the RISC-V variable definitions from \cref{sec:riscv-optimisations} to improve readability.

  \begin{mdframed}[backgroundcolor=backcolour,hidealllines=true,%
                   innerleftmargin=0,innerrightmargin=0,
                   innertopmargin=-0.2cm,innerbottommargin=-0.2cm]
  \begin{minipage}{0.36\textwidth}
\begin{lstlisting}[numbers=none,language=riscv]
105a4: addi sp, sp, -32
105a6: sd ra, 24(sp)
105a8: sd s0, 16(sp)
105aa: addi s0, sp, 32
105ac: lui a5, 16
105ae: addi a1, a5, 1536
105b2: lui a5, 16
105b4: addi a0, a5, 1544
105b8: jal -184
105bc: mv a5, a0
105be: sd a5, -24(s0)
105c2: jal -226
105c6: mv a5, a0
105c8: lw a5, 0(a5)
105ca: mv a1, a5
105cc: lui a5, 16
105ce: addi a0, a5, 1568
105d2: jal -226
105d6: nop
105d8: ld ra, 24(sp)
105da: ld s0, 16(sp)
105dc: addi sp, sp, 32
105de: ret
\end{lstlisting}
  \end{minipage}
  \hfill
  \begin{minipage}{0.63\textwidth}
\begin{lstlisting}[numbers=none,language=bil,escapechar=|]
X2 := X2 + (-32::64)
mem := mem with [X2 + (24::64),el]: 64 |$\leftarrow$| X1
mem := mem with [X2 + (16::64),el]: 64 |$\leftarrow$| X8
X8 := X2 + (32::64)
X15 := (65536::64)
X11 := X15 + (1536::64)
X15 := (65536::64)
X10 := X15 + (1544::64)
X1 := (67004::64), jmp (66816::64)
X15 := X10
mem := mem with [X8 + (-24::64),el]: 64 |$\leftarrow$| X15
X1 := (67014::64), jmp (66784::64)
X15 := X10
X15 := signed:64[mem[X15,el]:32)]
X11 := X15
X15 := (65536::64)
X10 := X15 + (1568::64)
X1 := (67030::64), jmp (66800::64)

X1 := mem[X2 + (24::64),el]:64
X8 := mem[X2 + (16::64),el]:64
X2 := X2 + (32::64)
jmp X1
\end{lstlisting}
  \end{minipage}
  \end{mdframed}

\newpage


\section*{Supplementary material}

**I  BIL MANUAL**

# A formal specification for BIL: BIL Instruction Language

October 5, 2018

## Contents

# 1 Introduction

This document describes the syntax and semantics of BAP Instruction Language. The language is used to represent a semantics of machine instructions. Each machine instruction is represented by a BIL program that captures side effect of the instruction.

# 2 Syntax

## 2.1 Metavariables

We define a small set of metavariables that are used to denote subscripts, numerals and string literals:

| | |
|---|---|
| *index*, *m*, *n* | subscripts |
| *id* | a literal for variable |
| *num* | number literal |
| *string*, *str* | quoted string literal |

## 2.2 BIL syntax

BIL program is reperesented as a sequence of statements. Each statement performs some side-effectful computation.

$$bil,\ seq \quad ::= $$
$$\quad\quad | \quad \{s_1;\ ..\ ;s_n\} \quad \mathsf{S}$$

$$stmt,\ s \quad ::= $$

|   |   |   |   |
|---|---|---|---|
| | $var := exp$ | | – assign $exp$ to $var$ |
| | **jmp** $e$ | | – transfer control to a given address $e$ |
| | **cpuexn** $(num)$ | | – interrupt CPU with a given interrupt $num$ |
| | **special** $(string)$ | | – instruction with unknown semantics |
| | **while** $(exp)\,seq$ | | – eval $seq$ while $exp$ is true |
| | **if** $(e)\,seq$ | $\mathsf{S}$ | – eval $seq$ if $e$ is true |
| | **if** $(e)\,seq_1$ **else** $seq_2$ | | – if $e$ is true then eval $seq_1$ else $seq_2$ |

BIL expressions are side-effect free. Expressions include a usual set of operations on bitvectors, like arithmetic operations and converting bitvectors of one size to bitvectors of another size (casting in BIL parlance). We write $[e_1/var]e_2$ for the capture-avoiding substitution of $e_1$ for free occurances of $var$ in $e_2$

$$exp,\ e \quad ::= $$

|   |   |   |   |
|---|---|---|---|
| | $(exp)$ | $\mathsf{S}$ | |
| | $var$ | | – a variable |
| | $word$ | | – an immediate value |
| | $v[w \leftarrow v' : sz]$ | | – a memory value |
| | $e_1[e_2, endian] : nat$ | | – load a value from address $e_2$ at storage $e_1$ |
| | $e_1\ \textbf{with}\ [e_2, endian] : nat \leftarrow e_3$ | | – update a storage $e_1$ with binding $e_2 \leftarrow e_3$ |
| | $e_1\ bop\ e_2$ | | – perform binary operation on $e_1$ and $e_2$ |
| | $uop\ e_1$ | | – perform an unary operation on $e_1$ |
| | $cast : nat[e]$ | | – extract or extend bitvector $e$ |
| | **let** $var = e_1$ **in** $e_2$ | | – bind $e_1$ to $var$ in expression $e_2$ |
| | **unknown** $[string] : type$ | | – unknown or undefined value of a given $type$ |
| | **ite** $e_1\ e_2\ e_3$ | | – evaluates to $e_2$ if $e_1$ is true else to $e_3$ |

|   **extract** : $nat_1$ : $nat_2[e]$      – extract or extend bitvector $e$
|   $e_1@e_2$      – concatenate two bitvector $e_1$ to $e_2$
|   $[e_1/var]e_2$      M      – the (capture avoiding) substitution of $e_1$ for $var$ in $e_2$

$var$      ::=
     |    $id$ : $type$    S

$bop$      ::=
     |    $aop$      – arithmetic operators
     |    $lop$      – logical operators

$aop$      ::=
     |    $+$      – plus
     |    $-$      – minus
     |    $*$      – times
     |    $/$      – divide
     |    $\overset{signed}{/}$      – signed divide
     |    $\%$      – modulo
     |    $\overset{signed}{\%}$      – signed modulo
     |    $\&$      – bitwise and
     |    $|$      – bitwise or
     |    **xor**      – bitwise xor
     |    $\ll$      – logical shift left
     |    $\gg$      – logical shift right
     |    $\ggg$      – arithmetic shift right

$lop$      ::=
     |    $=$      – equality
     |    $\neq$      – non-equality
     |    $<$      – less than
     |    $\leq$      – less than or equal
     |    $\overset{signed}{<}$      – signed less than
     |    $\overset{signed}{\leq}$      – signed less than or equal

$uop$      ::=
     |    $-$      – unary negation
     |    $\neg$      – bitwise complement

$endian$, $ed$      ::=
     |    **el**      – little endian
     |    **be**      – big endian

$cast$      ::=
     |    **low**      – extract lower bits
     |    **high**      – extract high bits
     |    **signed**      – extend with sign bit
     |    **unsigned**      – extend with zero

3

The type system of BIL consists of two type families - immediate values, indexed by a bitwidth, and storagies (aka memories), indexed with address bitwidth and values bitwidth.

$type,\ t$ ::=

| | $\mathbf{imm} < sz >$ | | – immediate of size $sz$ |
| | $\mathbf{mem} < sz_1, sz_2 >$ | | – memory with address size $sz_1$ and element size $sz_2$ |
| | $type(v)$ | M | a function that computes the type of a value |

## 2.3 Bitvector syntax

We define a type of *words*, which are concrete bitvectors represented by a pair of value and size. Many operations on words are required by the semantics and are listed below.

Operations marked with `sbv` are signed. All other operations are unsigned (if it does matter). Operations `ext` and `exts` performs extract/extend operation. The former is unsigned (i.e., it extends with zeros), the latter is signed. This operation extracts bits from a bitvector starting from $hi$ and ending with $lo$ bit (both ends included). If $hi$ is greater than the bitwidth of the bitvector, then it is extended with zeros (for `ext` operation) or with a sign bit (for `exts`) operation.

$word,\ w$ ::=

| | $num : nat$ | M | |
| | $(w)$ | S | |
| | $1 : nat$ | S | |
| | $\mathbf{true}$ | S | – sugar for 1:1 |
| | $\mathbf{false}$ | S | – sugar for 0:1 |
| | $w_1 \overset{bv}{+} w_2$ | S | – plus |
| | $w_1 \overset{bv}{-} w_2$ | S | – minus |
| | $w_1 \overset{bv}{*} w_2$ | S | – times |
| | $w_1 \overset{bv}{/} w_2$ | S | – division |
| | $w_1 \overset{sbv}{/} w_2$ | S | – signed division |
| | $w_1 \overset{bv}{\%} w_2$ | S | – modulo |
| | $w_1 \overset{sbv}{\%} w_2$ | S | – signed modulo |
| | $w_1 \overset{bv}{\ll} w_2$ | S | – logical shift left |
| | $w_1 \overset{bv}{\gg} w_2$ | S | – logical shift right |
| | $w_1 \overset{bv}{\ggg} w_2$ | S | – arithmetic shift right |
| | $w_1 \overset{bv}{\&} w_2$ | S | – bitwise and |
| | $w_1 \overset{bv}{\mid} w_2$ | S | – bitwise or |
| | $w_1 \overset{bv}{xor} w_2$ | S | – bitwise xor |
| | $w_1 \overset{bv}{<} w_2$ | S | – less than |
| | $w_1 \overset{sbv}{<} w_2$ | S | – signed less than |
| | $\overset{bv}{-} w$ | S | – integer negation |
| | $\overset{bv}{\sim} w$ | S | – logical negation |
| | $w_1 \overset{bv}{\cdot} w_2$ | S | – concatenation |
| | $\mathbf{ext}\ w \sim \mathbf{hi} : sz_1 \sim \mathbf{lo} : sz_2$ | S | – extract/extend |
| | $\mathbf{exts}\ w \sim \mathbf{hi} : sz_1 \sim \mathbf{lo} : sz_2$ | S | – signed extract/extend |

## 2.4  Value syntax

Values are syntactic subset of expressions. They are used to represent expressions that are not reducible.

We have three kinds of values — immediates, represented as bitvectors; unknown values and storages (memories in BIL parlance), represented symbolically as a list of assignments:

$val$, $v$      ::=
|  $word$
|  $v[w \leftarrow v' : sz]$
|  **unknown** $[string] : type$

## 2.5  Context syntax

Contexts are used in the typing judgments to specify the types of all variables. While each variable is annotated with its type, the context ensures that all uses of a given variable have the same type.

$\Gamma$      ::=
|  $[]$         – empty
|  $\Gamma, var$      – extend
|  $[var]$      S    – singleton list
|  $(\Gamma)$      S
|  $\mathsf{dom}(\Delta)$      M    – domain of a runtime binding context

## 2.6  Formula syntax

The following syntax is used to specify symbolic formulas in premises of judgments.

We use $\Delta$ to denote set of bindings of variables to values. The $\Delta$ context is represented as list of pairs. We write $(var, v) \in \Delta$ to indicate that the value $v$ is the right-most binding of $var$ in $\Delta$. Additionally, we write $\mathsf{dom}(\Delta)$ for $\Delta$'s *domain* (the set of variables for which it contains values).

We also add a small set of operations over natural numbers, like comparison and arithmetics. Natural numbers are mostly used to reason about sizes of bitvectors, that's why they are often referred as $sz$.

We also add syntax for equality comparison for values and variables.

$\Delta$      ::=
|  $[]$             – empty
|  $\Delta[var \leftarrow val]$      – extend

$formula$      ::=
|  $judgement$
|  $(formula)$      M
|  $v_1 \neq v_2$      M
|  $var_1 \neq var_2$      M
|  $w_1 <> w_2$      M
|  $nat_1 > nat_2$      M
|  $nat_1 = nat_2$      M
|  $nat_1 >= nat_2$      M
|  $nat_1 \% sz = 0$      M
|  $t_1 = t_2$      M
|  $e_1 \overset{\mathrm{def}}{:=} e_2$      M

| | $(var, val) \in \Delta$ | M |
| | $var \notin \mathsf{dom}(\Delta)$ | M |
| | $var \in \Gamma$ | M |
| | $id \notin \mathsf{dom}(\Gamma)$ | M |

$nat,\ sz$  ::=

| | | 0 | M |
| | | 1 | M |
| | | 8 | M |
| | | $nat_1 + nat_2$ | M |
| | | $nat_1 - nat_2$ | M |
| | | $(nat)$ | M |

## 2.7  Instruction syntax

To reason about the whole program we introduce a syntax for instruction. An instruction is a binary sequence of $w_2$ bytes, that was read by a decoder from an address $w_1$. The semantics of an instruction is described by the $bil$ program.

$insn$     ::=

| | | $\{\mathbf{addr} = w_1; \mathbf{size} = w2; \mathbf{code} = bil\}$ | S |

# 3 Typing

This section defines typing rules for BIL programs. We define four judgements: $\Gamma \vdash bil\,\textbf{is ok}$ for programs, $\Gamma \vdash s\,\textbf{is ok}$ for statements, $\Gamma \vdash e :: t$ for expressions, and $\Gamma\,\textbf{is ok}$ for contexts.

BIL statement-level variables represent global state, and are implicitly declared at their first use. While variables carry their type with them, it is still necessary to track them in a context during type checking. This is required to rule out programs like:

```
if (foo) then {x:imm<1> = 0} else {x:imm<32> = 42};
bar
```

Such a program would leave the type associated with `x` unclear in `bar`. This is ruled out because the typing rules are set up such that $\Gamma \vdash bil\,\textbf{is ok}$ implies $\Gamma\,\textbf{is ok}$, which requires that each variable has a single type.

$\boxed{\Gamma \vdash bil\,\textbf{is ok}}$

$$\frac{\Gamma \vdash stmt\,\textbf{is ok}}{\Gamma \vdash \{stmt\}\,\textbf{is ok}}\quad \text{T\_SEQ\_ONE}$$

$$\frac{\begin{array}{c}\Gamma \vdash s_1\,\textbf{is ok}\\ \Gamma \vdash \{s_2;\,..\,;s_n\}\,\textbf{is ok}\end{array}}{\Gamma \vdash \{s_1;\,s_2;\,..\,;s_n\}\,\textbf{is ok}}\quad \text{T\_SEQ\_REC}$$

$\boxed{\Gamma \vdash stmt\,\textbf{is ok}}$

$$\frac{\begin{array}{c}\Gamma \vdash var :: t\\ \Gamma \vdash exp :: t\end{array}}{\Gamma \vdash var := exp\,\textbf{is ok}}\quad \text{T\_MOVE}$$

$$\frac{\Gamma \vdash exp :: \textbf{imm} < nat >}{\Gamma \vdash \textbf{jmp}\,exp\,\textbf{is ok}}\quad \text{T\_JMP}$$

$$\frac{\Gamma\,\textbf{is ok}}{\Gamma \vdash \textbf{cpuexn}\,(num)\,\textbf{is ok}}\quad \text{T\_CPUEXN}$$

$$\frac{\Gamma\,\textbf{is ok}}{\Gamma \vdash \textbf{special}\,(str)\,\textbf{is ok}}\quad \text{T\_SPECIAL}$$

$$\frac{\begin{array}{c}\Gamma \vdash e :: \textbf{imm} < 1 >\\ \Gamma \vdash seq\,\textbf{is ok}\end{array}}{\Gamma \vdash \textbf{while}\,(e)seq\,\textbf{is ok}}\quad \text{T\_WHILE}$$

$$\frac{\begin{array}{c}\Gamma \vdash e :: \textbf{imm} < 1 >\\ \Gamma \vdash seq\,\textbf{is ok}\end{array}}{\Gamma \vdash \textbf{if}\,(e)seq\,\textbf{is ok}}\quad \text{T\_IFTHEN}$$

$$\frac{\begin{array}{c}\Gamma \vdash e :: \textbf{imm} < 1 >\\ \Gamma \vdash seq_1\,\textbf{is ok}\\ \Gamma \vdash seq_2\,\textbf{is ok}\end{array}}{\Gamma \vdash \textbf{if}\,(e)seq_1\,\textbf{else}\,seq_2\,\textbf{is ok}}\quad \text{T\_IF}$$

$\boxed{\Gamma \vdash exp :: type}$

$$\frac{\begin{array}{c}id : t \in \Gamma\\ \Gamma\,\textbf{is ok}\end{array}}{\Gamma \vdash id : t :: t}\quad \text{T\_VAR}$$

$$\frac{\begin{array}{c} sz > 0 \\ \Gamma \textbf{ is ok} \end{array}}{\Gamma \vdash num : sz :: \textbf{imm} < sz >} \quad \text{T\_INT}$$

$$\frac{\begin{array}{c} sz > 0 \\ nat > 0 \\ \Gamma \vdash v :: \textbf{mem} < nat, sz > \\ \Gamma \vdash v' :: \textbf{imm} < sz > \end{array}}{\Gamma \vdash v[num_1 : nat \leftarrow v' : sz] :: \textbf{mem} < nat, sz >} \quad \text{T\_MEM}$$

$$\frac{\begin{array}{c} sz'\%sz = 0 \\ sz' > 0 \\ \Gamma \vdash e_1 :: \textbf{mem} < nat, sz > \\ \Gamma \vdash e_2 :: \textbf{imm} < nat > \end{array}}{\Gamma \vdash e_1[e_2, ed] : sz' :: \textbf{imm} < sz' >} \quad \text{T\_LOAD}$$

$$\frac{\begin{array}{c} sz'\%sz = 0 \\ sz' > 0 \\ \Gamma \vdash e_1 :: \textbf{mem} < nat, sz > \\ \Gamma \vdash e_2 :: \textbf{imm} < nat > \\ \Gamma \vdash e_3 :: \textbf{imm} < sz' > \end{array}}{\Gamma \vdash e_1 \textbf{ with } [e_2, ed] : sz' \leftarrow e_3 :: \textbf{mem} < nat, sz >} \quad \text{T\_STORE}$$

$$\frac{\begin{array}{c} \Gamma \vdash e_1 :: \textbf{imm} < sz > \\ \Gamma \vdash e_2 :: \textbf{imm} < sz > \end{array}}{\Gamma \vdash e_1 \; aop \; e_2 :: \textbf{imm} < sz >} \quad \text{T\_AOP}$$

$$\frac{\begin{array}{c} \Gamma \vdash e_1 :: \textbf{imm} < sz > \\ \Gamma \vdash e_2 :: \textbf{imm} < sz > \end{array}}{\Gamma \vdash e_1 \; lop \; e_2 :: \textbf{imm} < 1 >} \quad \text{T\_LOP}$$

$$\frac{\Gamma \vdash e_1 :: \textbf{imm} < sz >}{\Gamma \vdash uop \; e_1 :: \textbf{imm} < sz >} \quad \text{T\_UOP}$$

$$\frac{\begin{array}{c} sz > 0 \\ sz >= nat \\ \Gamma \vdash e :: \textbf{imm} < nat > \end{array}}{\Gamma \vdash widen\_cast : sz[e] :: \textbf{imm} < sz >} \quad \text{T\_CAST\_WIDEN}$$

$$\frac{\begin{array}{c} sz > 0 \\ nat >= sz \\ \Gamma \vdash e :: \textbf{imm} < nat > \end{array}}{\Gamma \vdash narrow\_cast : sz[e] :: \textbf{imm} < sz >} \quad \text{T\_CAST\_NARROW}$$

$$\frac{\begin{array}{c} \Gamma \vdash e_1 :: t \\ \Gamma, id : t \vdash e_2 :: t' \end{array}}{\Gamma \vdash \textbf{let } id : t = e_1 \textbf{ in } e_2 :: t'} \quad \text{T\_LET}$$

$$\frac{\begin{array}{c} t \textbf{ is ok} \\ \Gamma \textbf{ is ok} \end{array}}{\Gamma \vdash \textbf{unknown} [str] : t :: t} \quad \text{T\_UNKNOWN}$$

$$\frac{\begin{array}{c} \Gamma \vdash e_1 :: \textbf{imm} < 1 > \\ \Gamma \vdash e_2 :: t \\ \Gamma \vdash e_3 :: t \end{array}}{\Gamma \vdash \textbf{ite } e_1 \; e_2 \; e_3 :: t} \quad \text{T\_ITE}$$

$$\frac{\begin{array}{c} \Gamma \vdash e :: \mathbf{imm} <sz> \\ sz_1 >= sz_2 \end{array}}{\Gamma \vdash \mathbf{extract} : sz_1 : sz_2[e] :: \mathbf{imm} <sz_1 - sz_2 + 1>} \ \text{\small T\_EXTRACT}$$

$$\frac{\begin{array}{c} \Gamma \vdash e_1 :: \mathbf{imm} <sz_1> \\ \Gamma \vdash e_2 :: \mathbf{imm} <sz_2> \end{array}}{\Gamma \vdash e_1 @ e_2 :: \mathbf{imm} <sz_1 + sz_2>} \ \text{\small T\_CONCAT}$$

$\boxed{t \, \mathbf{is \, ok}}$

$$\frac{sz > 0}{\mathbf{imm} <sz> \, \mathbf{is \, ok}} \ \text{\small TWF\_IMM}$$

$$\frac{\begin{array}{c} nat > 0 \\ sz > 0 \end{array}}{\mathbf{mem} <nat, sz> \, \mathbf{is \, ok}} \ \text{\small TWF\_MEM}$$

$\boxed{\Gamma \, \mathbf{is \, ok}}$

$$\frac{}{[] \, \mathbf{is \, ok}} \ \text{\small TG\_NIL}$$

$$\frac{\begin{array}{c} id \notin \mathsf{dom}(\Gamma) \\ t \, \mathbf{is \, ok} \\ \Gamma \, \mathbf{is \, ok} \end{array}}{(\Gamma, id : t) \, \mathbf{is \, ok}} \ \text{\small TG\_CONS}$$

# 4 Operational semantics

## 4.1 Model of a program

Program is coinductively defined as an infinite stream of program states, produced by a step rule. Each state is represented with a triplet $(\Delta, w, var)$, where $\Delta$ is a mapping from variables to values, $w$ is a program counter, and $var$ is a variable denoting currently active memory.

The **step** rule defines how a machine instruction is evaluated. We use "magic" rule `decode` that fetches instructions from the memory and decodes them to a BIL program.

The BIL code is evaluated using reduction rules of statements (see section 5). Then the program counter is updated with the $w_3$, that initially points to a byte following current instruction.

$$\boxed{\Delta, w, var \rightsquigarrow \Delta', w', var'}$$

$$\frac{delta, w, var \mapsto \{\, \mathbf{addr} = w_1;\ \mathbf{size} = w2;\ \mathbf{code} = bil\,\} \qquad \Delta, w_1 \overset{bv}{+} w_2 \vdash bil \rightsquigarrow \Delta', w_3, \{\,\}}{\Delta, w, var \rightsquigarrow \Delta', w_3, var} \quad \text{STEP}$$

$$\boxed{delta, w, var \mapsto insn}$$

$$\frac{}{delta, w, var \mapsto insn} \quad \text{DECODE}$$

# 5 Semantics of statements

The reduction rule defines transformation of a state for each statement. The state of the reduction rule consists of a pair $(\Delta, w)$, where $\Delta$ is a mapping from variables to values and $w$ is an address of a next instruction.

Two statements affect the state: `Move` statement introduces new $var \leftarrow v$ binding in $\Delta$, and `Jmp` affects program counter.

The `if` and `while` instructions introduce local control flow.

There is no special semantics associated with `special` and `cpuexn` statements.

$$\boxed{\Delta, word \vdash stmt \rightsquigarrow \Delta', word'}$$

$$\frac{\Delta \vdash e \rightsquigarrow^* v}{\Delta, w \vdash var := e \rightsquigarrow \Delta[var \leftarrow v], w} \quad \text{MOVE}$$

$$\frac{\Delta \vdash e \rightsquigarrow^* w'}{\Delta, w \vdash \mathbf{jmp}\, e \rightsquigarrow \Delta, w'} \quad \text{JMP}$$

$$\frac{}{\Delta, w \vdash \mathbf{cpuexn}\,(num) \rightsquigarrow \Delta, w} \quad \text{CPUEXN}$$

$$\frac{}{\Delta, w \vdash \mathbf{special}\,(str) \rightsquigarrow \Delta, w} \quad \text{SPECIAL}$$

$$\frac{\Delta \vdash e \rightsquigarrow^* \mathbf{true} \qquad \Delta, word \vdash seq \rightsquigarrow \Delta', word', \{\,\}}{\Delta, word \vdash \mathbf{if}\,(e) seq \rightsquigarrow \Delta', word'} \quad \text{IFTHEN\_TRUE}$$

$$\frac{\Delta \vdash e \rightsquigarrow^* \mathbf{true} \qquad \Delta, word \vdash seq \rightsquigarrow \Delta', word', \{\,\}}{\Delta, word \vdash \mathbf{if}\,(e) seq\, \mathbf{else}\, seq_1 \rightsquigarrow \Delta', word'} \quad \text{IF\_TRUE}$$

$$\frac{\Delta \vdash e \rightsquigarrow^* \mathbf{false} \qquad \Delta, word \vdash seq \rightsquigarrow \Delta', word', \{\,\}}{\Delta, word \vdash \mathbf{if}\,(e) seq_1\, \mathbf{else}\, seq \rightsquigarrow \Delta', word'} \quad \text{IF\_FALSE}$$

$$\dfrac{\begin{array}{l}\Delta_1 \vdash e \leadsto^* \textbf{true} \\ \Delta_1, word_1 \vdash seq \leadsto \Delta_2, word_2, \{\,\} \\ \Delta_2, word_2 \vdash \textbf{while}\,(e)\,seq \leadsto \Delta_3, word_3\end{array}}{\Delta_1, word_1 \vdash \textbf{while}\,(e)\,seq \leadsto \Delta_3, word_3} \quad \text{WHILE}$$

$$\dfrac{\Delta \vdash e \leadsto^* \textbf{false}}{\Delta, word \vdash \textbf{while}\,(e)\,seq \leadsto \Delta, word} \quad \text{WHILE\_FALSE}$$

$$\boxed{\Delta, word \vdash seq \leadsto \Delta', word', seq'}$$

$$\dfrac{\Delta, word \vdash s_1 \leadsto \Delta', word'}{\Delta, word \vdash \{s_1; s_2; \, .. \, ; s_n\} \leadsto \Delta', word', \{s_2; \, .. \, ; s_n\}} \quad \text{SEQ\_REC}$$

$$\dfrac{\Delta, word \vdash s_1 \leadsto \Delta', word'}{\Delta, word \vdash \{s_1; s_2\} \leadsto \Delta', word', \{s_2\}} \quad \text{SEQ\_LAST}$$

$$\dfrac{\Delta, word \vdash s_1 \leadsto \Delta', word'}{\Delta, word \vdash \{s_1\} \leadsto \Delta', word', \{\,\}} \quad \text{SEQ\_ONE}$$

$$\dfrac{}{\Delta, word \vdash \{\,\} \leadsto \Delta, word, \{\,\}} \quad \text{SEQ\_NIL}$$

# 6  Semantics of expressions

This section describes a small step operational semantics for expressions. A symbolic formula $\Delta \vdash e \rightarrow e'$ defines a step of transformation from expression $e$ to an expression $e'$ under given context $\Delta$.

A well formed (well typed) expression evaluates to a value expression, that is syntactic subset of expression grammar (see section 2.4).

A value can be either an immediate, represented by a bitvector, a unknown value, or a memory storage.

A memory storage is represented symbolically as a sequence of storages to the originally undefined memory. Each storage operation of size greater than 8 bits is desugared into a sequence of 8 bit storages in a big endian order.

A load operation will first reduce all sub expressions of a memory object to values and then recursively destruct the object until one of the following conditions is met:

**load-byte:** if the memory object is a storage of a `value` to an immediate (known) address that we're trying to load then the load expression is reduced to `value`.

**load-un-memory:** if the memory object is an `unknown` value, then the load expression evaluates to `unknown`.

**load-un-addr:** if the memory object is a storage to `unknown` value address then load expression evaluates to `unknown`.

This section also defines $\Delta \vdash e_1 \leadsto^* e_2$. This relation is the reflexive, transitive closure of $\leadsto$. It is useful to describe reductions that may take many steps. For example, in the evaluation rules for statements, it is often necessary to evaluate an expression completely to a value. The $\leadsto^*$ relation is used to allow reductions that take an indeterminate number of individual $\leadsto$ steps. Such a derivation can be built with repeated use of the REDUCE rule.

Some evaluation rules depend on the type of a value. Since there are two canonical forms for each type, we avoid duplicating each rule by defining the following metafunction:

$\boxed{type(v)}$ a function that computes the type of a value

$$type(v[num_1 : nat \leftarrow v' : sz]) \equiv \mathbf{mem} < nat, sz >$$
$$type(num : nat) \equiv \mathbf{imm} < nat >$$
$$type(\mathbf{unknown}\,[str] : t) \equiv t$$

$$\boxed{\Delta \vdash exp \rightsquigarrow exp'}$$

$$\frac{(var, v) \in \Delta}{\Delta \vdash var \rightsquigarrow v} \quad \text{VAR\_IN}$$

$$\frac{id : type \notin \mathsf{dom}(\Delta)}{\Delta \vdash id : type \rightsquigarrow \mathbf{unknown}\,[str] : type} \quad \text{VAR\_UNKNOWN}$$

$$\frac{\Delta \vdash e_2 \rightsquigarrow e_2'}{\Delta \vdash e_1[e_2, ed] : sz \rightsquigarrow e_1[e_2', ed] : sz} \quad \text{LOAD\_STEP\_ADDR}$$

$$\frac{\Delta \vdash e_1 \rightsquigarrow e_1'}{\Delta \vdash e_1[v_2, ed] : sz \rightsquigarrow e_1'[v_2, ed] : sz} \quad \text{LOAD\_STEP\_MEM}$$

$$\frac{}{\Delta \vdash v[w \leftarrow v' : sz][w, ed'] : sz \rightsquigarrow v'} \quad \text{LOAD\_BYTE}$$

$$\frac{w_1 \neq w_2}{\Delta \vdash v[w_1 \leftarrow v' : sz][w_2, ed] : sz \rightsquigarrow v[w_2, ed] : sz} \quad \text{LOAD\_BYTE\_FROM\_NEXT}$$

$$\frac{}{\Delta \vdash (\mathbf{unknown}\,[str] : t)[v, ed] : sz \rightsquigarrow \mathbf{unknown}\,[str] : \mathbf{imm} < sz >} \quad \text{LOAD\_UN\_MEM}$$

$$\frac{}{\Delta \vdash (v[w_1 \leftarrow v' : sz])[\mathbf{unknown}\,[str] : t, ed] : sz' \rightsquigarrow \mathbf{unknown}\,[str] : \mathbf{imm} < sz' >} \quad \text{LOAD\_UN\_ADDR}$$

$$\frac{\begin{array}{c} sz' > sz \\ \mathbf{succ}\,w = w' \\ type(v) = \mathbf{mem} < nat, sz > \end{array}}{\Delta \vdash v[w, \mathbf{be}] : sz' \rightsquigarrow v[w, \mathbf{be}] : sz@(v[w', \mathbf{be}] : (sz' - sz))} \quad \text{LOAD\_WORD\_BE}$$

$$\frac{\begin{array}{c} sz' > sz \\ \mathbf{succ}\,w = w' \\ type(v) = \mathbf{mem} < nat, sz > \end{array}}{\Delta \vdash v[w, \mathbf{el}] : sz' \rightsquigarrow v[w', \mathbf{el}] : (sz' - sz)@(v[w, \mathbf{be}] : sz)} \quad \text{LOAD\_WORD\_EL}$$

$$\frac{\Delta \vdash e_3 \rightsquigarrow e_3'}{\Delta \vdash e_1 \,\mathbf{with}\,[e_2, ed] : sz \leftarrow e_3 \rightsquigarrow e_1 \,\mathbf{with}\,[e_2, ed] : sz \leftarrow e_3'} \quad \text{STORE\_STEP\_VAL}$$

$$\frac{\Delta \vdash e_2 \rightsquigarrow e_2'}{\Delta \vdash e_1 \,\mathbf{with}\,[e_2, ed] : sz \leftarrow v_3 \rightsquigarrow e_1 \,\mathbf{with}\,[e_2', ed] : sz \leftarrow v_3} \quad \text{STORE\_STEP\_ADDR}$$

$$\frac{\Delta \vdash e_1 \rightsquigarrow e_1'}{\Delta \vdash e_1 \,\mathbf{with}\,[v_2, ed] : sz \leftarrow v_3 \rightsquigarrow e_1' \,\mathbf{with}\,[v_2, ed] : sz \leftarrow v_3} \quad \text{STORE\_STEP\_MEM}$$

$$\frac{\begin{array}{c} sz' > sz \\ \mathbf{succ}\,w = w' \\ type(v) = \mathbf{mem} < nat, sz > \\ e_1 \stackrel{\text{def}}{:=} (v \,\mathbf{with}\,[w, \mathbf{be}] : sz \leftarrow \mathbf{high} : sz[val]) \end{array}}{\Delta \vdash v \,\mathbf{with}\,[w, \mathbf{be}] : sz' \leftarrow val \rightsquigarrow e_1 \,\mathbf{with}\,[w', \mathbf{be}] : (sz' - sz) \leftarrow \mathbf{low} : (sz' - sz)[val]} \quad \text{STORE\_WORD\_BE}$$

$$\frac{\begin{array}{c} sz' > sz \\ \mathbf{succ}\,w = w' \\ type(v) = \mathbf{mem} < nat, sz > \\ e_1 \stackrel{\text{def}}{:=} (v \,\mathbf{with}\,[w, \mathbf{el}] : sz \leftarrow \mathbf{low} : sz[val]) \end{array}}{\Delta \vdash v \,\mathbf{with}\,[w, \mathbf{el}] : sz' \leftarrow val \rightsquigarrow e_1 \,\mathbf{with}\,[w', \mathbf{el}] : (sz' - sz) \leftarrow \mathbf{high} : (sz' - sz)[val]} \quad \text{STORE\_WORD\_EL}$$

$$\frac{type(v) = \mathbf{mem} <nat, sz>}{\Delta \vdash v \,\mathbf{with}\,[w, ed] : sz \leftarrow v' \rightsquigarrow v[w \leftarrow v' : sz]} \quad \text{STORE\_VAL}$$

$$\frac{type(v) = t}{\Delta \vdash v \,\mathbf{with}\,[\mathbf{unknown}\,[str] : t', ed] : sz' \leftarrow v_2 \rightsquigarrow \mathbf{unknown}\,[str] : t} \quad \text{STORE\_UN\_ADDR}$$

$$\frac{\Delta \vdash e_1 \rightsquigarrow e_1'}{\Delta \vdash \mathbf{let}\,var = e_1\,\mathbf{in}\,e_2 \rightsquigarrow \mathbf{let}\,var = e_1'\,\mathbf{in}\,e_2} \quad \text{LET\_STEP}$$

$$\frac{}{\Delta \vdash \mathbf{let}\,var = v\,\mathbf{in}\,e \rightsquigarrow [v/var]e} \quad \text{LET}$$

$$\frac{\Delta \vdash e_1 \rightsquigarrow e_1'}{\Delta \vdash \mathbf{ite}\,e_1\,v_2\,v_3 \rightsquigarrow \mathbf{ite}\,e_1'\,v_2\,v_3} \quad \text{ITE\_STEP\_COND}$$

$$\frac{\Delta \vdash e_2 \rightsquigarrow e_2'}{\Delta \vdash \mathbf{ite}\,e_1\,e_2\,v_3 \rightsquigarrow \mathbf{ite}\,e_1\,e_2'\,v_3} \quad \text{ITE\_STEP\_THEN}$$

$$\frac{\Delta \vdash e_3 \rightsquigarrow e_3'}{\Delta \vdash \mathbf{ite}\,e_1\,e_2\,e_3 \rightsquigarrow \mathbf{ite}\,e_1\,e_2\,e_3'} \quad \text{ITE\_STEP\_ELSE}$$

$$\frac{}{\Delta \vdash \mathbf{ite}\,\mathbf{true}\,v_2\,v_3 \rightsquigarrow v_2} \quad \text{ITE\_TRUE}$$

$$\frac{}{\Delta \vdash \mathbf{ite}\,\mathbf{false}\,v_2\,v_3 \rightsquigarrow v_3} \quad \text{ITE\_FALSE}$$

$$\frac{type(v_2) = t'}{\Delta \vdash \mathbf{ite}\,\mathbf{unknown}\,[str] : t\,v_2\,v_3 \rightsquigarrow \mathbf{unknown}\,[str] : t'} \quad \text{ITE\_UNK}$$

$$\frac{\Delta \vdash e_2 \rightsquigarrow e_2'}{\Delta \vdash v_1\,bop\,e_2 \rightsquigarrow v_1\,bop\,e_2'} \quad \text{BOP\_RHS}$$

$$\frac{\Delta \vdash e_1 \rightsquigarrow e_1'}{\Delta \vdash e_1\,bop\,e_2 \rightsquigarrow e_1'\,bop\,e_2} \quad \text{BOP\_LHS}$$

$$\frac{}{\Delta \vdash e\,aop\,\mathbf{unknown}\,[str] : t \rightsquigarrow \mathbf{unknown}\,[str] : t} \quad \text{AOP\_UNK\_RHS}$$

$$\frac{}{\Delta \vdash \mathbf{unknown}\,[str] : t\,aop\,e \rightsquigarrow \mathbf{unknown}\,[str] : t} \quad \text{AOP\_UNK\_LHS}$$

$$\frac{}{\Delta \vdash e\,lop\,\mathbf{unknown}\,[str] : t \rightsquigarrow \mathbf{unknown}\,[str] : \mathbf{imm} <1>} \quad \text{LOP\_UNK\_RHS}$$

$$\frac{}{\Delta \vdash \mathbf{unknown}\,[str] : t\,lop\,e \rightsquigarrow \mathbf{unknown}\,[str] : \mathbf{imm} <1>} \quad \text{LOP\_UNK\_LHS}$$

$$\frac{}{\Delta \vdash w_1 + w_2 \rightsquigarrow w_1 \overset{bv}{+} w_2} \quad \text{PLUS}$$

$$\frac{}{\Delta \vdash w_1 - w_2 \rightsquigarrow w_1 \overset{bv}{-} w_2} \quad \text{MINUS}$$

$$\frac{}{\Delta \vdash w_1 * w_2 \rightsquigarrow w_1 \overset{bv}{*} w_2} \quad \text{TIMES}$$

$$\frac{}{\Delta \vdash w_1 \,/\, w_2 \rightsquigarrow w_1 \overset{bv}{/} w_2} \quad \text{DIV}$$

$$\frac{}{\Delta \vdash w_1 \overset{signed}{/} w_2 \rightsquigarrow w_1 \overset{sbv}{/} w_2} \quad \text{SDIV}$$

$$\frac{}{\Delta \vdash w_1 \,\%\, w_2 \rightsquigarrow w_1 \overset{bv}{\%} w_2} \quad \text{MOD}$$

$$\frac{}{\Delta \vdash w_1 \overset{signed}{\%} w_2 \rightsquigarrow w_1 \overset{sbv}{\%} w_2} \quad \text{SMOD}$$

$$\frac{}{\Delta \vdash w_1 \ll w_2 \rightsquigarrow w_1 \overset{bv}{\ll} w_2} \quad \text{LSL}$$

$$\frac{}{\Delta \vdash w_1 \gg w_2 \rightsquigarrow w_1 \overset{bv}{\gg} w_2} \quad \text{LSR}$$

$$\frac{}{\Delta \vdash w_1 \ggg w_2 \rightsquigarrow w_1 \overset{bv}{\ggg} w_2} \quad \text{ASR}$$

$$\frac{}{\Delta \vdash w_1 \,\&\, w_2 \rightsquigarrow w_1 \overset{bv}{\&} w_2} \quad \text{LAND}$$

$$\frac{}{\Delta \vdash w_1 \mid w_2 \rightsquigarrow w_1 \overset{bv}{\mid} w_2} \quad \text{LOR}$$

$$\frac{}{\Delta \vdash w_1 \, \textbf{xor} \, w_2 \rightsquigarrow w_1 \overset{bv}{xor} w_2} \quad \text{XOR}$$

$$\frac{}{\Delta \vdash w = w \rightsquigarrow \textbf{true}} \quad \text{EQ\_SAME}$$

$$\frac{w_1 <> w_2}{\Delta \vdash w_1 = w_2 \rightsquigarrow \textbf{false}} \quad \text{EQ\_DIFF}$$

$$\frac{}{\Delta \vdash w \neq w \rightsquigarrow \textbf{false}} \quad \text{NEQ\_SAME}$$

$$\frac{w_1 <> w_2}{\Delta \vdash w_1 \neq w_2 \rightsquigarrow \textbf{true}} \quad \text{NEQ\_DIFF}$$

$$\frac{}{\Delta \vdash w_1 < w_2 \rightsquigarrow w_1 \overset{bv}{<} w_2} \quad \text{LESS}$$

$$\frac{}{\Delta \vdash w_1 \leq w_2 \rightsquigarrow (w_1 < w_2) \mid (w_1 = w_2)} \quad \text{LESS\_EQ}$$

$$\frac{}{\Delta \vdash w_1 \overset{signed}{<} w_2 \rightsquigarrow w_1 \overset{sbv}{<} w_2} \quad \text{SIGNED\_LESS}$$

$$\frac{}{\Delta \vdash w_1 \overset{signed}{\leq} w_2 \rightsquigarrow (w_1 = w_2) \,\&\, (w_1 \overset{signed}{<} w_2)} \quad \text{SIGNED\_LESS\_EQ}$$

$$\frac{\Delta \vdash e \rightsquigarrow e'}{\Delta \vdash uop\ e \rightsquigarrow uop\ e'} \quad \text{UOP}$$

$$\frac{}{\Delta \vdash uop\ \textbf{unknown}\,[str] : t \rightsquigarrow \textbf{unknown}\,[str] : t} \quad \text{UOP\_UNK}$$

$$\frac{}{\Delta \vdash \neg w \overset{bv}{\rightsquigarrow} \sim w} \quad \text{NOT}$$

$$\frac{}{\Delta \vdash - w \overset{bv}{\rightsquigarrow} - w} \quad \text{NEG}$$

$$\frac{\Delta \vdash e_2 \rightsquigarrow e_2'}{\Delta \vdash e_1 @ e_2 \rightsquigarrow e_1 @ e_2'} \quad \text{CONCAT\_RHS}$$

$$\frac{\Delta \vdash e_1 \rightsquigarrow e_1'}{\Delta \vdash e_1 @ v_2 \rightsquigarrow e_1' @ v_2} \quad \text{CONCAT\_LHS}$$

$$\frac{type(v_2) = \textbf{imm} < sz_2 >}{\Delta \vdash \textbf{unknown}\,[str] : \textbf{imm} < sz_1 > @ v_2 \rightsquigarrow \textbf{unknown}\,[str] : \textbf{imm} < sz_1 + sz_2 >} \quad \text{CONCAT\_LHS\_UN}$$

$$\frac{type(v_1) = \mathbf{imm} < sz_1 >}{\Delta \vdash v_1@\mathbf{unknown}\,[str] : \mathbf{imm} < sz_2 > \rightsquigarrow \mathbf{unknown}\,[str] : \mathbf{imm} < sz_1 + sz_2 >} \quad \text{CONCAT\_RHS\_UN}$$

$$\frac{}{\Delta \vdash w_1@w_2 \rightsquigarrow w_1 \overset{bv}{\cdot} w_2} \quad \text{CONCAT}$$

$$\frac{\Delta \vdash e \rightsquigarrow e'}{\Delta \vdash \mathbf{extract} : sz_1 : sz_2[e] \rightsquigarrow \mathbf{extract} : sz_1 : sz_2[e']} \quad \text{EXTRACT\_REDUCE}$$

$$\frac{}{\Delta \vdash \mathbf{extract} : sz_1 : sz_2[\mathbf{unknown}\,[str] : t] \rightsquigarrow \mathbf{unknown}\,[str] : \mathbf{imm} < (sz_1 - sz_2) + 1 >} \quad \text{EXTRACT\_UN}$$

$$\frac{}{\Delta \vdash \mathbf{extract} : sz_1 : sz_2[w] \rightsquigarrow \mathbf{ext}\,w \sim \mathbf{hi} : sz_1 \sim \mathbf{lo} : sz_2} \quad \text{EXTRACT}$$

$$\frac{\Delta \vdash e \rightsquigarrow e'}{\Delta \vdash cast : sz[e] \rightsquigarrow cast : sz[e']} \quad \text{CAST\_REDUCE}$$

$$\frac{}{\Delta \vdash cast : sz[\mathbf{unknown}\,[str] : t] \rightsquigarrow \mathbf{unknown}\,[str] : \mathbf{imm} < sz >} \quad \text{CAST\_UNK}$$

$$\frac{}{\Delta \vdash \mathbf{low} : sz[w] \rightsquigarrow \mathbf{ext}\,w \sim \mathbf{hi} : (sz - 1) \sim \mathbf{lo} : 0} \quad \text{CAST\_LOW}$$

$$\frac{}{\Delta \vdash \mathbf{high} : sz[num : sz'] \rightsquigarrow \mathbf{ext}\,num : sz' \sim \mathbf{hi} : (sz' - 1) \sim \mathbf{lo} : (sz' - sz)} \quad \text{CAST\_HIGH}$$

$$\frac{}{\Delta \vdash \mathbf{signed} : sz[w] \rightsquigarrow \mathbf{exts}\,w \sim \mathbf{hi} : (sz - 1) \sim \mathbf{lo} : 0} \quad \text{CAST\_SIGNED}$$

$$\frac{}{\Delta \vdash \mathbf{unsigned} : sz[w] \rightsquigarrow \mathbf{ext}\,w \sim \mathbf{hi} : (sz - 1) \sim \mathbf{lo} : 0} \quad \text{CAST\_UNSIGNED}$$

$\boxed{\mathbf{succ}\,w_1 = exp}$

$$\frac{}{\mathbf{succ}\,num : sz = num \overset{bv}{+} 1 : sz} \quad \text{SUCC}$$

$\boxed{\Delta \vdash exp \rightsquigarrow^* exp'}$

$$\frac{}{\Delta \vdash e \rightsquigarrow^* e} \quad \text{REFL}$$

$$\frac{\Delta \vdash e_1 \rightsquigarrow e_2 \quad \Delta \vdash e_2 \rightsquigarrow^* e_3}{\Delta \vdash e_1 \rightsquigarrow^* e_3} \quad \text{REDUCE}$$



\end{document}